\documentclass[%
preprint,
amsmath,amssymb,
]{revtex4-1}
\usepackage[normalem]{ulem} 
\usepackage{xcolor, soul} 
\sethlcolor{green}
\usepackage{graphicx}
\usepackage{mathrsfs}
\usepackage{amsmath}
\usepackage{placeins}
\usepackage{epstopdf, epsfig}
\usepackage{subcaption}
\usepackage[export]{adjustbox}
\usepackage[%
colorlinks=true,
pdfborder={0 0 0},
linkcolor=blue,
citecolor=blue,
urlcolor=blue
]{hyperref}

\usepackage{dcolumn}
\usepackage{bm}
\newcommand{\upi}{\pi}

\begin{document}

\title{Local analysis of the clustering, velocities and accelerations of particles settling in turbulence}

	\author{Mohammadreza Momenifar}
	\author{Andrew D. Bragg}
	\email{andrew.bragg@duke.edu}
	\affiliation{Department of Civil and Environmental Engineering, Duke University, Durham, North Carolina 27708, USA}


\begin{abstract}
Using 3D Vorono\text{\"i} analysis, we explore the local dynamics of small, settling, inertial particles in isotropic turbulence using Direct Numerical Simulations (DNS). We independently vary the Taylor Reynolds number $R_\lambda \in[90,398]$, Froude number $Fr\equiv a_\eta/g\in[0.052,\infty]$ (where $a_\eta$ is the Kolmogorov acceleration, and $g$ is the acceleration due to gravity), and Kolmogorov scale Stokes number $St\equiv\tau_p/\tau_\eta\in[0,3]$. In agreement with previous results using global measures of particle clustering, such as the Radial Distribution Function (RDF), we find that for small Vorono\text{\"i} volumes (corresponding to the most clustered particles), the behavior is strongly dependent upon $St$ and $Fr$, but only weakly dependent upon $R_\lambda$, unless $St>1$. However, larger Vorono\text{\"i} volumes (void regions) exhibit a much stronger dependence on $R_\lambda$, even when $St\leq 1$, and we show that this, rather than the behavior at small volumes, is the cause of the sensitivity of the standard deviation of the Vorono\text{\"i} volumes that has been previously reported. We also show that the largest contribution to the particle settling velocities is associated with increasingly larger Vorono\text{\"i} volumes as the settling parameter $Sv\equiv St/Fr$ is increased.

Our local analysis of the acceleration statistics of settling inertial particles shows that clustered particles experience a net acceleration in the direction of gravity, while particles in void regions experience the opposite. The particle acceleration variance, however, is a convex function of the Vorono\text{\"i} volumes, with or without gravity, which seems to indicate a non-trivial relationship between the Vorono\text{\"i} volumes and the sizes of the turbulent flow scales. Results for the variance of the fluid acceleration at the inertial particle positions are of the order of the square of the Kolmogorov acceleration and depend only weakly on Vorono\text{\"i} volumes. These results call into question the ``sweep-stick'' mechanism for particle clustering in turbulence which would lead one to expect that clustered particles reside in the special regions where the fluid acceleration is zero (or at least small). 

We then consider the properties of particles in clusters, which are regions of connected Vorono\text{\"i} cells whose volume is less than a certain threshold. The results show self-similarity of the clusters, and that the statistics of the cluster volumes depends only weakly on $St$, with a stronger dependance on $Fr$ and $R_\lambda$. Finally, we compare the average settling velocities of all particles in the flow with those in clusters, and show that those in the clusters settle much faster, in agreement with previous work. However, we also find that this difference grows significantly with increasing $R_\lambda$ and exhibits a non-monotonic dependence on $Fr$. The kinetic energy of the particles, however, are almost the same for particles whether they are in clusters or not.
\end{abstract}


%
\maketitle

\section{Introduction}
Turbulent multiphase flows have been the subject of a numerous studies for many years because of their broad range of applications in nature and industrial contexts such as distribution of greenhouse gases \cite{d2014turbulence}, cloud formation \cite{beard1993warm}, volcanic eruptions \cite{ongaro2007parallel}, contaminant transport \cite{maxey87}, heat exchangers \cite{bianco2011numerical}, boiling heat transfer \cite{momenifar2015effect}, combustion processes \cite{kuo2012fundamentals}, air-lift pumps \cite{hanafizadeh2014void}, cavitating flows \cite{arabnejad2019numerical}, sedimentation process in settling tanks \cite{vahidifar2018introducing}, thermosyphons \cite{hanafizadeh2016thermosyphon} and fluidized bed reactors \cite{bi2000state} to name a few. The present work explores the class of multi-phase flows involving the motion of small, settling inertial particles in turbulent flows. 


A well-known and striking effect of particle inertia is that it leads to the spatial clustering of the particles in turbulent flows \cite{maxey87,bec07}. This has important implications for the rate at which particles collide in turbulent flows \cite{sundaram4,toschi2009lagrangian,shaw2003particle,ireland2016effecta}. It is also important since regions with a high concentration of particles have the potential to significantly modify the underlying turbulent flow in those same regions, through momentum coupling with the fluid \cite{balachandar10}. Several different methods and techniques have been used to analyze the particle clustering, such as box-counting (\cite{aliseda2002effect}), Vorono\text{\"i} analysis (\cite{monchaux2010preferential}), Lyapunov exponents (\cite{bec2006lyapunov}), Minkowski functionals (\cite{calzavarini2008dimensionality}) and radial distribution functions  (RDFs) (\cite{sundaram4,ireland2016effecta,ireland2016effectb}). These methods can provide different insights and perspectives into the spatial clustering, and two of the most commonly used methods are the RDF and Vorono\text{\"i} analysis. 

The RDF, defined by the probability of finding a particle at a given distance from a reference particle, is unaffected by variation in the number of particles in the domain (except for statistical noise) and is directly related to the particle collision kernel (\cite{sundaram4}). However, since it is constructed using spatial averages, the RDF cannot be used to consider the properties of individual clusters of particles, and only provides global information on the clustering. 
In Vorono\text{\"i} analysis, the domain is divided into cells associated with each individual particle, where any point in a given cell is closer to that particle than any other. This approach gives local insight into the particle clustering, with the local particle concentration field given by the inverse of the volume of the Vorono\text{\"i} cell. Using this technique enables characterization of the topology, kinematics and dynamics of individual clusters in the turbulent flow. However, a disadvantage is that unlike the RDF, the results of Vorono\text{\"i} analysis are sensitive to the number of particles in the flow, at least quantitatively (\cite{tagawa_mercado_prakash_calzavarini_sun_lohse_2012}). As discussed in \cite{petersen2019experimental}, since the RDF and Vorono\text{\"i} analysis provide different information on the particle clustering (e.g. global vs. local information), they can both be used to provide a more complete understanding of the problem.


Recently, extensive investigations into the effects of the flow Taylor Reynolds number, $R_\lambda$, and the Froude number $Fr$ (quantifying the effect of gravity) on the particle clustering have been performed using the RDF computed from Direct Numerical Simulations (DNS) \cite{ireland2016effecta,ireland2016effectb,momenifar2019influence}. These studies have shown that when the Kolmogorov-scale Stokes number of the particle, $St$, is $\lesssim 1$, there is virtually no effect of $R_\lambda$ on the clustering, whereas the clustering increases with increasing $R_\lambda$ when $St>1$. Concerning the effect of gravity, it was found that for $St\lesssim 1$, gravity leads to a reduction in the clustering, while for $St>1$ it enhances the clustering. In contrast, it appears that such a systematic study of the parametric dependence of clustering as quantified by a Vorono\text{\"i} analysis has not been undertaken. We now summarize what has been done on this.

The experimental study of \cite{obligado2014preferential} considered the clustering of small water droplets in isotropic turbulence using Vorono\text{\"i} analysis. They examined the range $200\leq R_\lambda \leq 400$ and $2\leq St\leq 10$, and reported that the level of clustering was enhanced with increasing $R_\lambda$. The experimental study of \cite{sumbekova2017preferential} considered $170\leq R_\lambda \leq 450$,  $0.1\leq St \leq 5$ and found that the clustering is strongly enhanced with increasing $R_\lambda$, but with weak dependence on $St$. These studies focused on the regime of large $Fr$, where the effects of gravity on the particle motion are weak. The recent experimental study of \cite{petersen2019experimental} also used Vorono\text{\"i} analysis to quantify the clustering and explored the range $200 \leq R_\lambda\leq 500$, $0.37 \leq St\leq 20.8$, and $0.73\leq Fr\leq 3.7$. They found that the clusters exhibit a range of sizes (up to the integral scale of the flow), posses self-similarity, and are elongated and aligned with the gravity direction \cite{petersen2019experimental}. They also found that the clustering is enhanced with increasing $R_\lambda$, and in agreement with the RDF analysis in \cite{ireland2016effecta,ireland2016effectb,momenifar2019influence}, that gravity can enhance the clustering of the inertial particles. A limitation of these experimental results is that their Vorono\text{\"i} analysis is only two-dimensional, and it is known that at least in the context of the RDF, two-dimensional analysis of particle clustering can differ significantly from the full three-dimensional results \cite{holtzer02}. Moreover, and perhaps more importantly, in these experiments it was not possible to independently vary $R_\lambda$ and $Fr$, and so a systematic understanding of their individual effects on the clustering as quantified by Vorono\text{\"i} analysis is lacking. DNS investigations into the effect of $R_\lambda$ on the Vorono\text{\"i} analysis of particle clustering have been undertaken, and in agreement with the experimental results, they show that the clustering is enhanced with increasing $R_\lambda$ \cite{tagawa_mercado_prakash_calzavarini_sun_lohse_2012}. However, only the limited range $75\leq R_\lambda\leq 180$ was considered, and they did not consider the effect of gravity.

In addition to analysis of the local particle concentration using Vorono\text{\"i} analysis, the DNS study of \cite{baker2017coherent} explored the idea of coherent clusters of inertial particles, with $Fr= 0.1$ in a low Reynolds number flow, $R_\lambda=65$. They defined a coherent cluster as a group of connected Vorono\text{\"i} cells whose volumes are smaller than a certain threshold. Moreover, they specified that the total volume of the cluster of connected cells must exceed $8\eta^3$ (where $\eta$ is the Kolmogorov length scale) in order to be considered a coherent cluster. They found that the coherent clusters exhibit a tendency to align with the direction of the local vorticity vector, and that the cluster volume depends strongly on $St$. Furthermore, in the presence of gravity, coherent clusters were found to be elongated and aligned with the gravity direction, as also found experimentally in \cite{petersen2019experimental}. Since they only considered a single set of values for $R_\lambda$ and $Fr$, the effects of these parameters on the coherent clusters is not known.

In addition to using Vorono\text{\"i} analysis to quantify the particle clustering, has also been used to understand the settling rates of particles in turbulence and how this is related to the local particle concentration and turbulence properties \cite{aliseda2002effect,dejoan2013preferential, monchaux2017settling,bosse2006small,dejoan2011dns}. Again, however, the effects of $R_\lambda$ and $Fr$ on the results was not systematically explored. It has also been recently shown theoretically and numerically how the settling of particles in turbulence is affected by a range of flow scales that depends on $St, R_\lambda$, and $Fr$ \cite{tom2019}. It is therefore of interest to explore the effect of  $St, R_\lambda$, and $Fr$ on the particle settle speeds using Vorono\text{\"i} analysis in order to gain further insights into the multiscale nature of the problem.

A recent striking finding concerning the effect of gravitational settling is that it can dramatically enhance the fluctuating accelerations of inertial particles in turbulence \cite{parishani2015effects,ireland2016effectb}, which in turn can profoundly affect collision and mixing rates of settling particles in turbulence \cite{dhariwal2018small,dhariwal2019,   momenifar2019influence}. An analysis of how these enhanced accelerations might be related to the local particle concentration has not yet been undertaken, but which would provide insights concerning the physical mechanism responsible for the effect, and also the scales at which it occurs.

In summary of this brief review, it is clear that there is a need to systematically explore the effects of $ R_\lambda$, and $Fr$ on particle clustering and geometry of clusters as analyzed using Vorono\text{\"i} analysis, and to understand how this might differ from the perspective gained using the RDF analysis in \cite{ireland2016effecta,ireland2016effectb,momenifar2019influence}. Moreover, there is a need to explore how the enhanced fluctuating particle accelerations due to gravitational settling might be connected to the properties of the local particle concentration in order to gain further insights into the mechanisms responsible for this effect. In the present work we address these issues.

\section{Computational Details}

Since one of the main goals of this work is to understand how the $R_\lambda, Fr$ dependence of particle clustering as determined by Vorono\text{\"i} analysis might differ from that based on an RDF analysis in \cite{ireland2016effecta,ireland2016effectb,momenifar2019influence}, we therefore consider the same description for the particles and fluid as that in \cite{ireland2016effecta,ireland2016effectb,momenifar2019influence}. In particular, we focus on the motion of a dilute suspension of one-way coupled, small, heavy, spherical inertial particles whose motion is governed by a simplified version of the equation of Maxey \& Riley (\cite{maxey1983equation}) 
\begin{equation}
\ddot{\boldsymbol{x}}^p(t)\equiv\dot{\boldsymbol{v}}^p(t)=\frac{\boldsymbol{u}(\boldsymbol{x}^p(t),t)-\boldsymbol{v}^p(t)}{\tau_p}+\boldsymbol{g},
\label{MR_inertial}
\end{equation}
where ${\boldsymbol{x}}^p(t),{\boldsymbol{v}}^p(t)$ are the particle position and velocity vectors, $\boldsymbol{u}(\boldsymbol{x}^p(t),t)$ is the fluid velocity at the particle position, $\boldsymbol{g}$ is the gravitational acceleration, and $\tau_p$ is the particle response time. Furthermore, fluid particles are tracked by solving $\dot{\boldsymbol{x}}^p(t)\equiv\boldsymbol{u}(\boldsymbol{x}^p(t),t)$. 
\begin{table}
	\centering
	\renewcommand{\arraystretch}{0.7}
	\setlength{\tabcolsep}{12pt}
	\begin{tabular}{ccccc}
		$\mathrm{Parameter}$ & $\mathrm{DNS} \,1 $ & $\mathrm{DNS} \,2$  & $\mathrm{DNS} \,3$  & $\mathrm{DNS} \,4$   \\
		$N$ & 128 & 128 & 1024 & 512  \\
		$R_\lambda$ &  93 &  94 &  90 &  224  \\
		$Fr$ &  $ \infty $ &  0.3 &  0.052 &  $ \infty $  \\
		$\mathscr{L}$ & 2$\upi$ & 2$\upi$ & 16$\upi$ & 2$\upi$   \\       
		$\nu$ & 0.005 & 0.005 & 0.005 & 0.0008289  \\
		$\epsilon$ & 0.324 & 0.332 & 0.257 & 0.253  \\
		$l$ & 1.48 & 1.49 & 1.47 & 1.40 \\
		$l/\eta$ & 59.6 & 60.4 & 55.6 & 204 \\
		$u'$ & 0.984 & 0.996 & 0.912 & 0.915  \\
		$u'/u_\eta$ & 4.91 & 4.92 & 4.82 & 7.60  \\
		$T_L$ & 1.51 & 1.50 & 1.61 & 1.53  \\
		$T_L/\tau_\eta$ & 12.14 & 12.24 & 11.52 & 26.8 \\
		$\kappa_{{\rm max}}\eta$ & 1.5 & 1.48 & 1.61 & 1.66 \\
		$N_p$ & 262,144 & 262,144 & 16,777,216 & 2,097,152  \\
	\end{tabular}
	\begin{tabular}{ccccc}
		\hline
		$\mathrm{Parameter}$ & $\mathrm{DNS} \,5$ & $\mathrm{DNS} \,6$  & $\mathrm{DNS}\,7 $  & $\mathrm{DNS} \,8$   \\
		$N$ & 512 & 1024 & 1024 & 1024  \\
		$R_\lambda$ &  237 &  230 &  398 &  398 \\
		$Fr$ &  0.3 &  0.052 & $ \infty $  & 0.052 \\
		$\mathscr{L}$ & 2$\upi$ & 4$\upi$ & 2$\upi$ & 2$\upi$  \\       
		$\nu$ & 0.0008289  & 0.0008289 & 0.0003 & 0.0003 \\
		$\epsilon$ & 0.2842 & 0.239 & 0.223 & 0.223  \\
		$l$ & 1.43 & 1.49 & 1.45 & 1.45  \\
		$l/\eta$ & 214 & 213 & 436 & 436  \\
		$u'$ & 0.966 & 0.914 & 0.915 & 0.915  \\
		$u'/u_\eta$ & 7.82 & 7.7 & 10.1 & 10.1  \\
		$T_L$ & 1.48 & 1.63 & 1.58 & 1.58  \\
		$T_L/\tau_\eta$ & 27.36 & 27.66 & 43.0 & 43.0  \\
		$\kappa_{{\rm max}}\eta$ & 1.62 & 1.68 & 1.60 & 1.60 \\
		$N_p$ & 2,097,152 & 16,777,216 & 2,097,152 & 2,097,152  \\
	\end{tabular}
	
	\caption{Simulation parameters for the DNS study of isotropic turbulence (arbitrary units).
		$N$ is the number of grid points in each direction, 
		$R_\lambda \equiv u'\lambda/\nu$ is the Taylor micro-scale
		Reynolds number ($R_\lambda \equiv \sqrt{15Re}$ for homogeneous and isotropic flows), $\lambda\equiv u'/\langle(\boldsymbol{\nabla} \boldsymbol{u})^2\rangle^{1/2} $ is the Taylor micro-scale,
		$\mathscr{L} $ is the box size, $\nu$ is the fluid kinematic viscosity, $\epsilon \equiv 2\nu \int_0^{\kappa_{\rm max}}\kappa^2 E(\kappa) {\rm d}\kappa $ is the mean
		turbulent kinetic energy dissipation rate,
		$l \equiv 3\upi/(2k)\int_0^{\kappa_{\rm max}}E(\kappa)/\kappa {\rm d}\kappa $  is the integral length scale, $\eta \equiv \nu^{3/4}/\epsilon^{1/4}$ is the Kolmogorov length scale, 
		$u' \equiv \sqrt{(2k/3)}$ is the fluid r.m.s. fluctuating 
		velocity, $k$ is the turbulent kinetic energy, 
		$u_\eta$ is the Kolmogorov velocity scale, 
		$T_L \equiv l/u^\prime$ is the large-eddy turnover
		time, $\tau_\eta \equiv \sqrt{(\nu/\epsilon)}$ is the Kolmogorov time scale, 
		$\kappa_{\rm max}=\sqrt{2}N/3$ is the maximum
		resolved wavenumber, and $N_p$ is  
		the number of particles per Stokes number.}
	{\label{tab:parameters}}
\end{table}
\FloatBarrier
The relevant non-dimensional parameters in this problem are the Stokes number ($St\equiv\tau_p/\tau_\eta$), the settling parameter ($Sv\equiv \tau_p g/u_{\eta}$, where the numerator represents particle's terminal velocity in the laminar flow) and the Froude number ($Fr\equiv a_\eta/{g}=\epsilon^{3/4}/(\nu^{1/4}{g})$) which characterize the particle's inertia, the effect of gravity on inertial particles and the effect of gravity on the flow, respectively. Here
$a_\eta$ is the Kolmogorov acceleration,
$\tau_{\eta}$ is the Kolmogorov timescale,
$u_\eta$ is the Kolmogorov velocity scale, and
$\epsilon$ is the mean turbulent kinetic energy dissipation rate.

Direct Numerical Simulations (DNS) of the incompressible Navier-Stokes equation are performed on a triply periodic cube of length $\mathscr {L}$, using a pseudo-spectral method on a uniform mesh with $N^3$ grid points. A deterministic forcing scheme is used to generate statistically stationary, isotropic turbulence. Details of numerical setup and simulations can be found in \cite{ireland2013highly,ireland2016effecta,ireland2016effectb}, and information on the simulations are provided in Table~\ref{tab:parameters}. 

As mentioned in the introduction, Vorono\text{\"i} analysis can be sensitive to the number of particles used in the analysis. Therefore, when comparing results from different DNS with different $R_\lambda$ and $Fr$ we choose the number of particles such that the average distance between the particles is the same in each study, $\overline{r}\equiv\mathscr {L}/N_p^{1/3}=7.9 \eta$. This number is comparable to that in the studies of \cite{tagawa_mercado_prakash_calzavarini_sun_lohse_2012,monchaux2012measuring,monchaux2010preferential}. A smaller value for $\overline{r}$ would be desirable, especially for performing the coherent cluster analysis \cite{baker2017coherent}. However, the computational expense for the higher $R_\lambda$ cases we consider make it unfeasible to consider significantly smaller values of $\overline{r}$.

Using our DNS data, we perform a 3D Vorono\text{\"i} analysis of the particles, and then compute a range of statistical quantities using the information from the analysis. In order to address the issue of open Vorono\text{\"i} cells at the boundaries of the domain, the particles across the boundaries are periodically repeated to ensure that all the cells are closed and the sum of Vorono\text{\"i} volumes is equal to the domain volume (\cite{nilsen2013voronoi,rabencov2015voronoi,yuan2018three}). Therefore, the Vorono\text{\"i} volumes associated with the particles adjacent to the boundaries are ill-defined and so are ignored in our analysis. 
\section{Results and discussion}
\subsection{Vorono\text{\"i} volume distributions}

\begin{figure}
	\vspace{-0.7in}
	\centering
	\begin{subfigure}[b]{0.5\linewidth}
		\includegraphics[width=\linewidth]{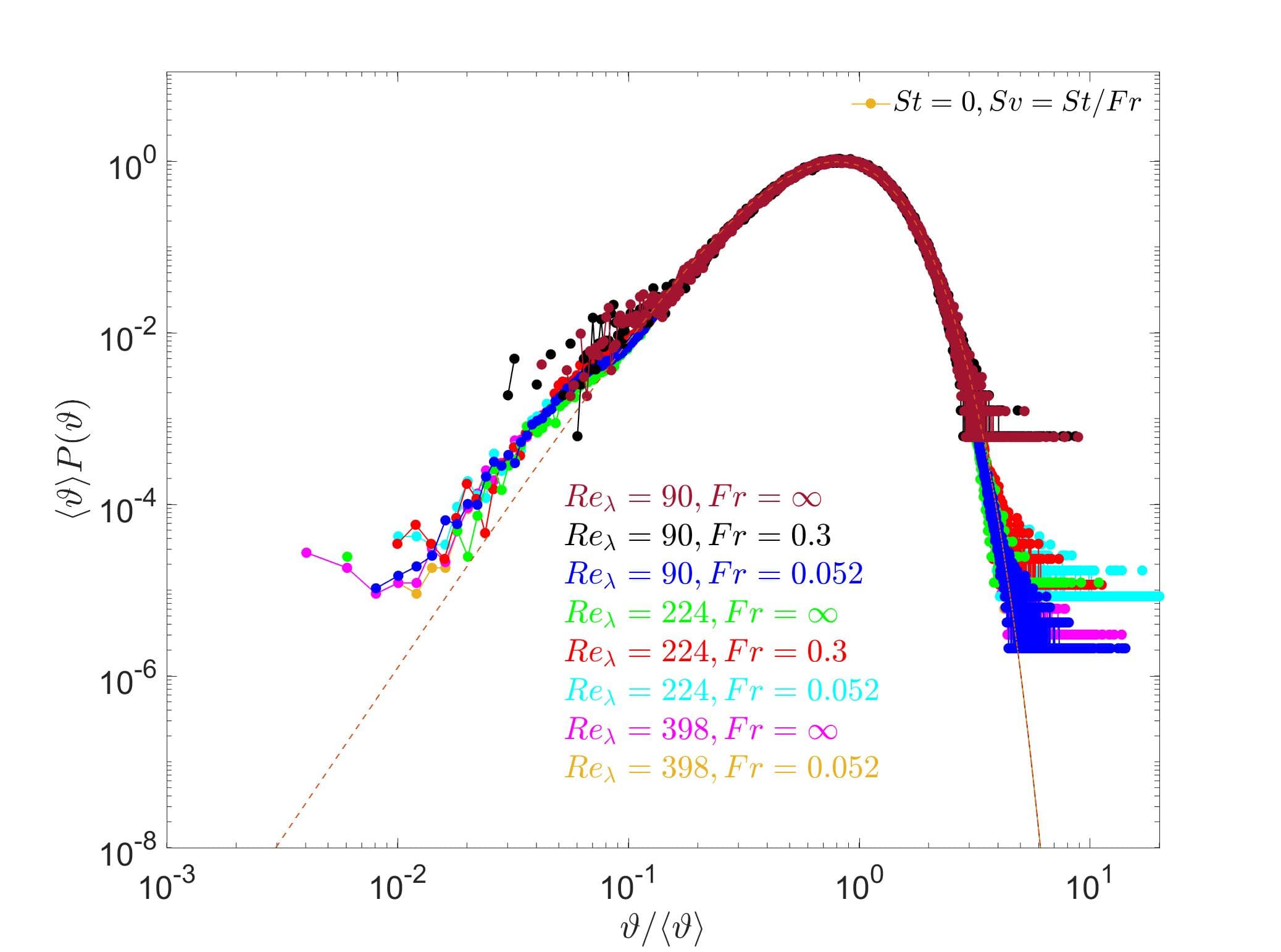}
		\caption{$St=0$}
	\end{subfigure}%
	\begin{subfigure}[b]{0.5\linewidth}
		\includegraphics[width=\linewidth]{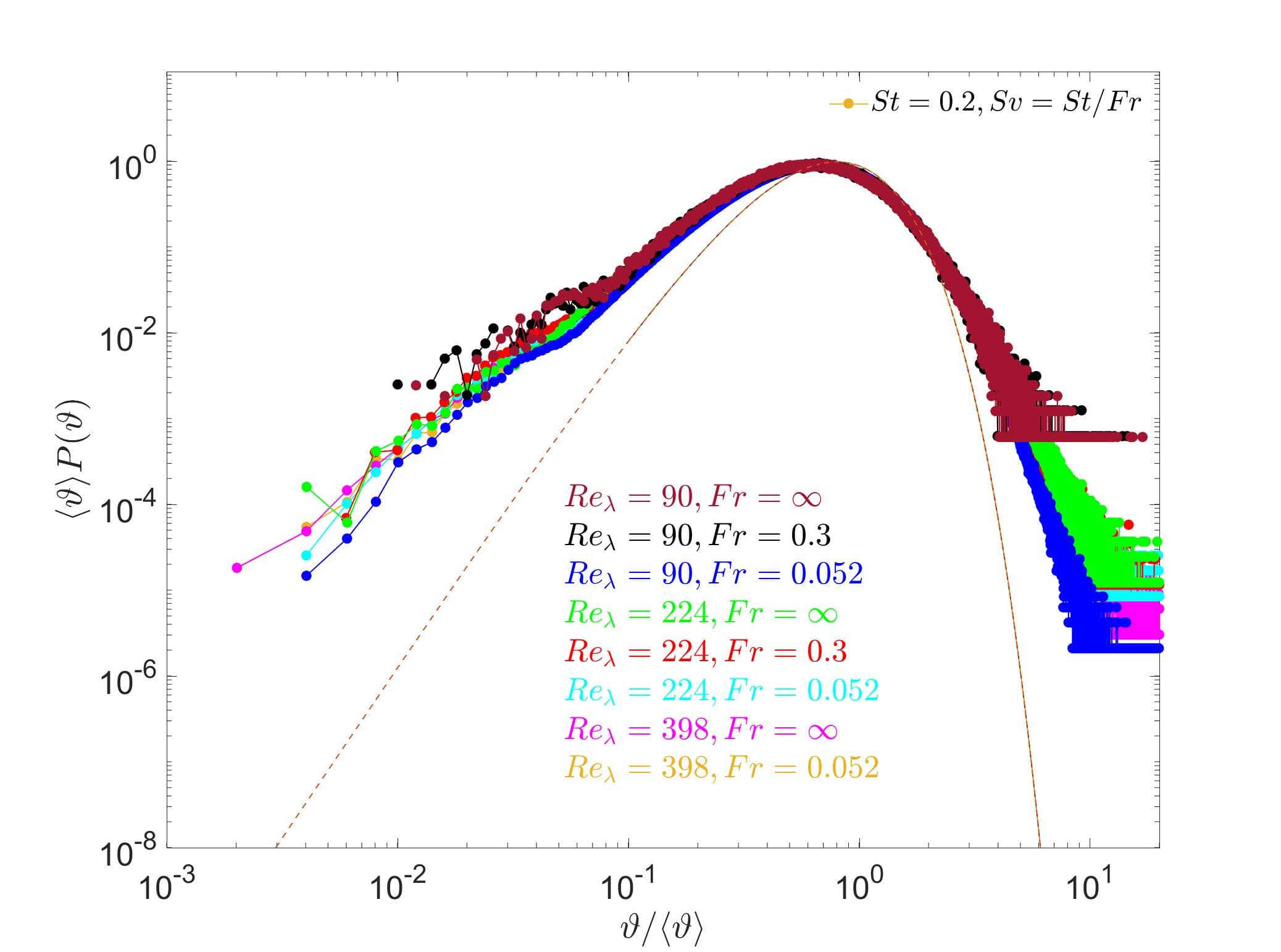}
		\caption{$St=0.2$ }
	\end{subfigure}
	
	\begin{subfigure}[b]{0.5\linewidth}
		\includegraphics[width=\linewidth]{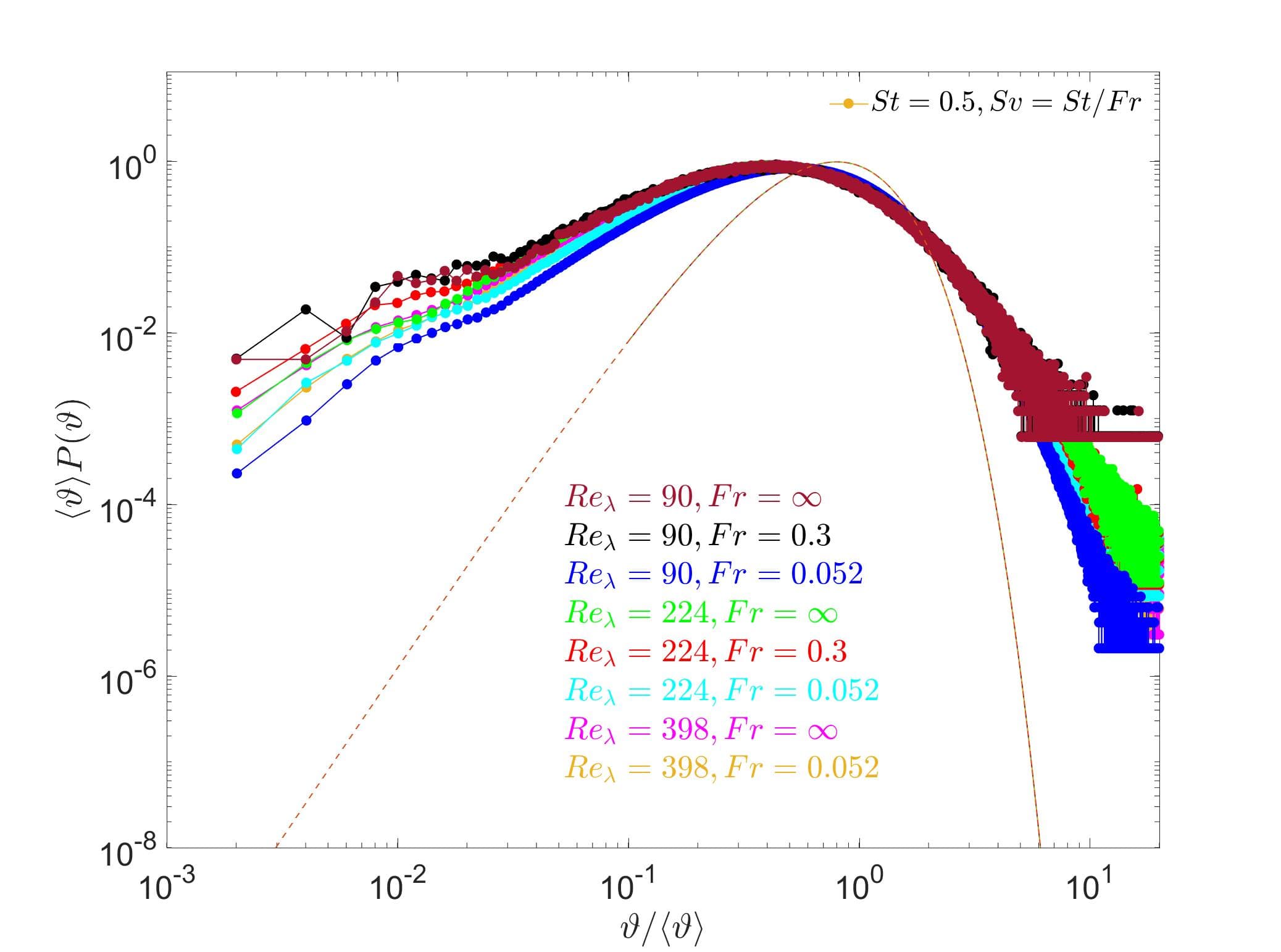}
		\caption{$St=0.5$}
	\end{subfigure}%
	\begin{subfigure}[b]{0.5\linewidth}
		\includegraphics[width=\linewidth]{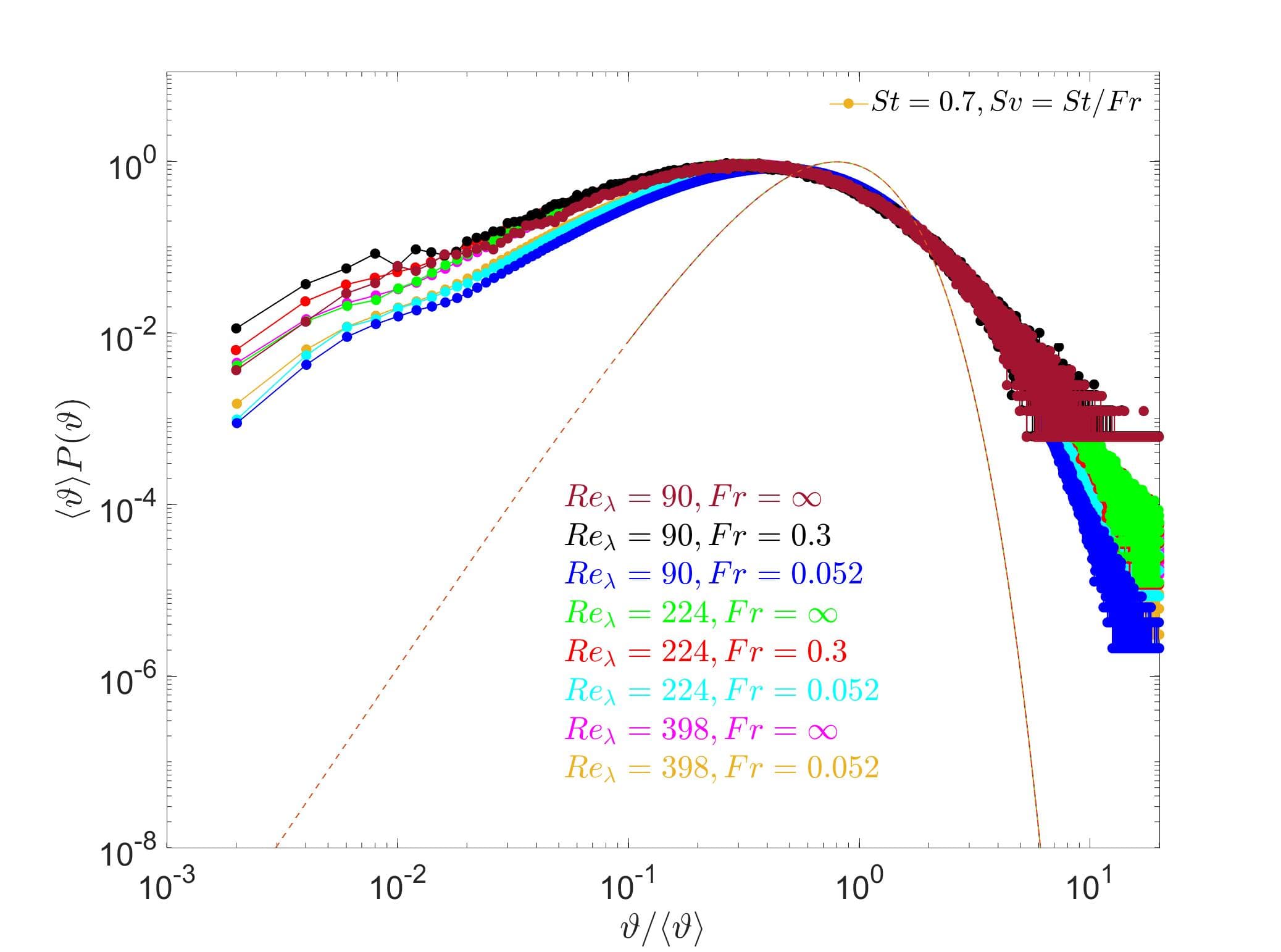}
		\caption{$St=0.7$ }
	\end{subfigure}
	\begin{subfigure}[b]{0.5\linewidth}
		\includegraphics[width=\linewidth]{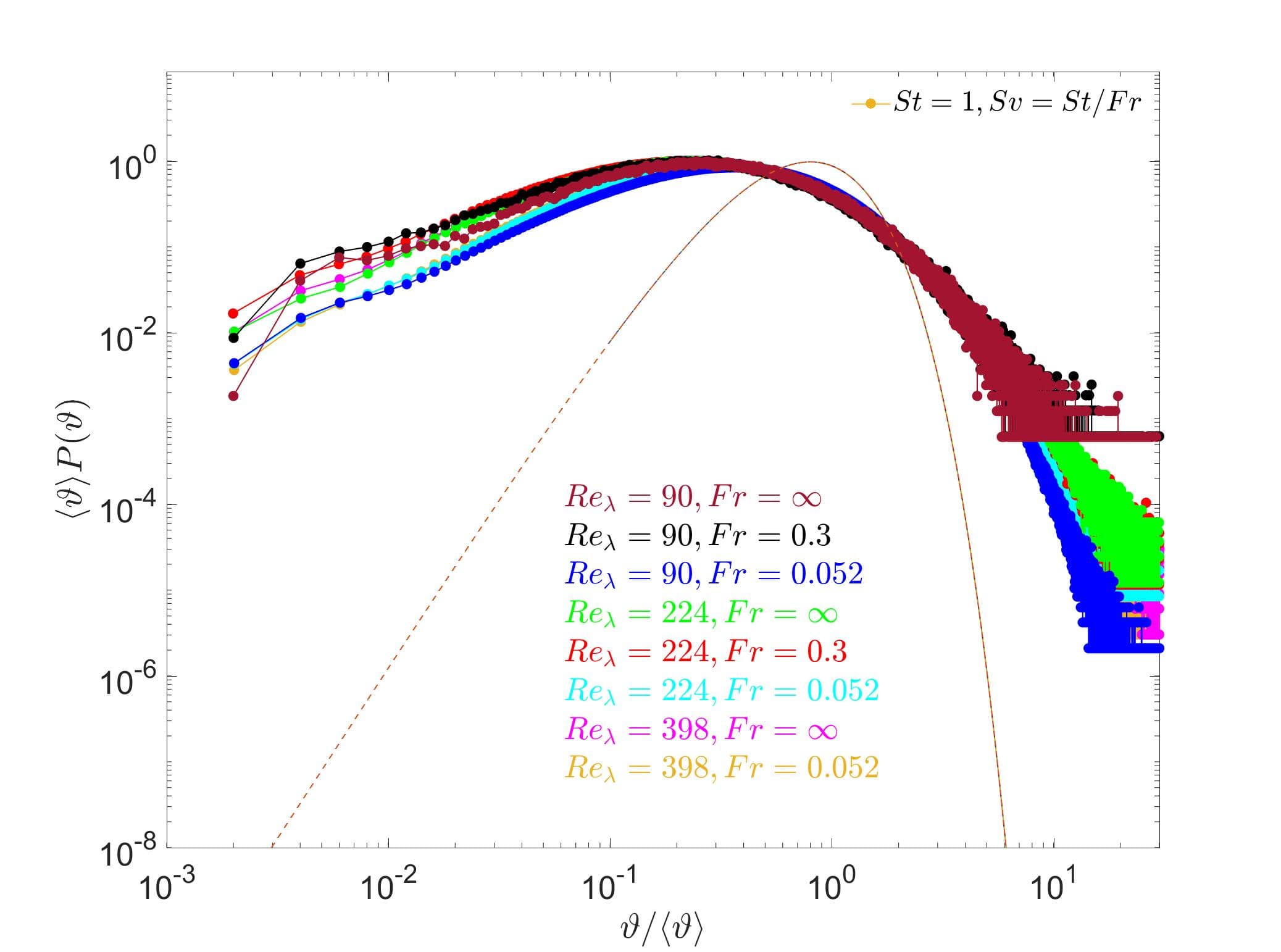}
		\caption{$St=1$ }
	\end{subfigure}%
	\begin{subfigure}[b]{0.5\linewidth}
		\includegraphics[width=\linewidth]{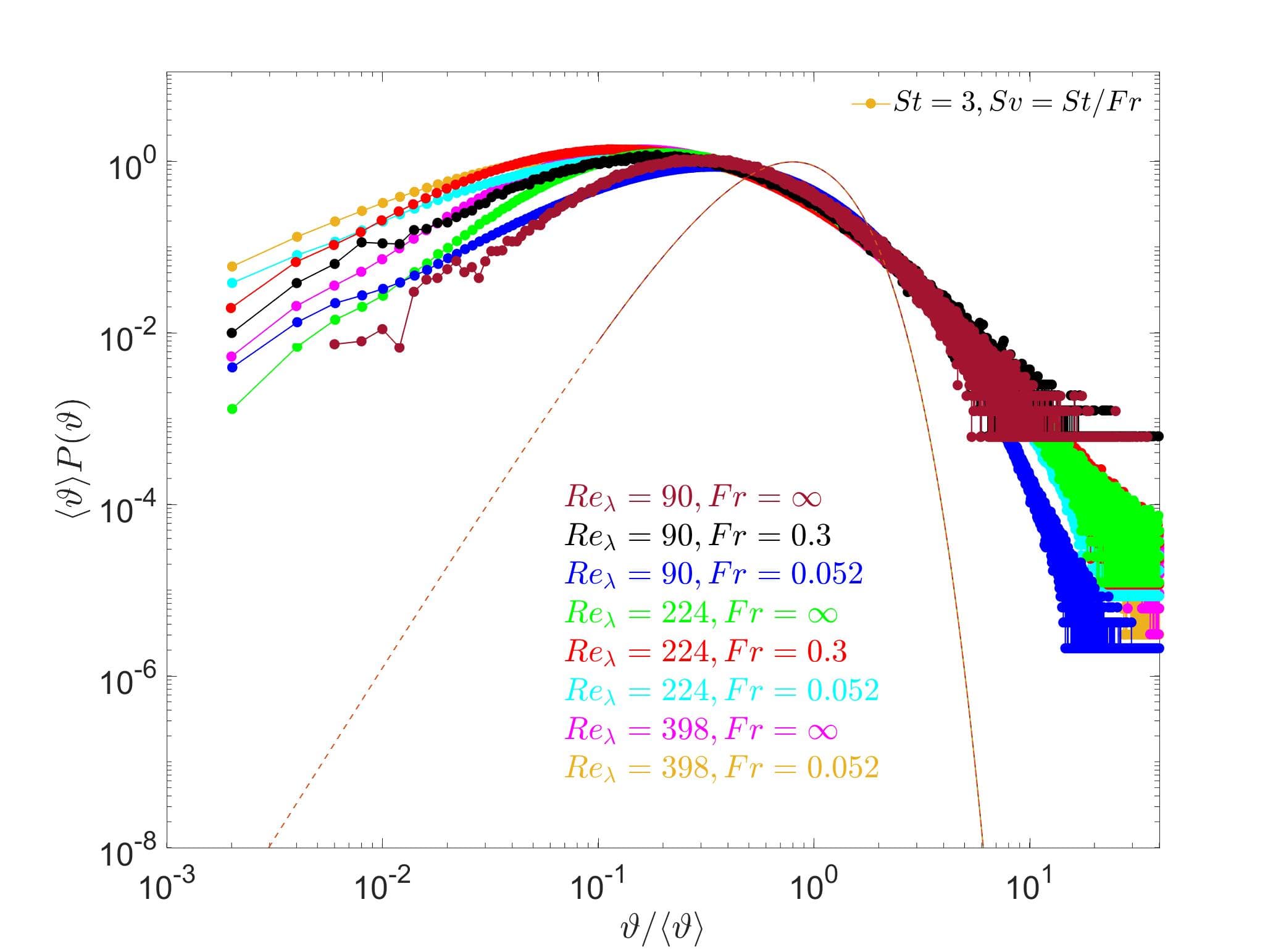}
		\caption{$St=3$ }
	\end{subfigure}
	\caption{PDF of the Vorono\text{\"i} volumes (normalized by the mean volume) at different cases of $Fr$ and $R_\lambda$ combinations for (a) $St=0$, (b) $St=0.2$, (c) $St=0.5$, (d) $St=0.7$,(e) $St=1$, and (f) $St=3$. Different colors represents different cases. The dashed line represents the Random Poisson Distribution.}\label{fig:VT_PDF_Vol_norm_eta_loglog_St}
\end{figure}

In figure \ref{fig:VT_PDF_Vol_norm_eta_loglog_St}, the Probability Distribution Function (PDF) of the Vorono\text{\"i} cell volumes (normalized by the mean Vorono\text{\"i} cells volume) for different $St$ and with different $Fr$ and $R_\lambda$ combinations are shown. The small/large Vorono\text{\"i} volumes represent the high/low concentration regions of the flow. For $St=0$ the particles are randomly distributed, and their PDF follows that of a random Poisson process
(RPP), described by \cite{ferenc2007size}. Compared to the $St=0$ case, the PDFs for $St>0$ are significantly higher at small volumes, indicating an enhanced probability of finding regions with high local particle concentration, i.e. clustering. In the experimental study of \cite{obligado2011reynolds} they observed a crossover, such that at sufficiently small volumes, the PDFs of the Vorono\text{\"i} volumes for inertial particles drops below that for fluid particles. They conjectured that this effect was due to interactions between the particles when they are sufficiently close. In our results we do not observe this crossover, and neither did the numerical studies of \cite{tagawa_mercado_prakash_calzavarini_sun_lohse_2012,baker2017coherent}. While this may be due to our neglect of the effect of particle interactions, we note that the crossover was also not observed in the recent experimental work of \cite{petersen2019experimental}. 

Concerning the effect of $R_\lambda$, the results in figure \ref{fig:VT_PDF_Vol_norm_eta_loglog_St} indicate that with respect to the behavior at small volumes, the effect of $R_\lambda$ depends on both $St$ and $Fr$. When $St<3$, the effect of $R_\lambda$ is quite weak, sometimes leading to a slight increase of the PDF at small volumes that then saturates as $R_\lambda$ is increased (e.g. for the case $St=0.5$, $Fr=0.052$), while in other cases it leads to a slight decrease of the PDF at small volumes (e.g. for the case $St=0.7$, $Fr=0.3$). However, for $St=3$, the sensitivity of the Vorono\text{\"i} volume PDF to $R_\lambda$ becomes much more apparent, with the clustering becoming stronger as $R_\lambda$ is increased. This dependence of the clustering on $R_\lambda$ is similar to that found in \cite{ireland2016effecta,ireland2016effectb} where the RDF was used to analyze the clustering. Arguments in \cite{ireland2016effecta,ireland2016effectb,bragg14d} suggest that this behavior arises because unless $St$ is sufficiently large, particles in the dissipation range are not able to remember their interaction with the inerital-range turbulence along their path-history. In this case, their motion is dominated by the dissipate range dynamics of the flow, and their dissipation range motion is not affected by the changing size of the inertial range as $R_\lambda$ is increased. Nevertheless, even if their dissipation range clustering is not affected by the inertial range, the clustering could still be affected by the strong intermittency of the dissipation range turbulence \cite{frisch}. In \cite{ireland2016effecta,bragg14b} it was argued that because the RDF is a low-order measure of the particle phase-space dynamics, it is not very sensitive to the flow intermittency that as most apparent in the high-order moments. However, the PDF of the Vorono\text{\"i} volumes is not confined to low-order information, and could in principle be sensitive to intermittent fluctuations in the particle motion, yet our results imply that the particle clustering is only weakly affected by the flow intermittency. One possible explanation for this surprising behavior is that due to the preferential sampling of the flow by the inertial particles, the inertial particles avoid the intermittent regions of the flow \cite{bec06a,ireland2016effecta,ireland2016effectb}, and as a result their spatial clustering is not affected by the increased intermittency in the underlying turbulence as $R_\lambda$ is increased.

Concerning the effect of $Fr$, the results in figure \ref{fig:VT_PDF_Vol_norm_eta_loglog_St} indicate that with respect to the behavior at small volumes, the clustering initially becomes stronger as $Fr$ is reduced from $\infty$ (no gravity case) to $0.3$, but then becomes weaker as $Fr$ is further reduced. Therefore, gravity has a non-monotonic effect on the clustering, and whether it enhances or reduces the clustering depends on $St$ and $Fr$. 
\FloatBarrier
This effect of gravity on clustering agrees with the conclusions based on the RDF analysis in \cite{ireland2016effectb}, although in that analysis the non-monotonic dependance of the clustering on $Fr$ was not so clear since only the values $Fr=\infty, 0.052$ were considered, and not the intermediate value $Fr=0.3$. However, such non-monotonic dependence was observed using an RDF analysis in our previous study \cite{momenifar2019influence}. Physical arguments for the effect of gravity on clustering can be found in \cite{ireland2016effectb}. 

\begin{figure}
\hspace{15mm}
	\begin{subfigure}[b]{\linewidth}
		\includegraphics[width=0.75\linewidth]{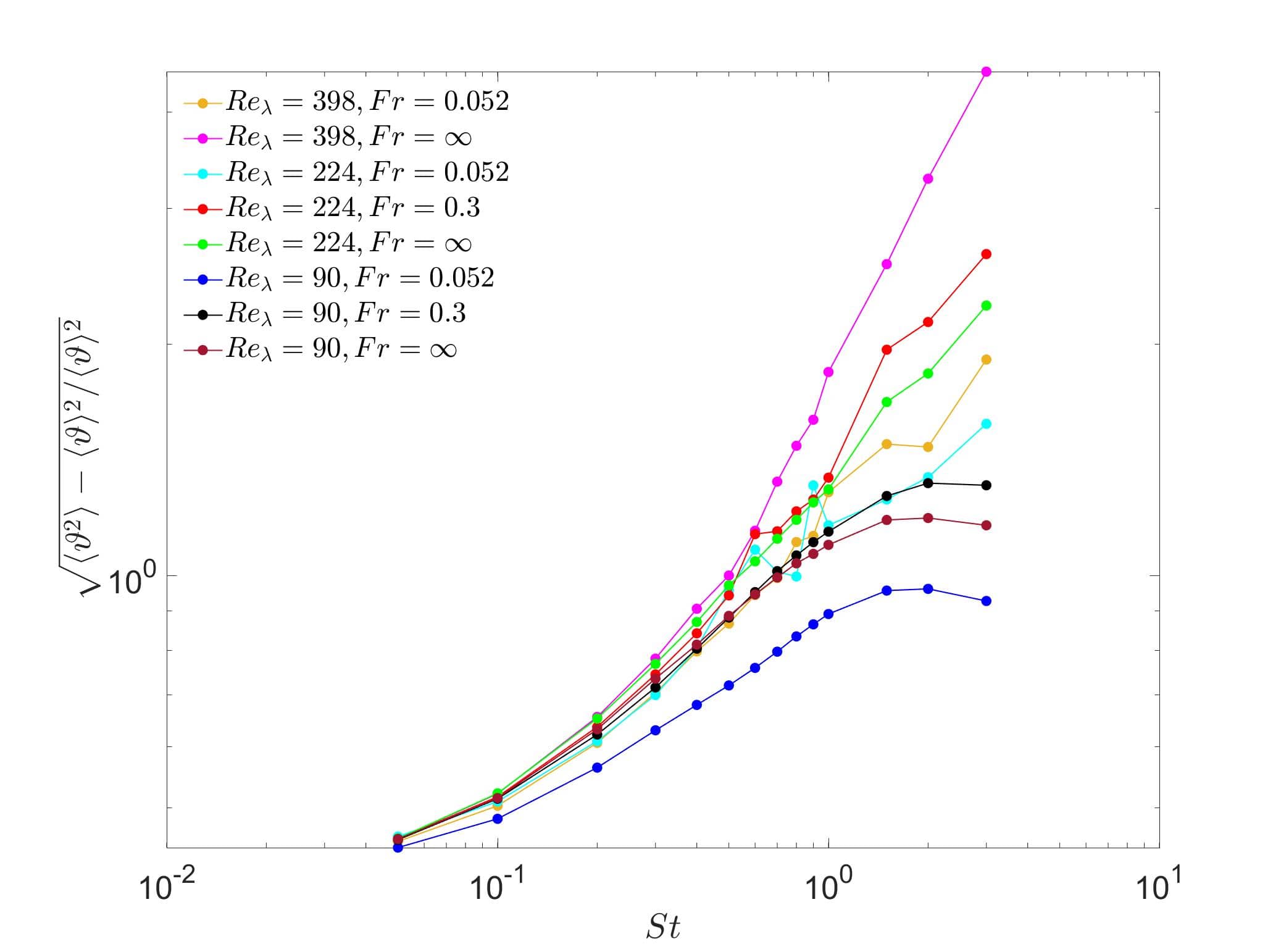}
	\end{subfigure}%
	\caption{Standard deviation of Vorono\text{\"i} cells normalized by their mean as a function of $St$.}
	\label{fig:VT_Var_Vol_norm_mean}
\end{figure}

\begin{figure}
	\vspace{-0.7in}
	\centering
	\begin{subfigure}[b]{0.5\linewidth}
		\includegraphics[width=\linewidth]{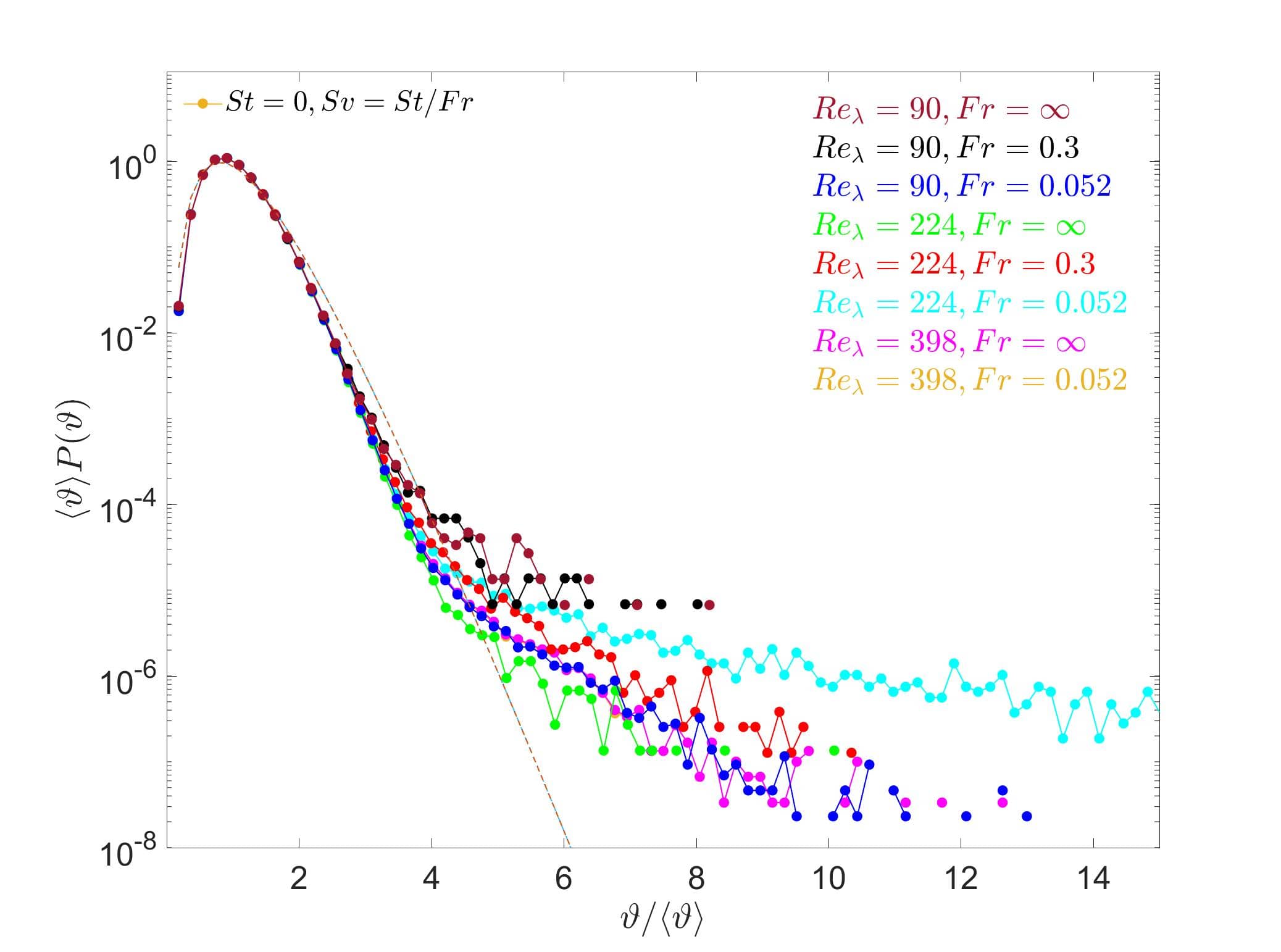}
		\caption{$St=0$}
	\end{subfigure}%
	\begin{subfigure}[b]{0.5\linewidth}
		\includegraphics[width=\linewidth]{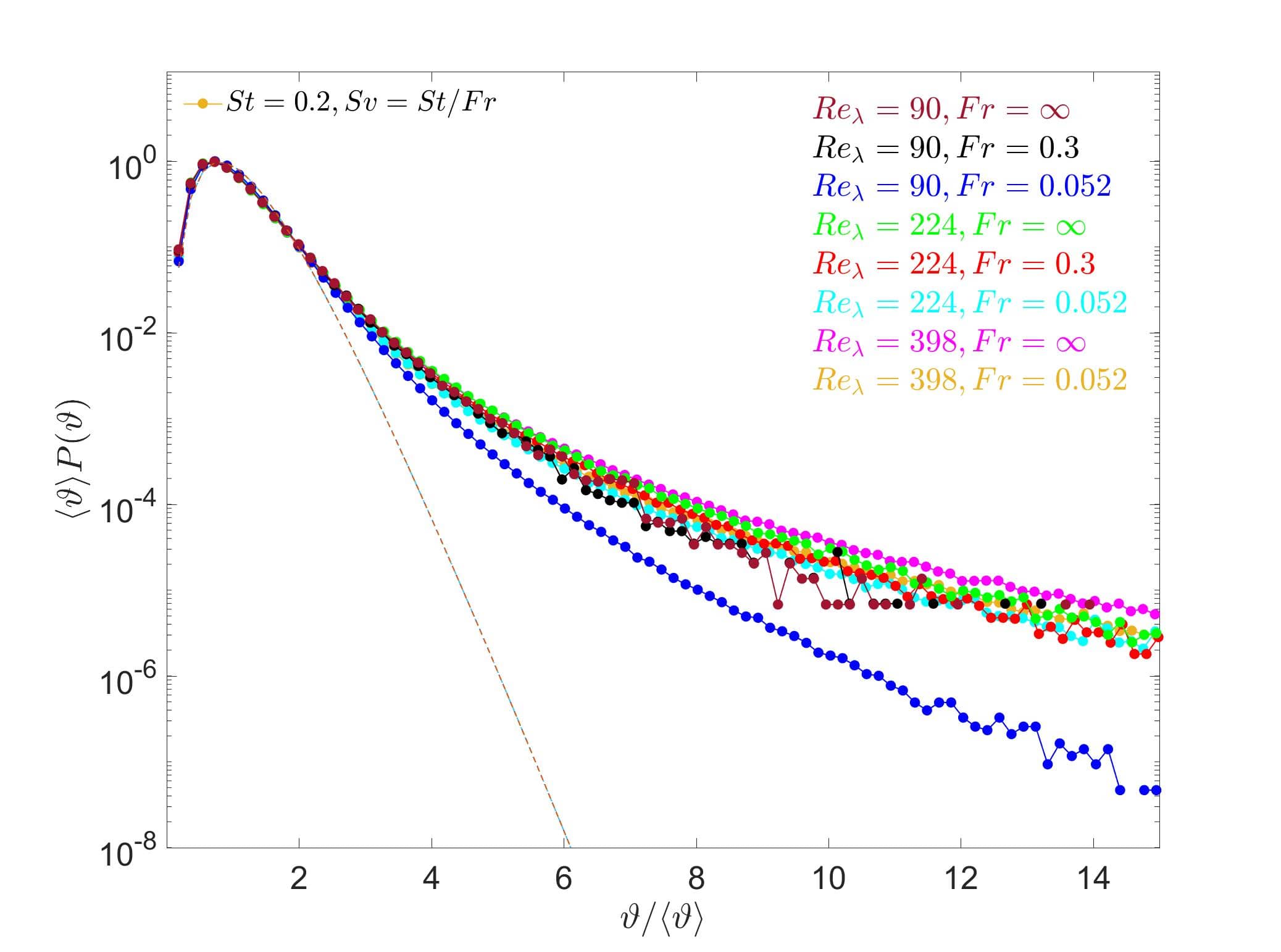}
		\caption{$St=0.2$ }
	\end{subfigure}
	
	\begin{subfigure}[b]{0.5\linewidth}
		\includegraphics[width=\linewidth]{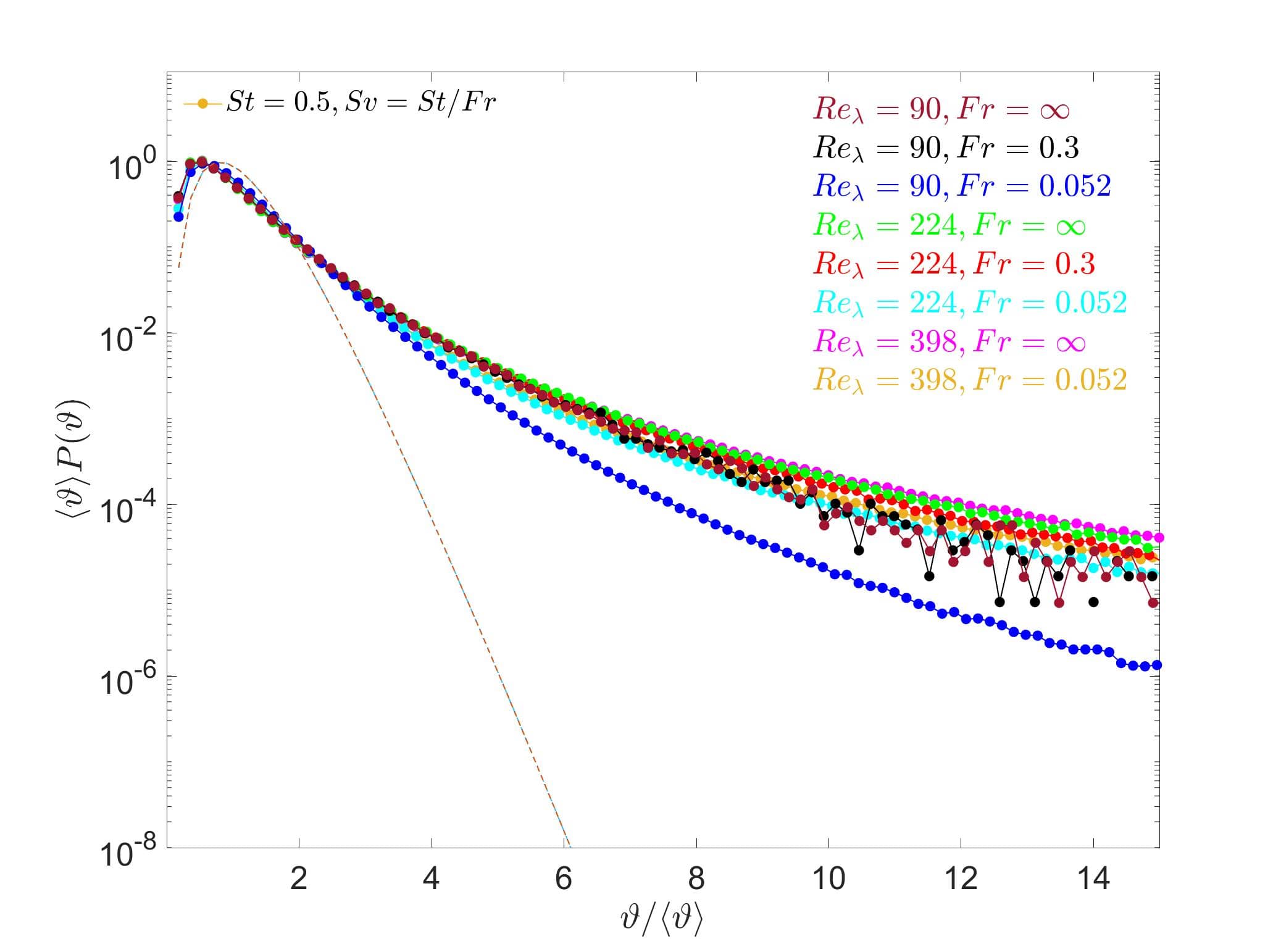}
		\caption{$St=0.5$}
	\end{subfigure}%
	\begin{subfigure}[b]{0.5\linewidth}
		\includegraphics[width=\linewidth]{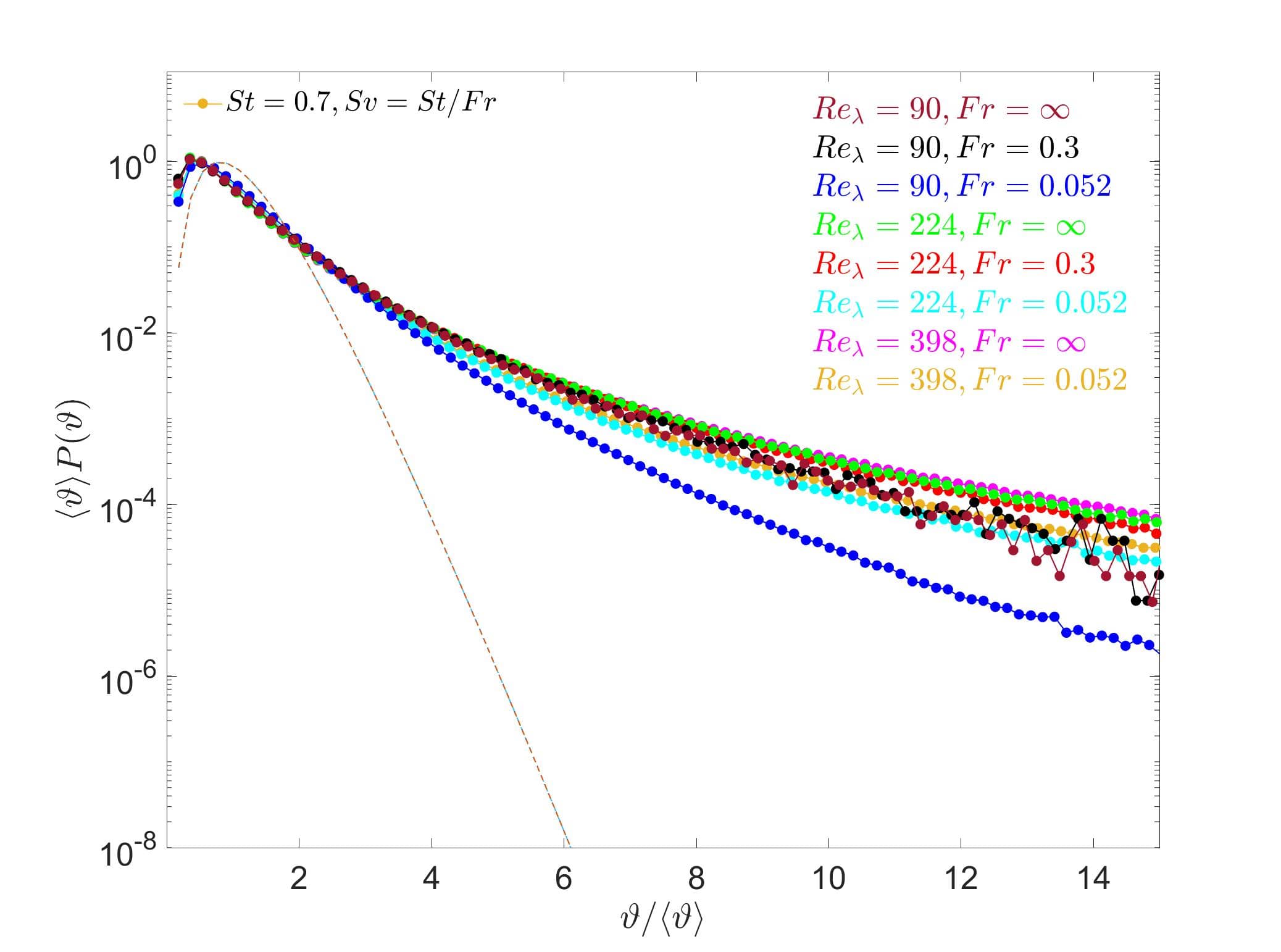}
		\caption{$St=0.7$ }
	\end{subfigure}
	\begin{subfigure}[b]{0.5\linewidth}
		\includegraphics[width=\linewidth]{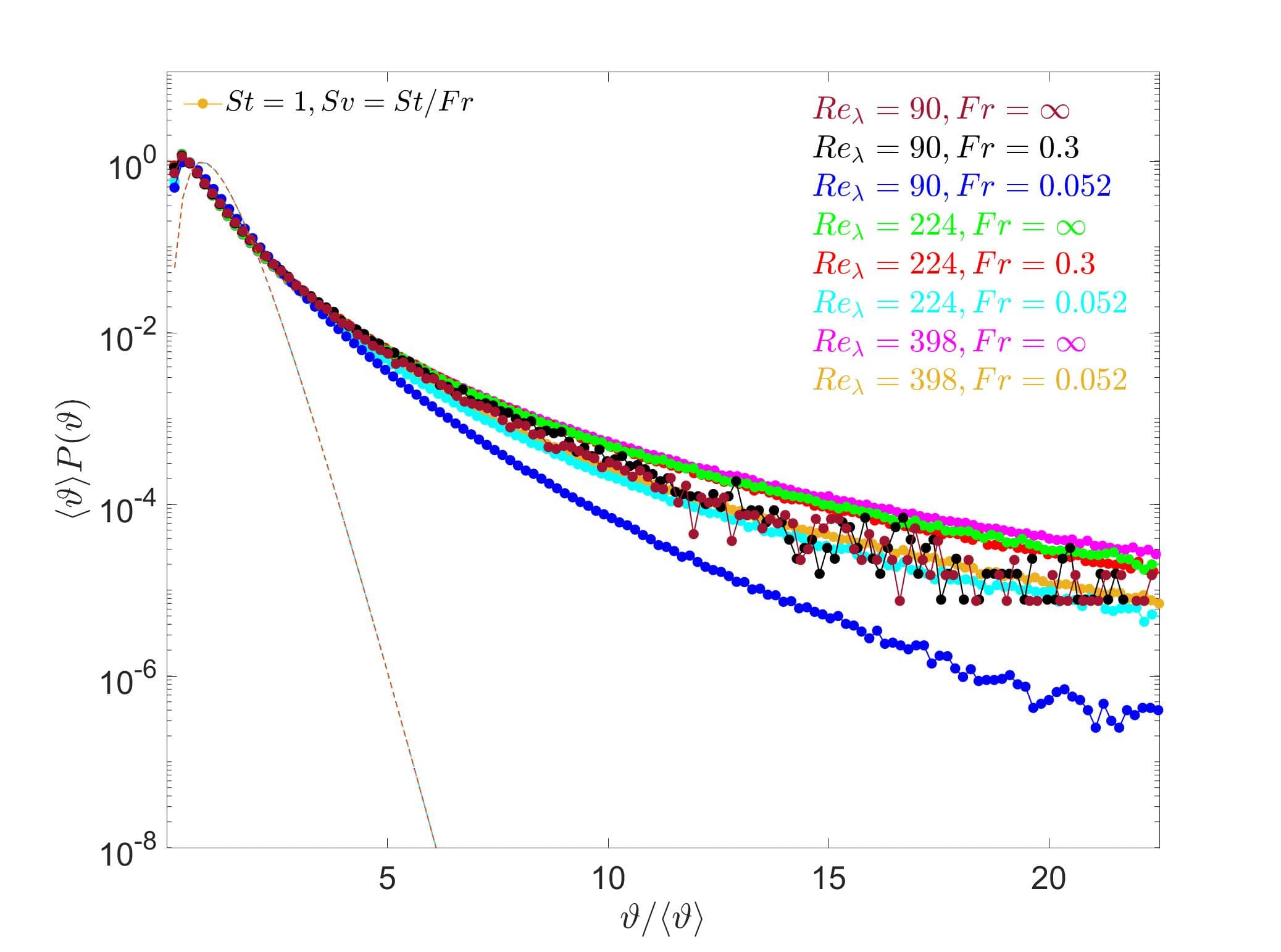}
		\caption{$St=1$ }
	\end{subfigure}%
	\begin{subfigure}[b]{0.5\linewidth}
		\includegraphics[width=\linewidth]{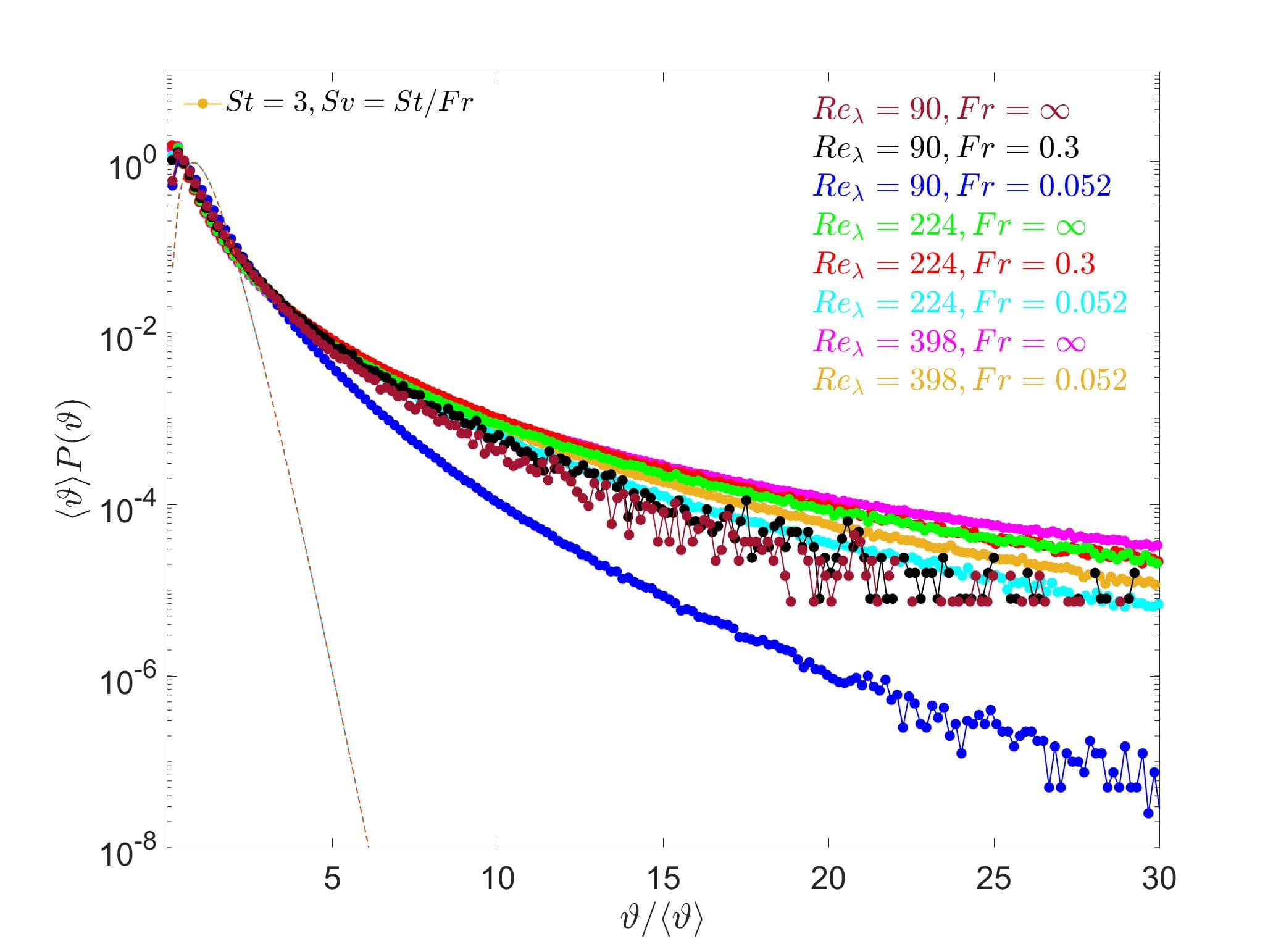}
		\caption{$St=3$ }
	\end{subfigure}
	\caption{PDF of the Vorono\text{\"i} volumes (normalized by the mean volume) at different cases of $Fr$ and $R_\lambda$ combinations for (a) $St=0$, (b) $St=0.2$, (c) $St=0.5$, (d) $St=0.7$,(e) $St=1$, and (f) $St=3$. Different colors represents different cases. The dashed line represents the Random Poisson Distribution.}\label{fig:VT_PDF_Vol_norm_meanVol_semilogy_St}
\end{figure}

To obtain more insight into the $R_\lambda$ dependence, in figure \ref{fig:VT_Var_Vol_norm_mean} we plot the standard deviation (s.d.) of the Vorono\text{\"i} volumes as a function of $St$. In agreement with the experimental study of \cite{obligado2011reynolds}, we observe that for $R_\lambda=90$, the s.d. increases with increasing $St$ until it reaches a peak around $St\approx 2$, after which it decreases. Also in agreement with their results we find that the s.d. becomes larger and the peak shifts to higher values of $St$ as $R_\lambda$ is increased. This is essentially due to the fact that as $R_\lambda$ is increased, more flow scales are present in the turbulence at which the particles can cluster. However, for any finite $St$, the growth of the s.d. with increasing $R_\lambda$ must saturate, since at sufficiently large scales the particles behave as fluid particles and do not cluster \cite{bragg15}. The results in figure \ref{fig:VT_Var_Vol_norm_mean} reveal a stronger sensitivity to $R_\lambda$ than is implied by the results in figure \ref{fig:VT_PDF_Vol_norm_eta_loglog_St} which may indicate that the sensitivity of the s.d. to $R_\lambda$ is mainly in the behavior of the PDF of Vorono\text{\"i} cells with large volume, whose behavior is not clear in a loglog plot. To consider this, in figure \ref{fig:VT_PDF_Vol_norm_meanVol_semilogy_St} we plot the PDF of Vorono\text{\"i} volumes in a log-lin plot. Comparing the results with those in figure \ref{fig:VT_PDF_Vol_norm_eta_loglog_St} confirms that it is the Vorono\text{\"i} cells with larger volumes that exhibit the strongest sensitivity to $R_\lambda$ and that the sensitivity to $R_\lambda$ is generally stronger in the presence of gravity. 

The data in figure \ref{fig:VT_PDF_Vol_norm_eta_loglog_St} shows that the PDF for inertial particles intersects that for fluid particles at two points, denoted by $\vartheta_C$ and $\vartheta_v$, respectively. These points are often used to define thresholds for detecting clusters and voids, with volumes less than $\vartheta_C$ denoting clustered regions, and volumes greater than $\vartheta_v$ denoting void regions. In view of these definitions, it is then apparent that the $R_\lambda$ dependence of the s.d. of the Vorono\text{\"i} volumes comes mainly from the $R_\lambda$ of the void regions, and not the clustered regions. In this way our conclusions are consistent with the findings in \cite{ireland2016effecta,ireland2016effectb,momenifar2019influence} based on the RDF that the small-scale clustering is quite insensitive to $R_\lambda$ when $St\lesssim 1$, and the strong $R_\lambda$ dependence of the s.d. of the Vorono\text{\"i} volumes does not contradict this.

A related point is that the results in figure \ref{fig:VT_Var_Vol_norm_mean} indicate that gravity suppresses the s.d. of the Vorono\text{\"i} volumes, while for the same cases (e.g. $St=3, R_\lambda=398$), figure \ref{fig:VT_PDF_Vol_norm_eta_loglog_St} indicates that gravity enhances the small-scale clustering. This is again because the s.d. is dominated by the behavior of the void regions rather than the clustered regions, and the void regions tend to be suppressed by gravity. This shows again the non-trivial effect of gravity on the particle clustering; while the probability of clustered and void regions both increase with increasing $St$, for a given $St$, decreasing $Fr$ can enhance the probability of clustered regions while lowering the probability of void regions. These points illustrate that using the s.d. Vorono\text{\"i} volumes as a measure of clustering should be used with caution, since its properties are dominated by void, rather than clustered regions of the flow.

\FloatBarrier
\FloatBarrier

\begin{figure}
	\vspace{-0.7in}
	\centering
	\begin{subfigure}[b]{0.5\linewidth}
		\includegraphics[width=\linewidth]{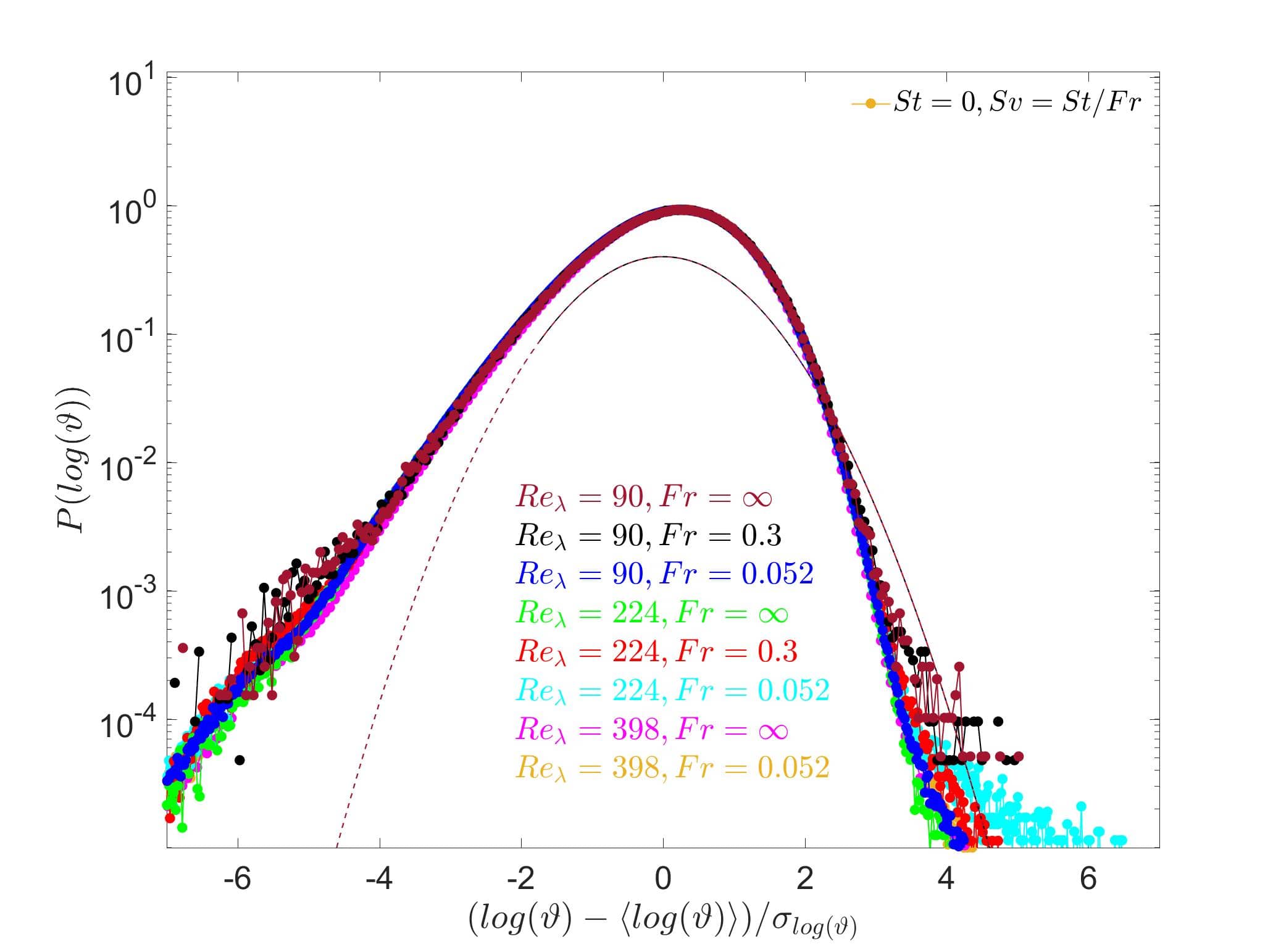}
		\caption{$St=0$}
	\end{subfigure}%
	\begin{subfigure}[b]{0.5\linewidth}
		\includegraphics[width=\linewidth]{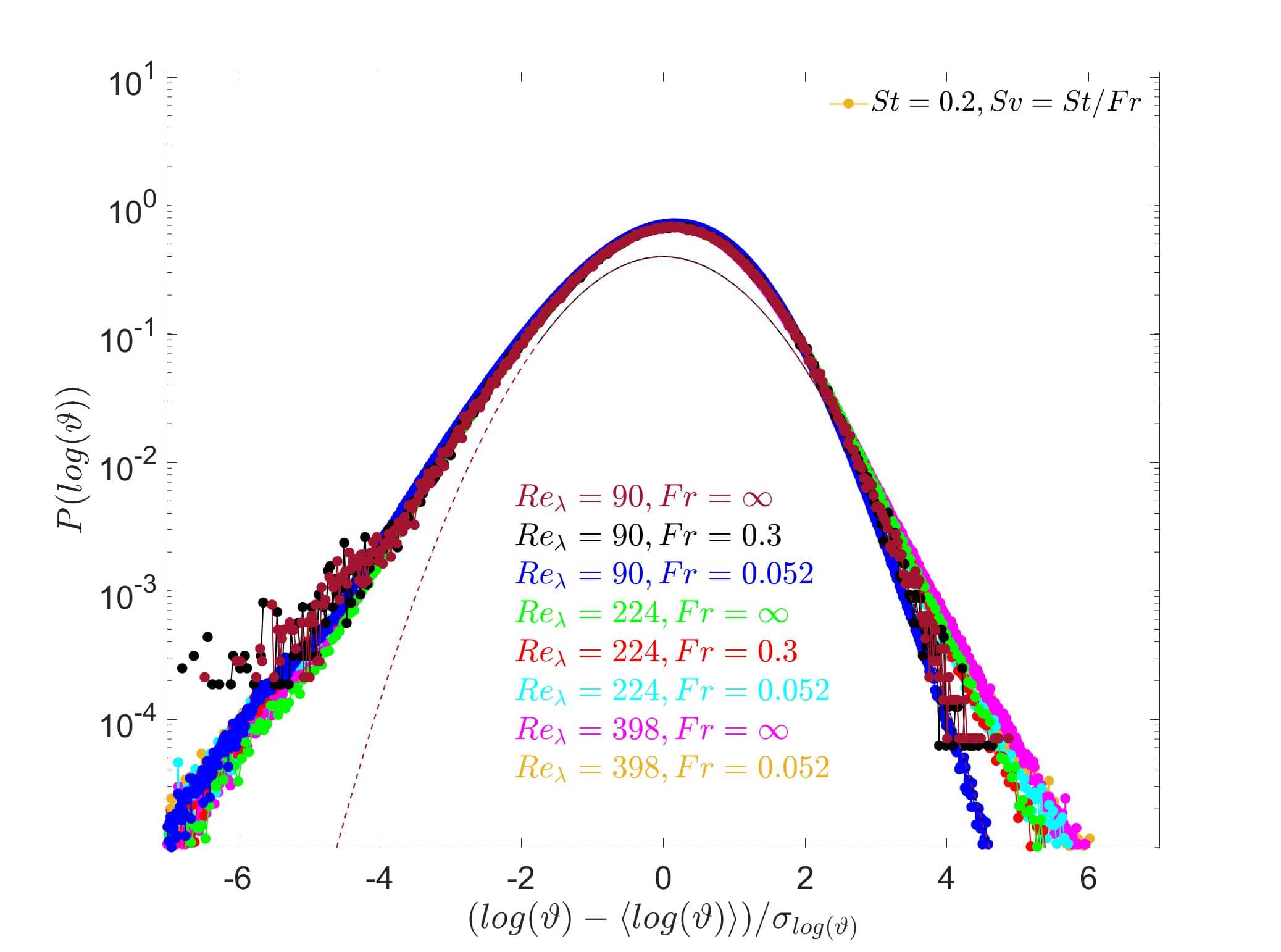}
		\caption{$St=0.2$ }
	\end{subfigure}
	
	\begin{subfigure}[b]{0.5\linewidth}
		\includegraphics[width=\linewidth]{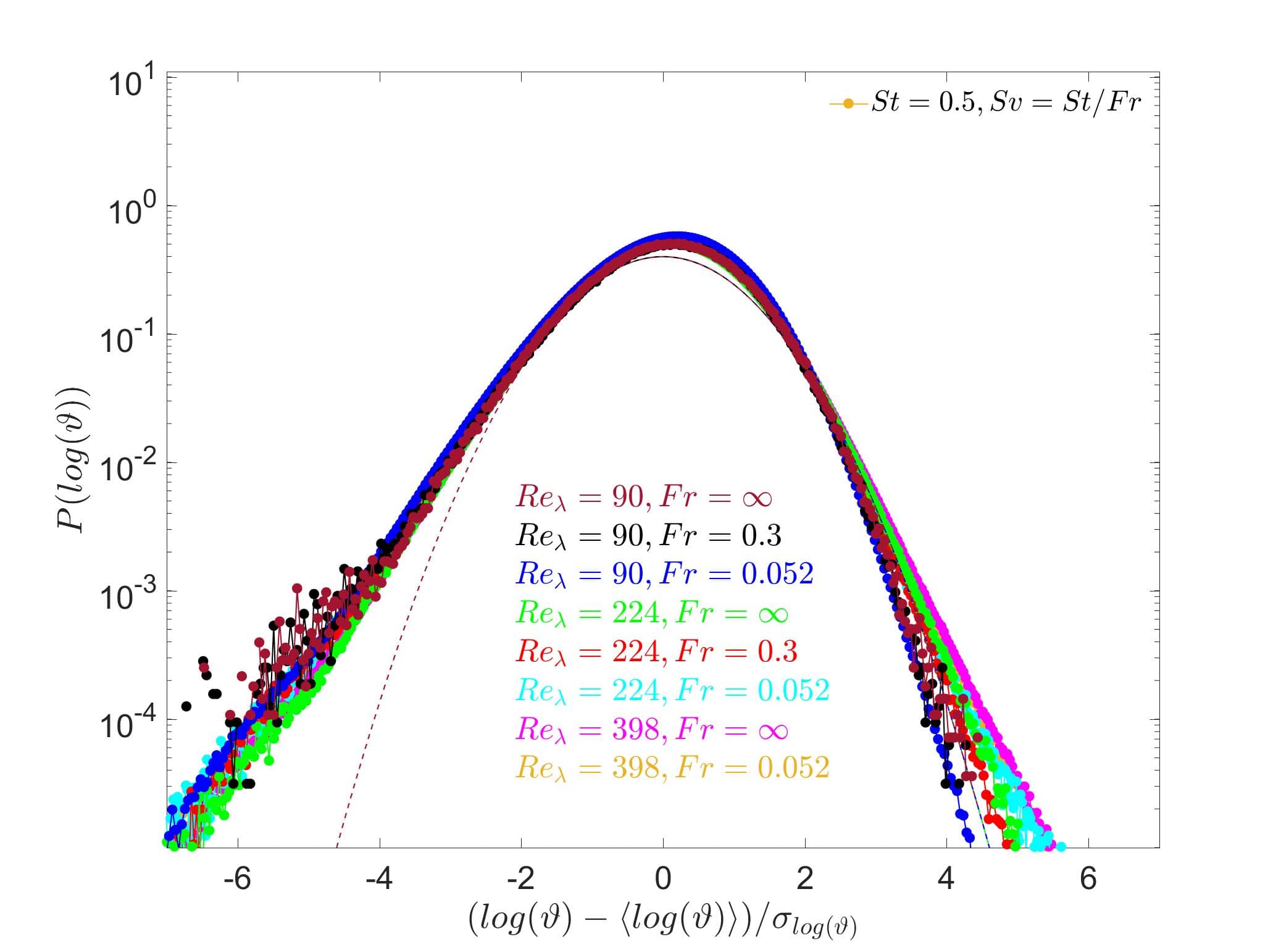}
		\caption{$St=0.5$}
	\end{subfigure}%
	\begin{subfigure}[b]{0.5\linewidth}
		\includegraphics[width=\linewidth]{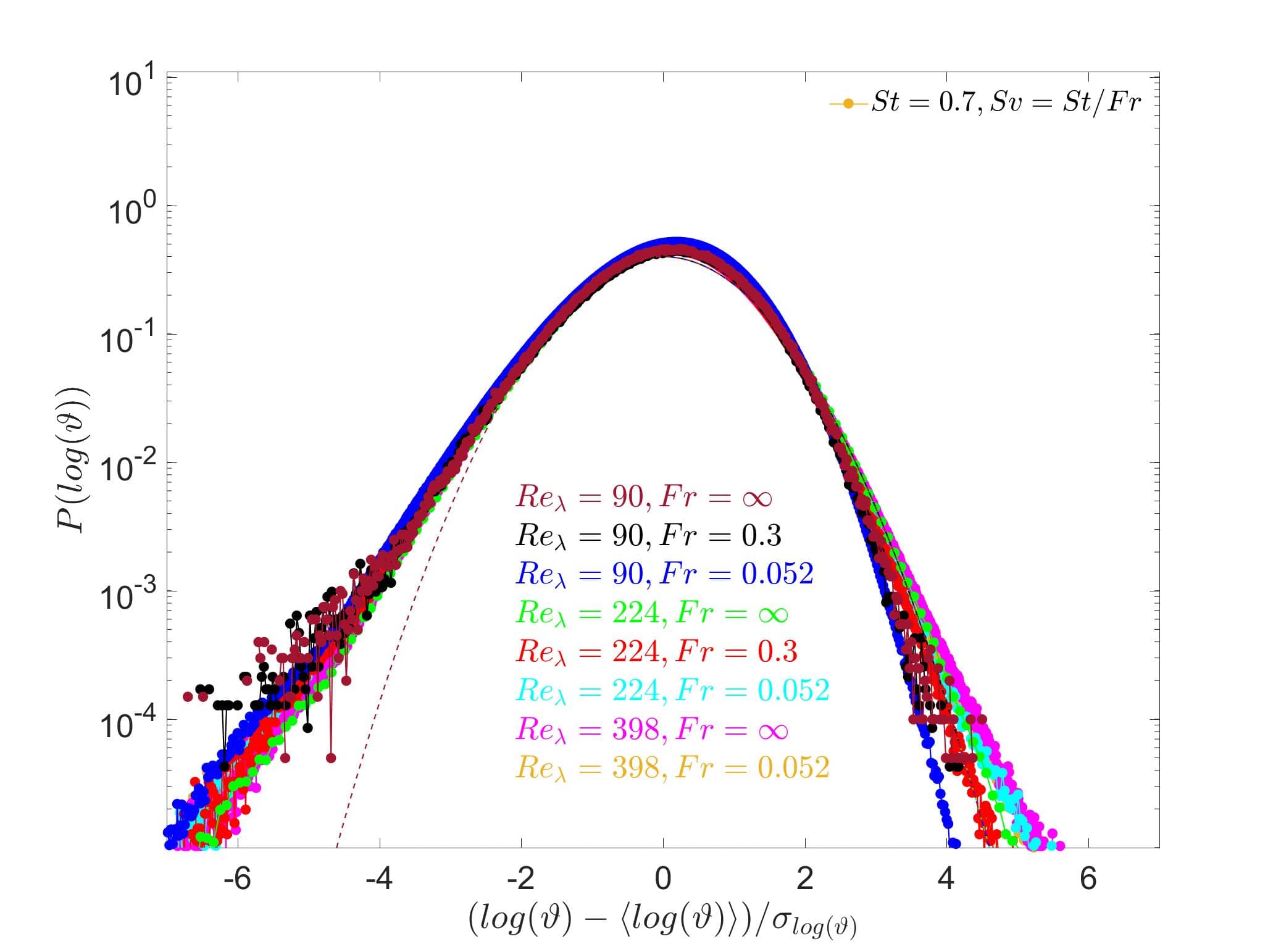}
		\caption{$St=0.7$ }
	\end{subfigure}
	\begin{subfigure}[b]{0.5\linewidth}
		\includegraphics[width=\linewidth]{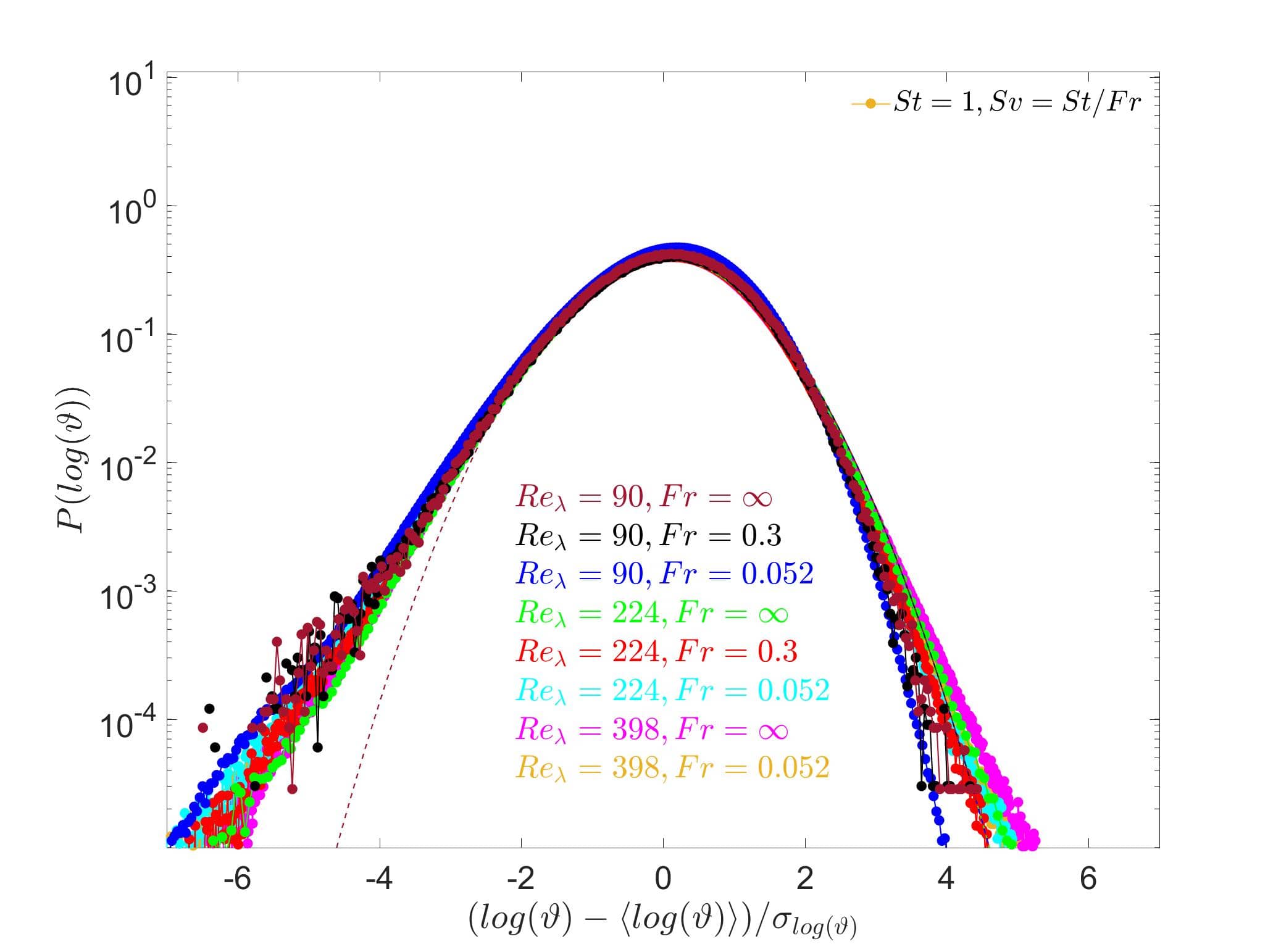}
		\caption{$St=1$ }
	\end{subfigure}%
	\begin{subfigure}[b]{0.5\linewidth}
		\includegraphics[width=\linewidth]{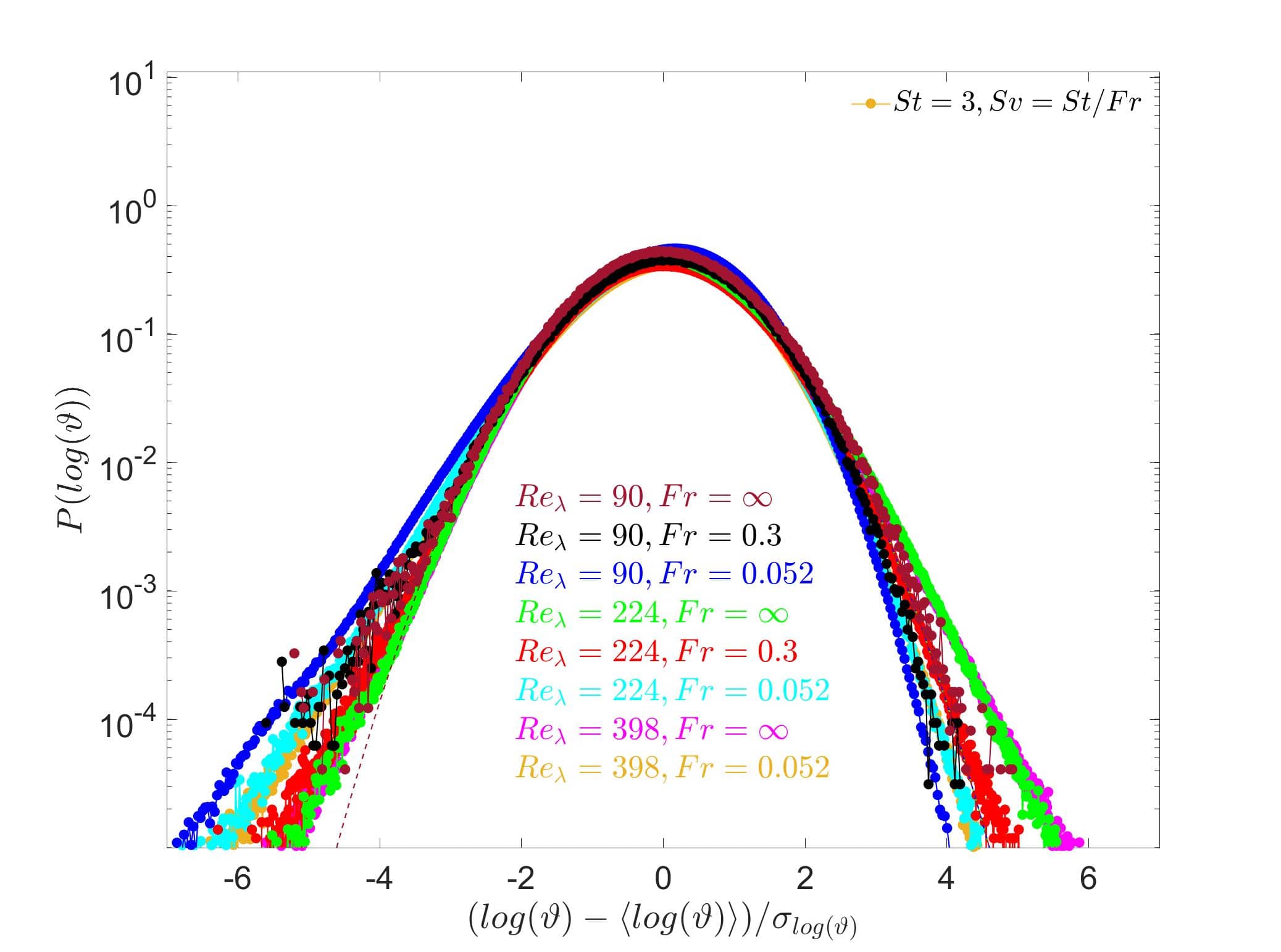}
		\caption{$St=3$ }
	\end{subfigure}
	\caption{ PDF of the log-normal distribution of Vorono\text{\"i} volumes (centered at the mean and normalized by standard deviation) at different cases of $Fr$ and $R_\lambda$ combinations for (a) $St=0$, (b) $St=0.2$, (c) $St=0.5$, (d) $St=0.7$,(e) $St=1$, and (f) $St=3$. Different colors represents different cases and dashed line denotes the Gaussian distribution.}\label{fig:CF_mixed_VT_PDF_Vol_logN_semilogy_St}
\end{figure}

In a number of previous studies it has been reported that the PDF of Vorono\text{\"i} volumes follows a log-normal distribution \cite{obligado2014preferential,monchaux2010preferential,sumbekova2017preferential,petersen2019experimental,dejoan2013preferential}. 
In figure \ref{fig:CF_mixed_VT_PDF_Vol_logN_semilogy_St} the centered and normalized PDFs of the logarithm of the Vorono\text{\"i} volumes are presented. The results show that for $St\ge0.5$ a Gaussian distribution can approximate the shape of $log(\vartheta)$ in the interval $\pm2\sigma_{log(\vartheta)}$, while large deviations from the Gaussian PDF are found outside of this range. In particular, the tails of the PDF are heavier than a Gaussian distribution, as also observed in \cite{dejoan2013preferential}. However, the PDF seems to approach a Gaussian shape as $St$ increases. For smaller $St$ the Gaussian PDF is not followed, which is to be expected since in the limit $St\to0$ the PDF follows that for a RPP, a shown earlier. Therefore, the log-normal shape for the PDF of Vorono\text{\"i} volumes is only a reasonable approximation in certain regimes. Departures from the log-normal behavior exhibit different (and not entirely clear) dependencies on $R_\lambda$ and $Fr$ for the left and right tails of the PDF, an explanation for which is not clear.


%
\FloatBarrier

\subsection{Particle settling velocities}

\begin{figure}
	\centering
	\begin{subfigure}[b]{0.5\linewidth}
		\includegraphics[width=\linewidth]{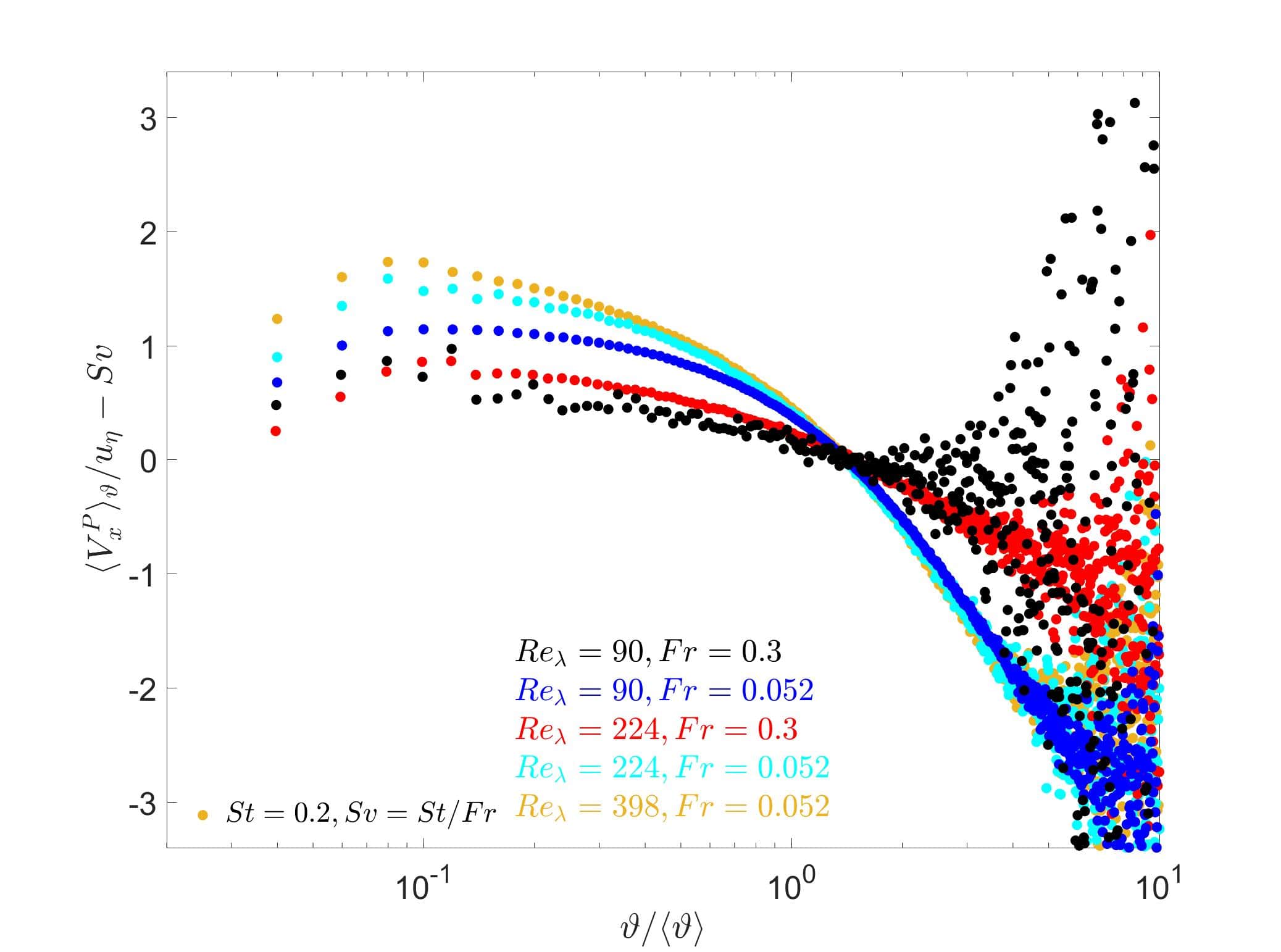}
		\caption{$St=0.2$ }
	\end{subfigure}%
	\begin{subfigure}[b]{0.5\linewidth}
		\includegraphics[width=\linewidth]{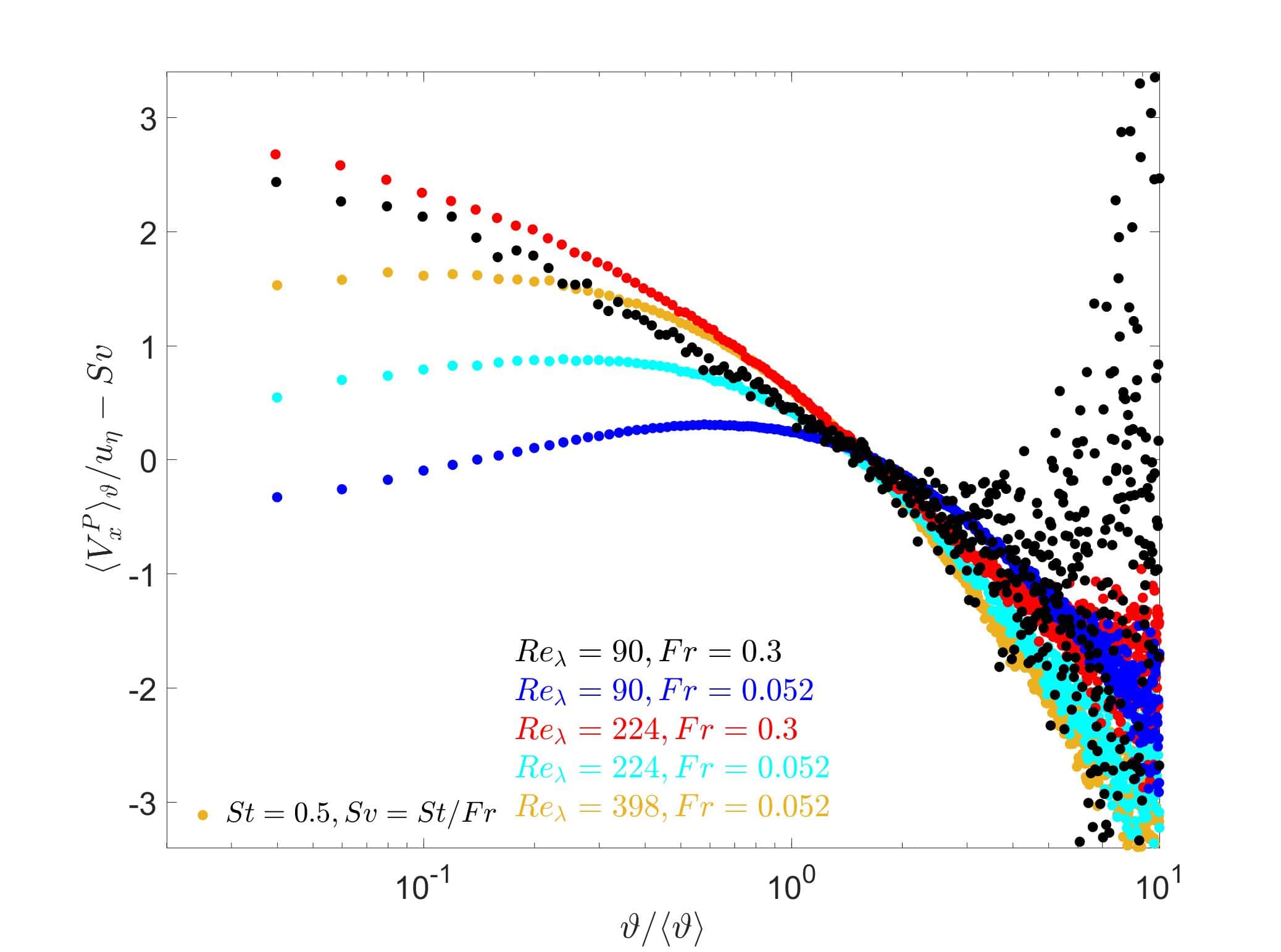}
		\caption{$St=0.5$}
	\end{subfigure}
	\begin{subfigure}[b]{0.5\linewidth}
		\includegraphics[width=\linewidth]{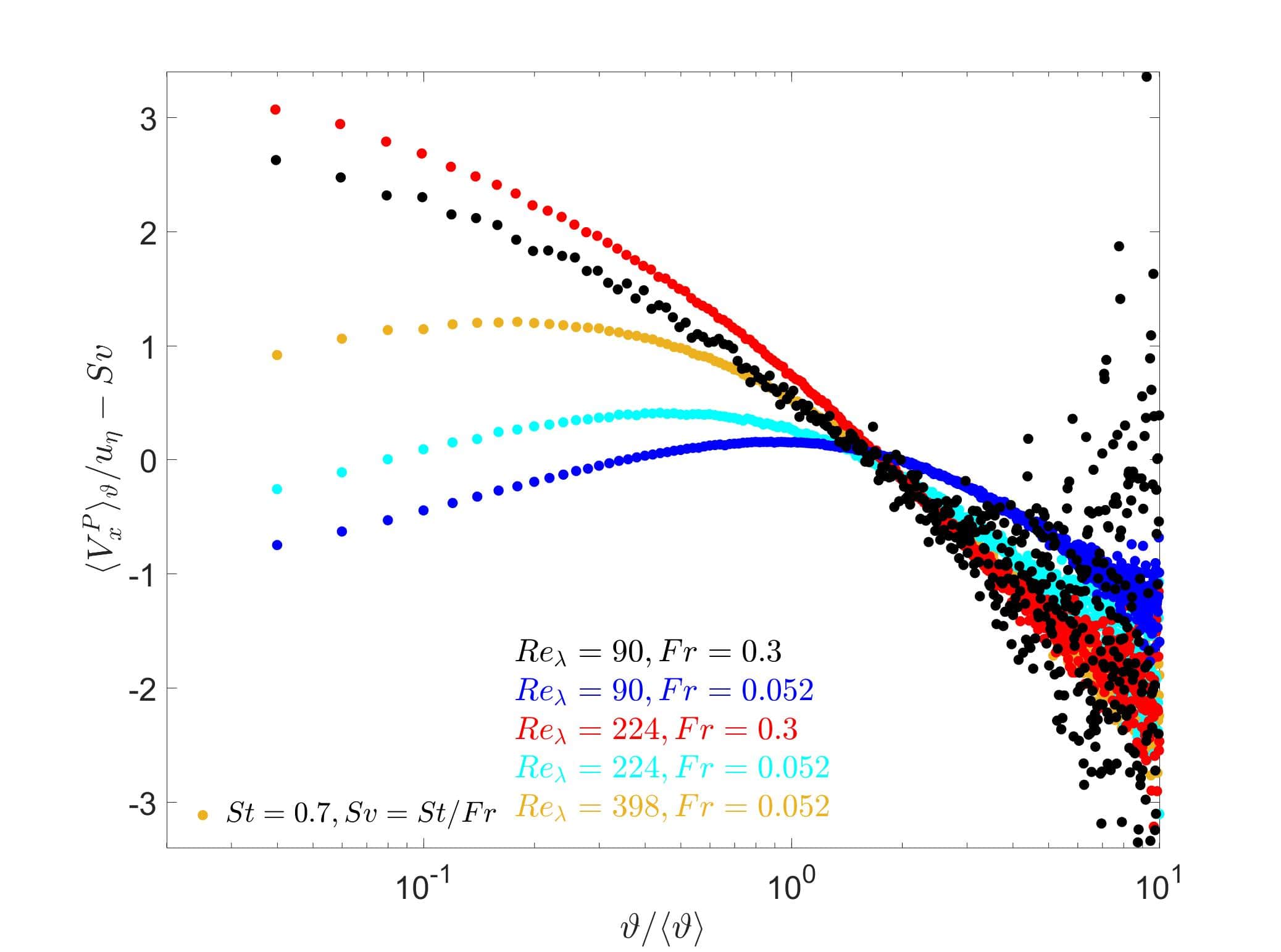}
		\caption{$St=0.7$ }
	\end{subfigure}%
	\begin{subfigure}[b]{0.5\linewidth}
		\includegraphics[width=\linewidth]{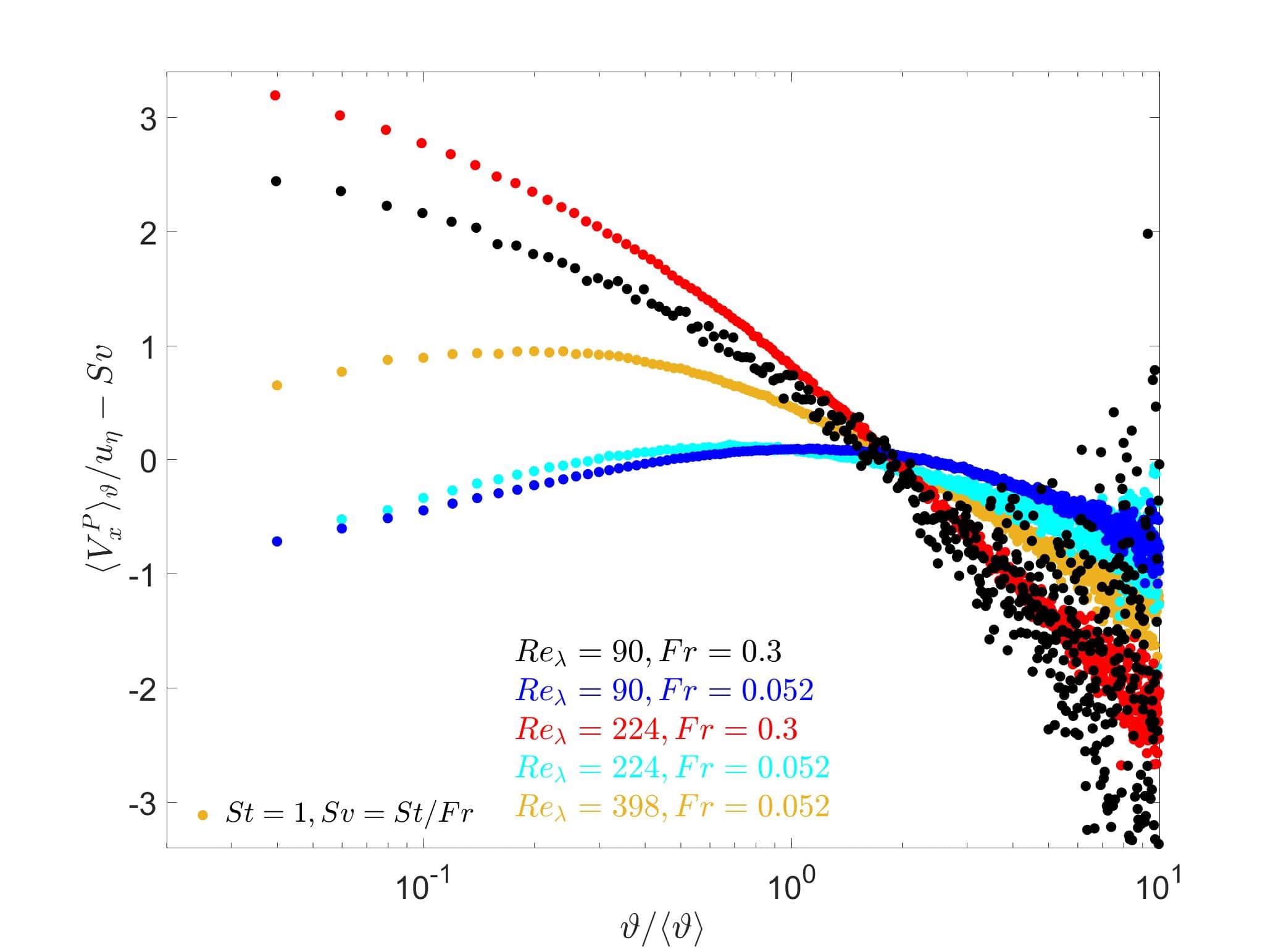}
		\caption{$St=1$ }
	\end{subfigure}%
	\caption{Average velocity of particles conditioned on the PDF of the Vorono\text{\"i} volumes at different cases of $Fr$ and $R_\lambda$ combinations for (a) $St=0$, (b) $St=0.2$, (c) $St=0.5$, (d) $St=0.7$, and (e) $St=1$. Different colors represents different cases.}\label{fig:VT_PDF_AvgVel_St}
\end{figure}

We now turn to consider the average velocity of particles conditioned on the Vorono\text{\"i} volumes to explore how the particle settling velocities depend upon the local particle concentration. The results in figure \ref{fig:VT_PDF_AvgVel_St} clearly illustrate the correlation between the particle settling velocity and the local concentration (inverse of the Vorono\text{\"i} volumes), as also observed over smaller ranges of $R_\lambda$ and $Fr$ in the DNS studies of \cite{monchaux2017settling,frankel2016settling,baker2017coherent} as well as the experimental study of \cite{aliseda2002effect}. The recent experiments of \cite{petersen2019experimental} also considered this quantity under a wide range of particle loadings, to examine the influence of two-way coupling between the particles and fluid on the particle settling velocities and its relation to the local concentration. Their results at low particle loadings are similar to ours, suggesting, as expected, that the one-way coupled assumption yields physically realistic results in that regime.

The results show that the $Fr$ for which the settling velocity reaches the largest value depends upon $St$, which is simply because in the particle equation of motion, the important non-dimensional number is the settling number $Sv\equiv St/Fr$. However, for $St>0.2$, the settling enhancement is strongest for intermediate values of $Fr$, in agreement with the DNS study of \cite{dejoan2013preferential}. For $Fr=0.3$, the average settling velocity enhances almost monotonically with increasing particle concentration (the decrease observed at small volumes for $St=0.2$ may be due to noise). However, for $Fr=0.052$, there is a local concentration at which the mean settling velocity is maximum, and then decreases for higher concentrations. If we assume that larger Vorono\text{\"i} volumes are associated with larger flow scales, then this is consistent with the arguments in \cite{tom2019} that as $Fr$ is decreased, the scales of the flow responsible for enhancing the particle settling velocity increase. The results also show that increasing $R_\lambda$ can significantly increase the settling velocities, especially for $St\geq 0.5$. This is explained by the recent theoretical analysis of \cite{tom2019} that shows that as $R_\lambda$ is increased, more and more energetic flow scales are introduced to the flow that are able to contribute to the particle settling velocity enhancement through the preferential sweeping mechanism described below.

Concerning the physical mechanism underlying the correlation between the settling velocity and concentration/clustering, since our DNS are one-way coupled the effect cannot be explained in terms of the particles modifying the local flow field, but in terms of the way the particles passively sample the underlying flow field. \cite{maxey87} provided a theoretical analysis and on the basis of this proposed that in the regime $St\ll1$, the particles are preferentially swept around the downward (direction of gravity) moving side of vortices in the flow, and that as a result of this preferential sweeping, their average settling velocity exceeds the Stokes settling velocity. This is the so-called ``preferential sweeping mechanism'' \cite{wang93}. This theoretical analysis was recently extended by \cite{tom2019} to arbitrary $St$, wherein the basic mechanism is the same as that proposed by Maxey, except that now the size of the vortices around which the particles are preferentially swept is shown to depend essentially on $St, Fr$ and $R_\lambda$, and is a multiscale preferential sweeping mechanism. However, it was pointed out in \cite{tom2019} that unlike the regime $St\ll1$, in the regime $St\geq O(1)$ the mechanism generating the clustering is distinct from the preferential sweeping mechanism. Indeed, for $St\geq O(1)$ the clustering is generated by a non-local mechanism and not the preferential sampling of the local flow field \cite{bragg14d,bragg15}. As a result, while the results in figure \ref{fig:VT_PDF_AvgVel_St} for $St\ll1$ can be directly explained in terms of the preferential sweeping mechanism, the same does not apply for $St\geq O(1)$. A direct and unambiguous test of the preferential sweeping mechanism would be to condition the average particle velocity not on the concentration, but on some measure of the local preferential sampling of the flow.

\FloatBarrier

\subsection{Local Analysis of Particle Accelerations}

In \cite{parishani2015effects,ireland2016effectb} it was shown that settling can significantly enhance the accelerations of inertial particles in turbulent flows, and in \cite{momenifar2019influence} it was also shown that the variance of the particle accelerations generally increase with increasing $R_\lambda$ and/or with decreasing $Fr$. To gain further insight into the behavior of the particle accelerations, we perform a local analysis by conditioning the particle acceleration behavior on the local Vorono\text{\"i} volume, to understand how the accelerations are linked to the concentration field.

\begin{figure}
	\vspace{-0.7in}
	\centering
	\begin{subfigure}[b]{0.5\linewidth}
			\includegraphics[width=\linewidth]{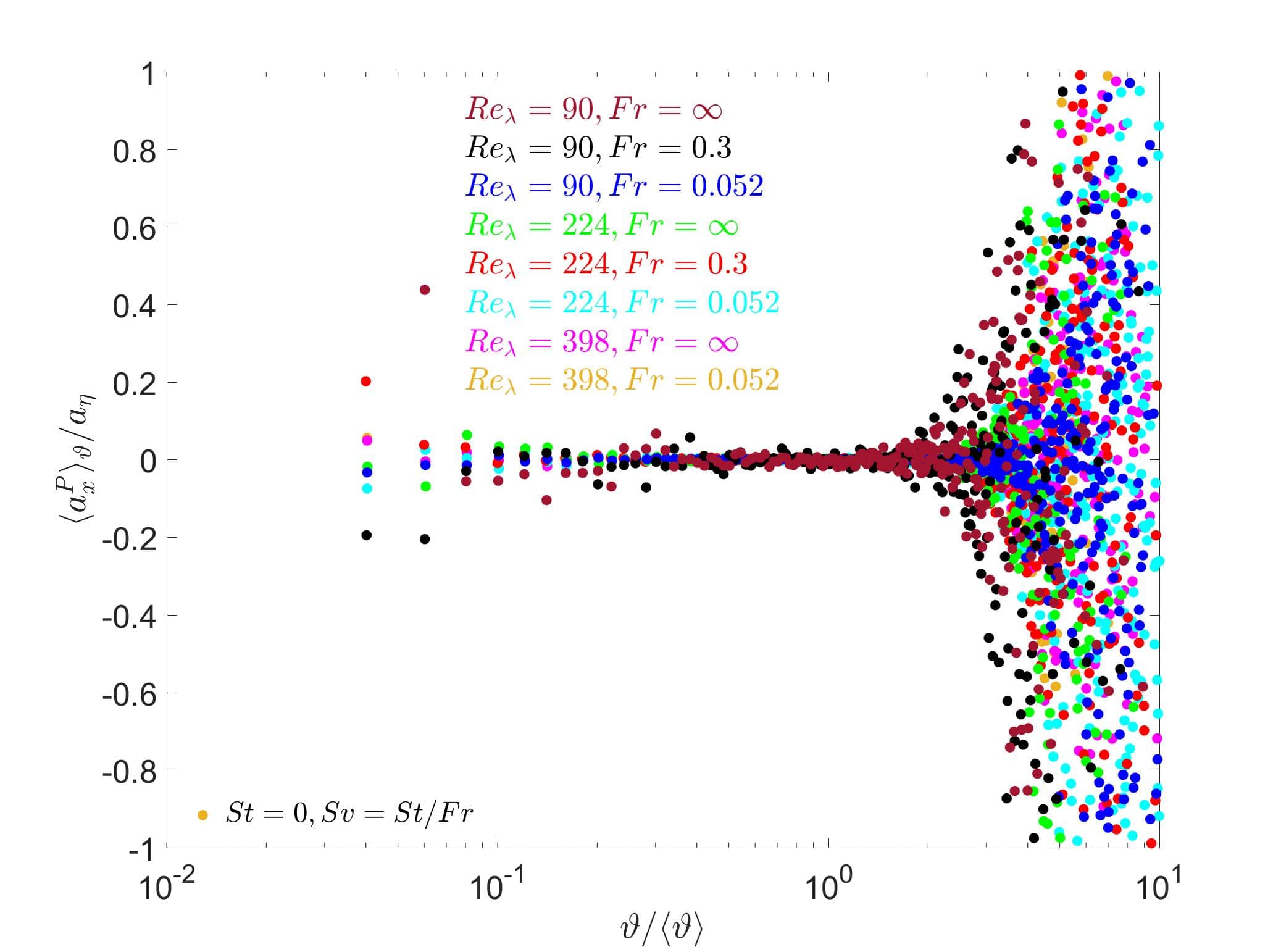}
			\caption{$St=0$}
	\end{subfigure}%
	\begin{subfigure}[b]{0.5\linewidth}
		\includegraphics[width=\linewidth]{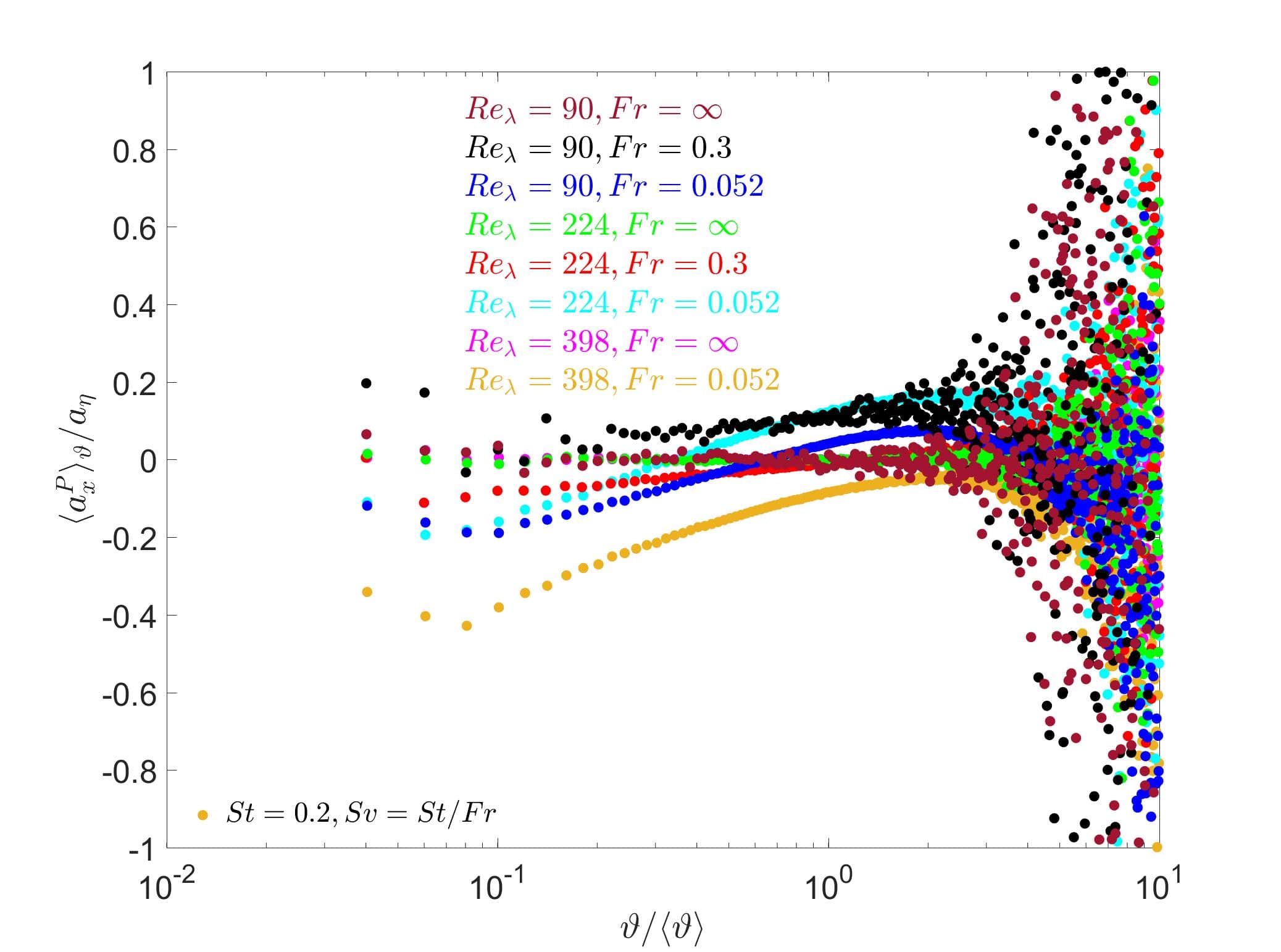}
		\caption{$St=0.2$ }
	\end{subfigure}	
	\begin{subfigure}[b]{0.5\linewidth}
		\includegraphics[width=\linewidth]{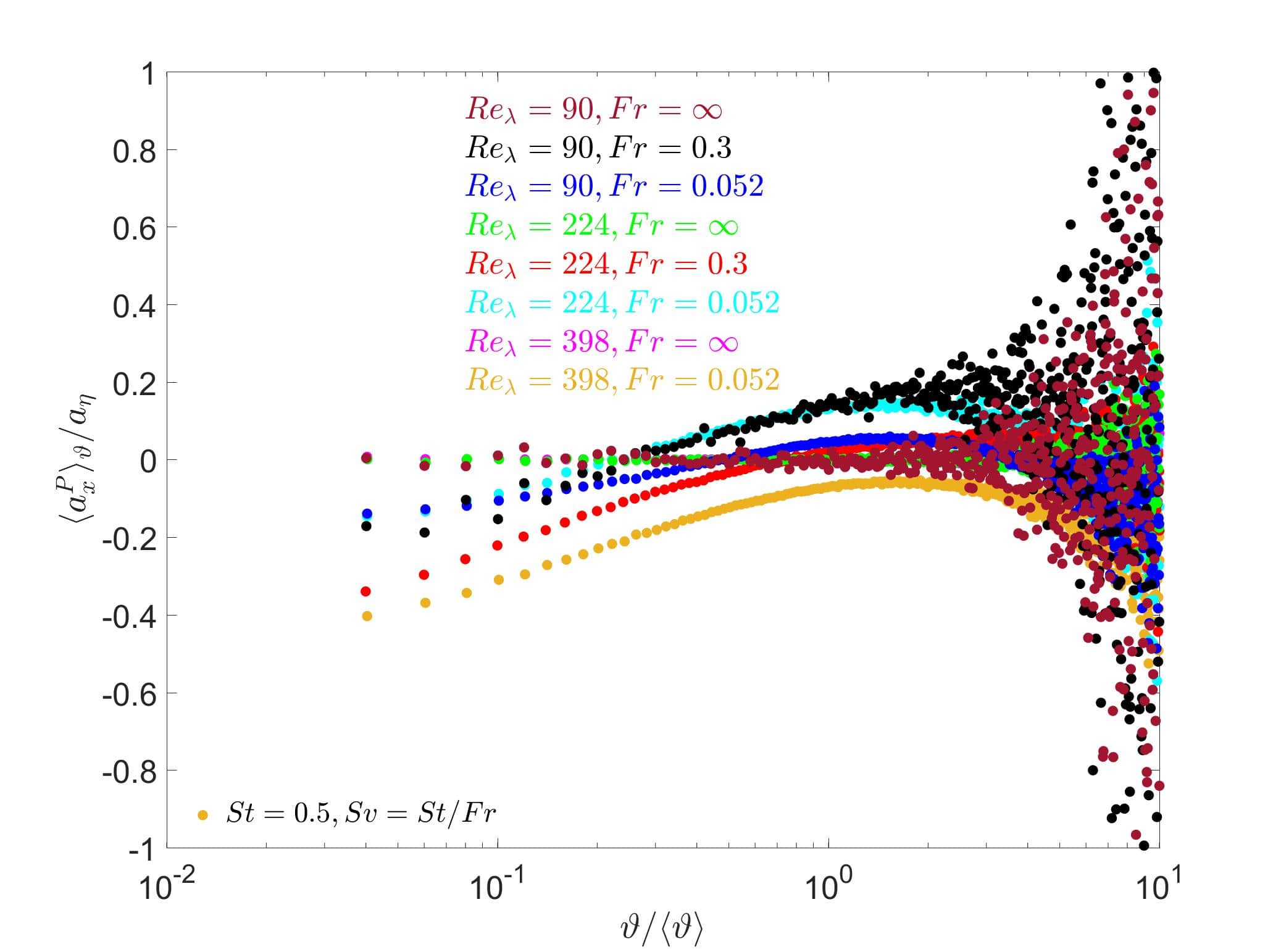}
		\caption{$St=0.5$}
	\end{subfigure}%
	\begin{subfigure}[b]{0.5\linewidth}
		\includegraphics[width=\linewidth]{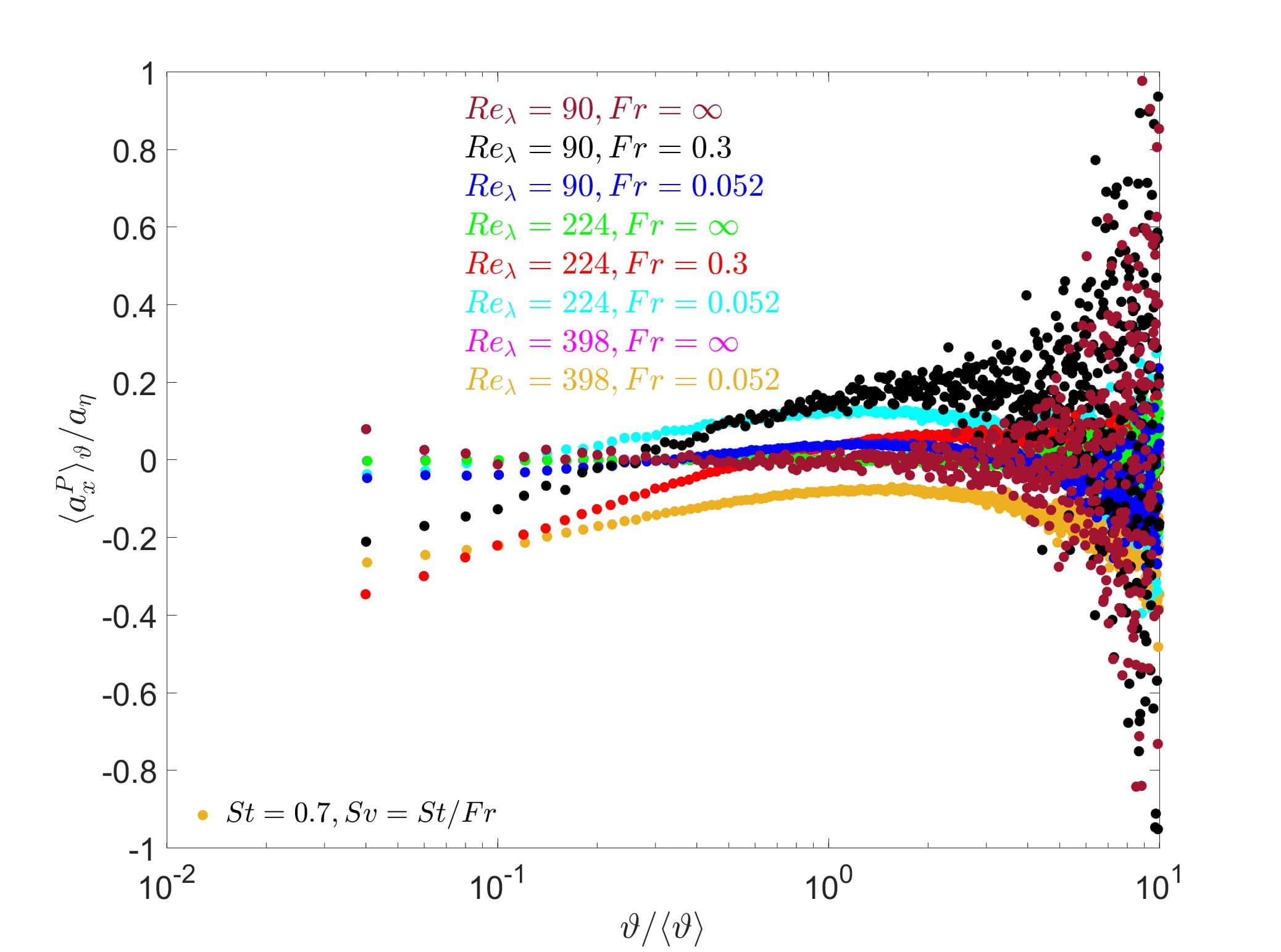}
		\caption{$St=0.7$ }
	\end{subfigure}
	\begin{subfigure}[b]{0.5\linewidth}
		\includegraphics[width=\linewidth]{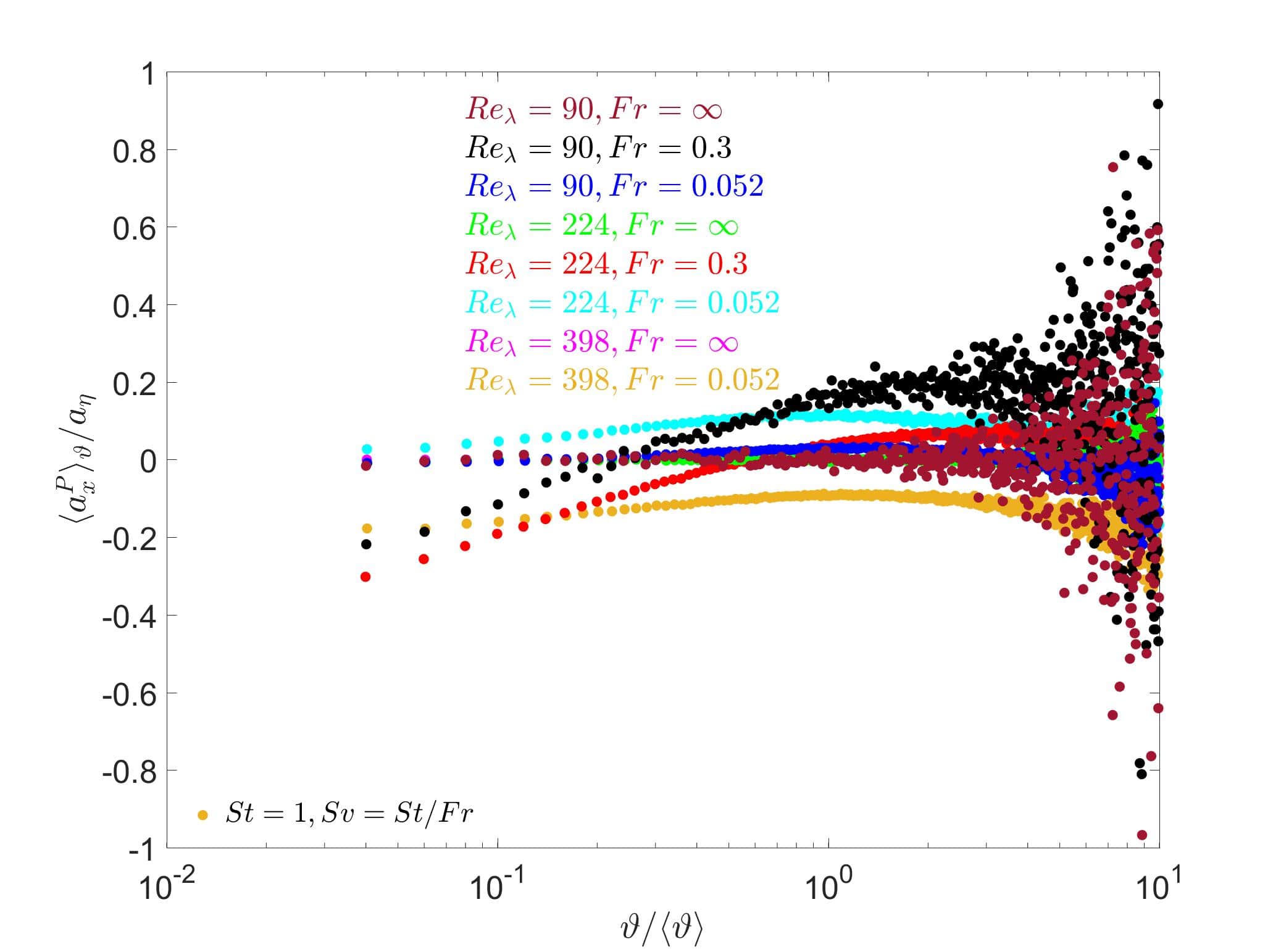}
		\caption{$St=1$ }
	\end{subfigure}%
		\begin{subfigure}[b]{0.5\linewidth}
			\includegraphics[width=\linewidth]{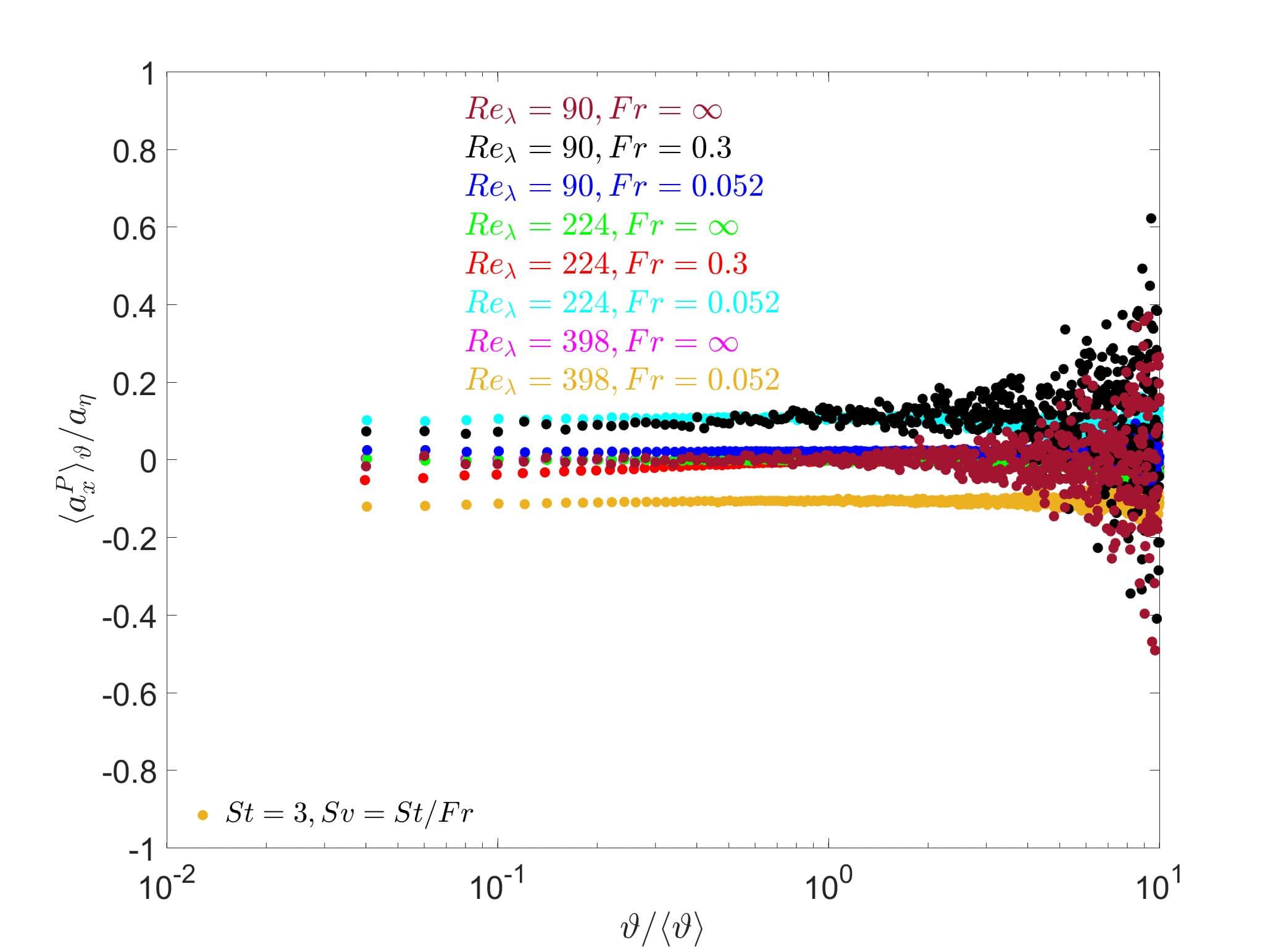}
			\caption{$St=3$ }
		\end{subfigure}
	\caption{ Average acceleration of particles, in the gravity direction, conditioned on the PDF of the Vorono\text{\"i} volumes at different cases of $Fr$ and $R_\lambda$ combinations for  (a) $St=0$, (b) $St=0.2$, (c) $St=0.5$, (d) $St=0.7$,(e) $St=1$, and (f) $St=3$. Different colors represents different cases.}\label{fig:VT_PDF_PAccel_x_Cnd_St}
\end{figure}

\begin{figure}
	\vspace{-0.7in}
	\centering
	\begin{subfigure}[b]{0.5\linewidth}
		\includegraphics[width=\linewidth]{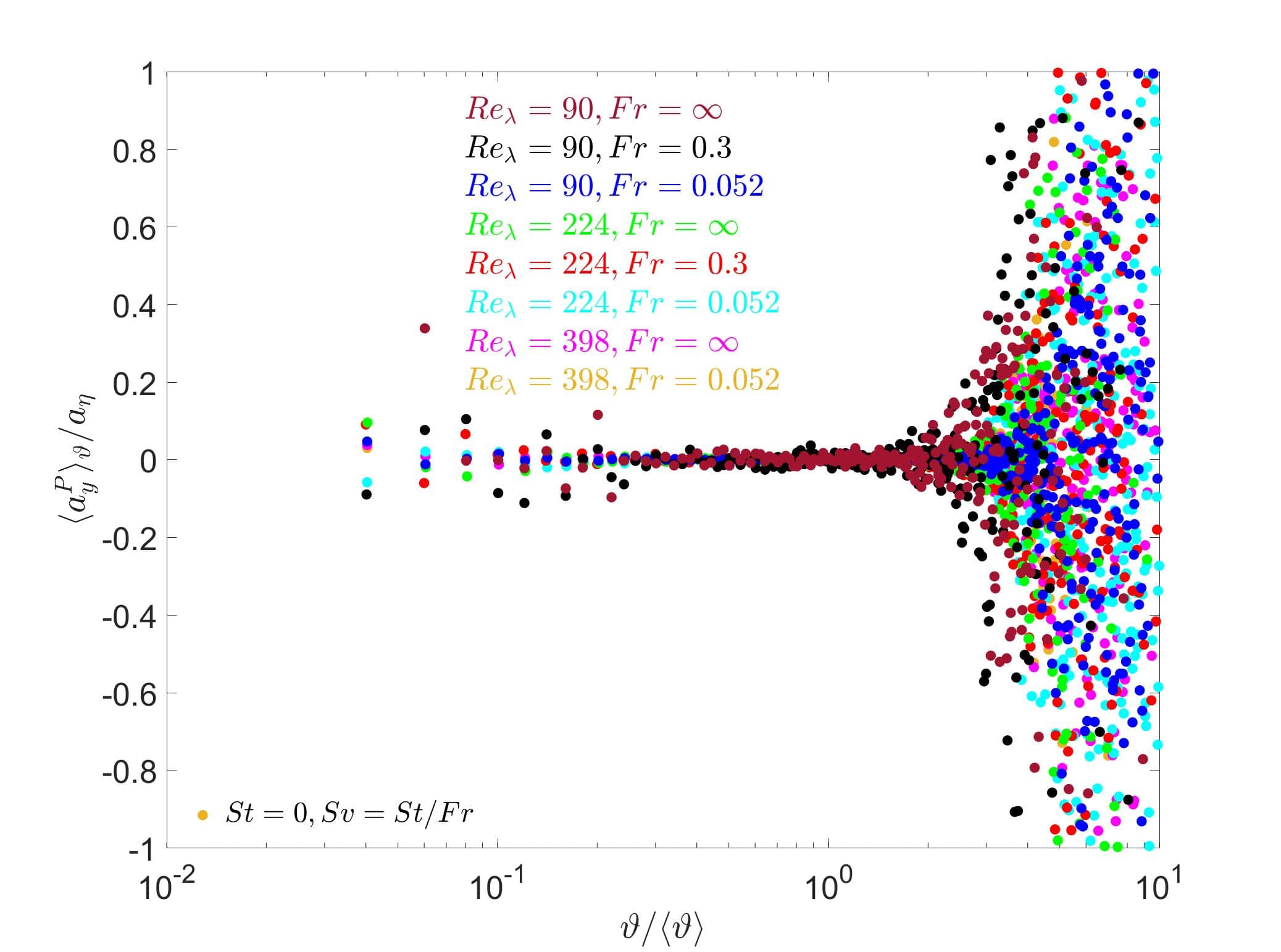}
		\caption{$St=0$}
	\end{subfigure}%
	\begin{subfigure}[b]{0.5\linewidth}
		\includegraphics[width=\linewidth]{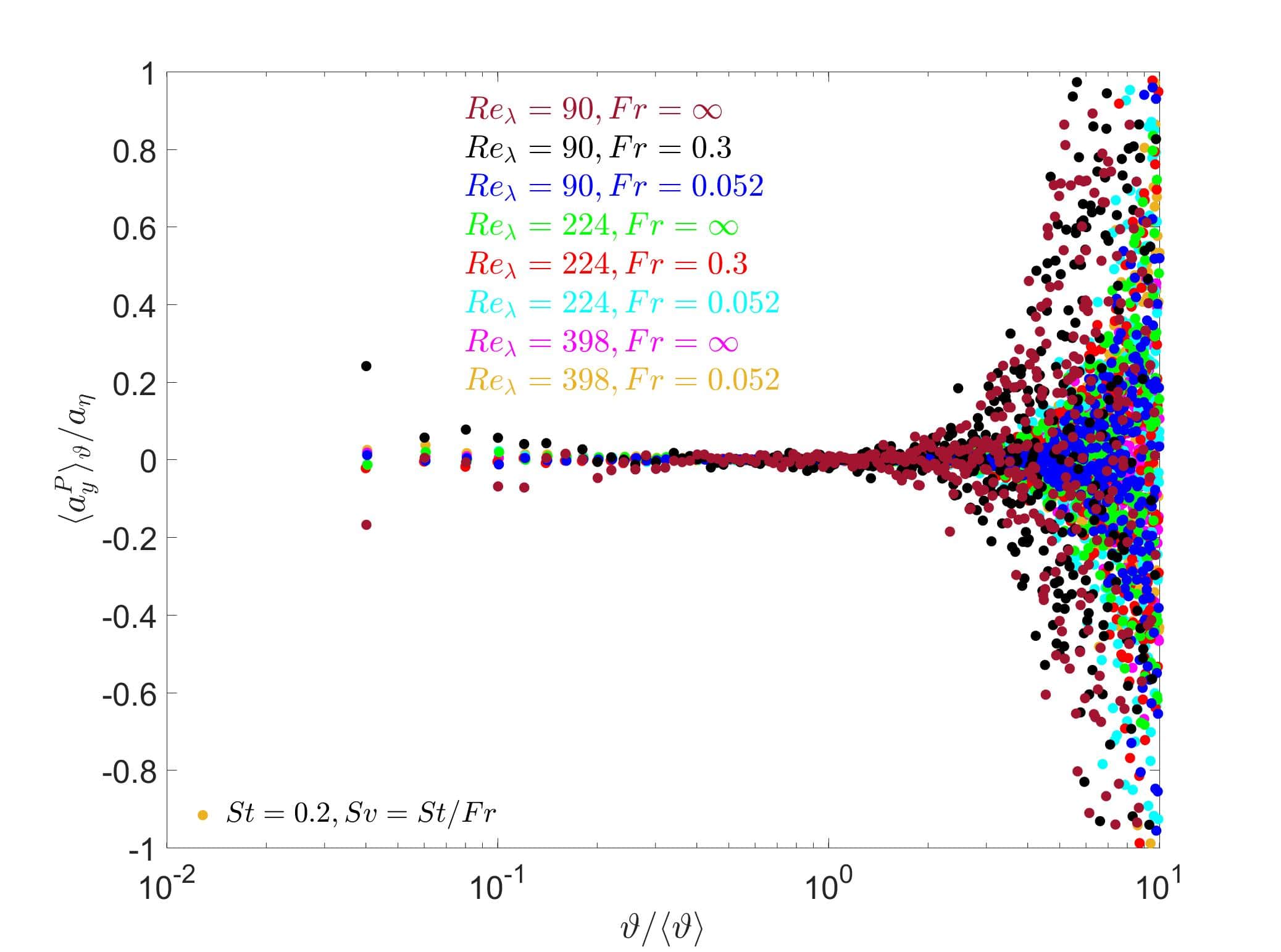}
		\caption{$St=0.2$ }
	\end{subfigure}	
	\begin{subfigure}[b]{0.5\linewidth}
		\includegraphics[width=\linewidth]{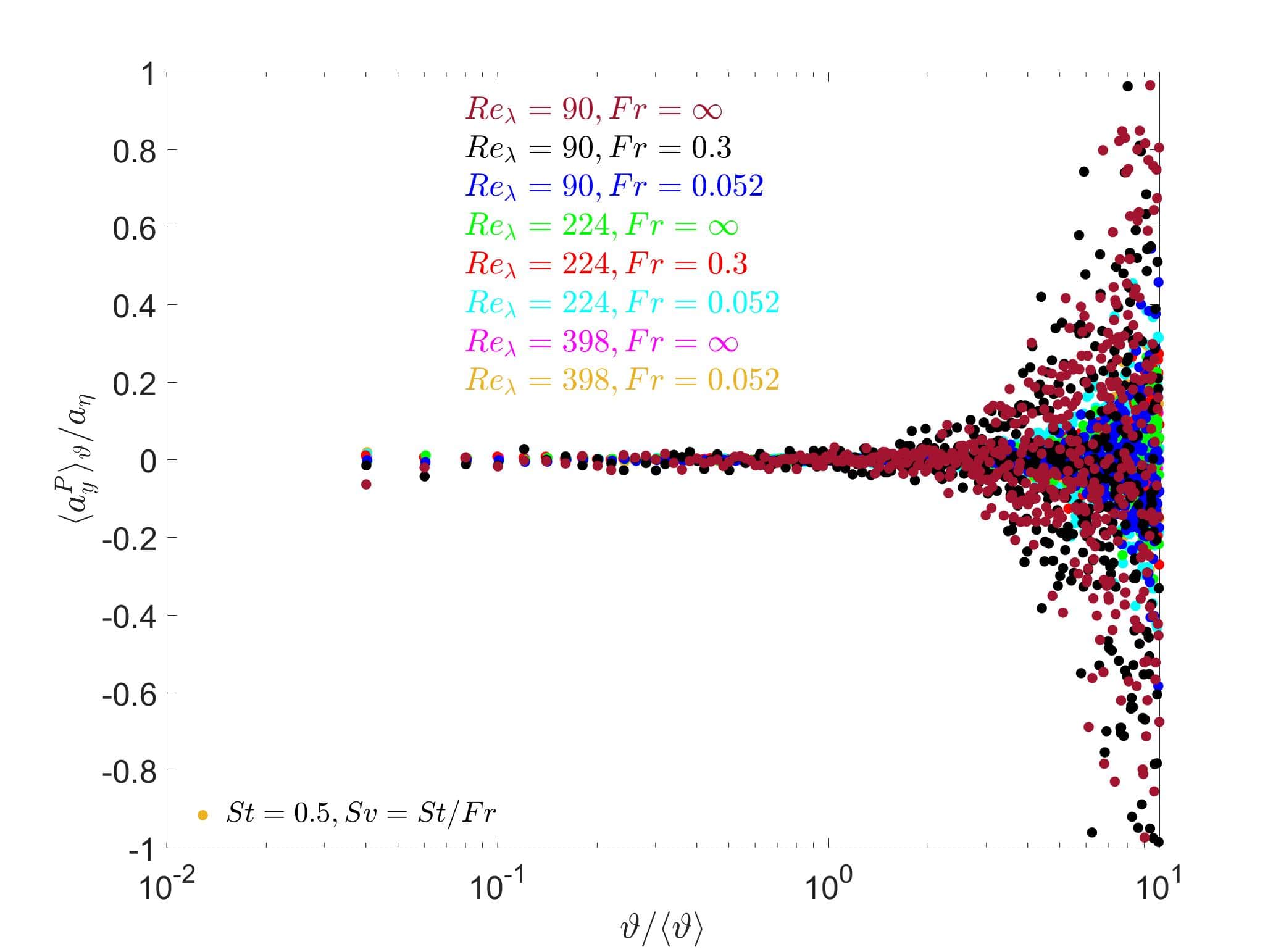}
		\caption{$St=0.5$}
	\end{subfigure}%
	\begin{subfigure}[b]{0.5\linewidth}
		\includegraphics[width=\linewidth]{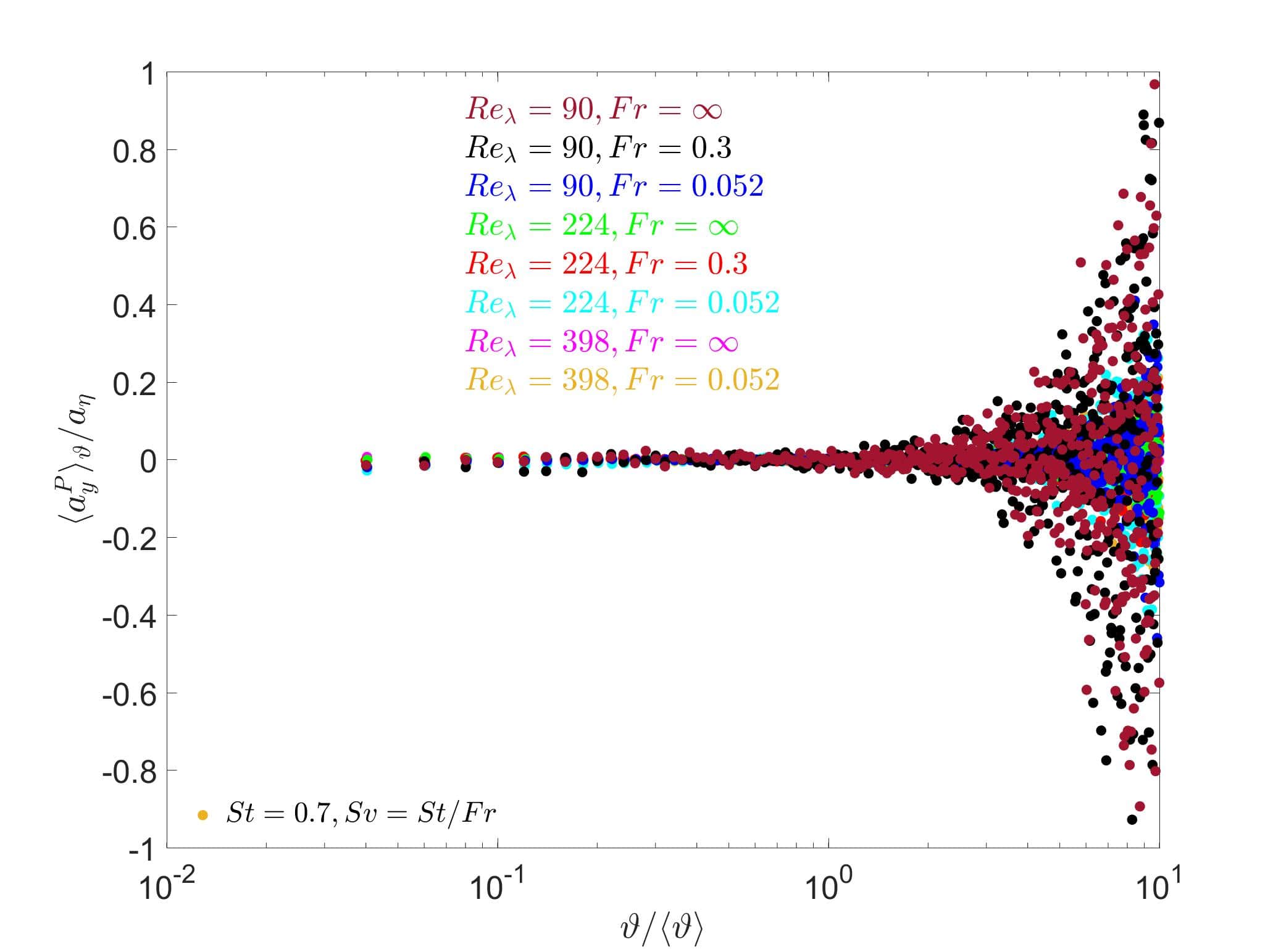}
		\caption{$St=0.7$ }
	\end{subfigure}
	\begin{subfigure}[b]{0.5\linewidth}
		\includegraphics[width=\linewidth]{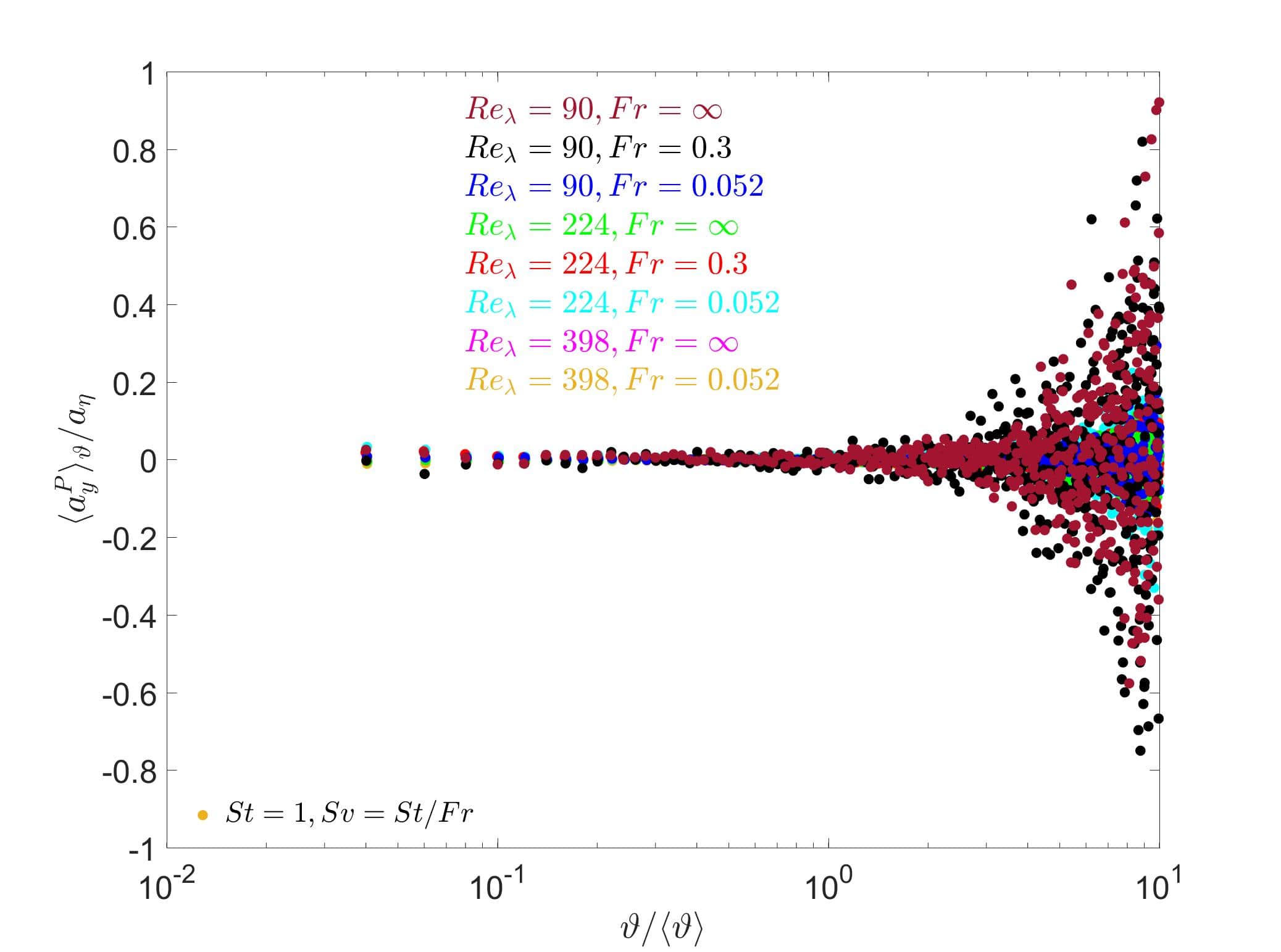}
		\caption{$St=1$ }
	\end{subfigure}%
	\begin{subfigure}[b]{0.5\linewidth}
		\includegraphics[width=\linewidth]{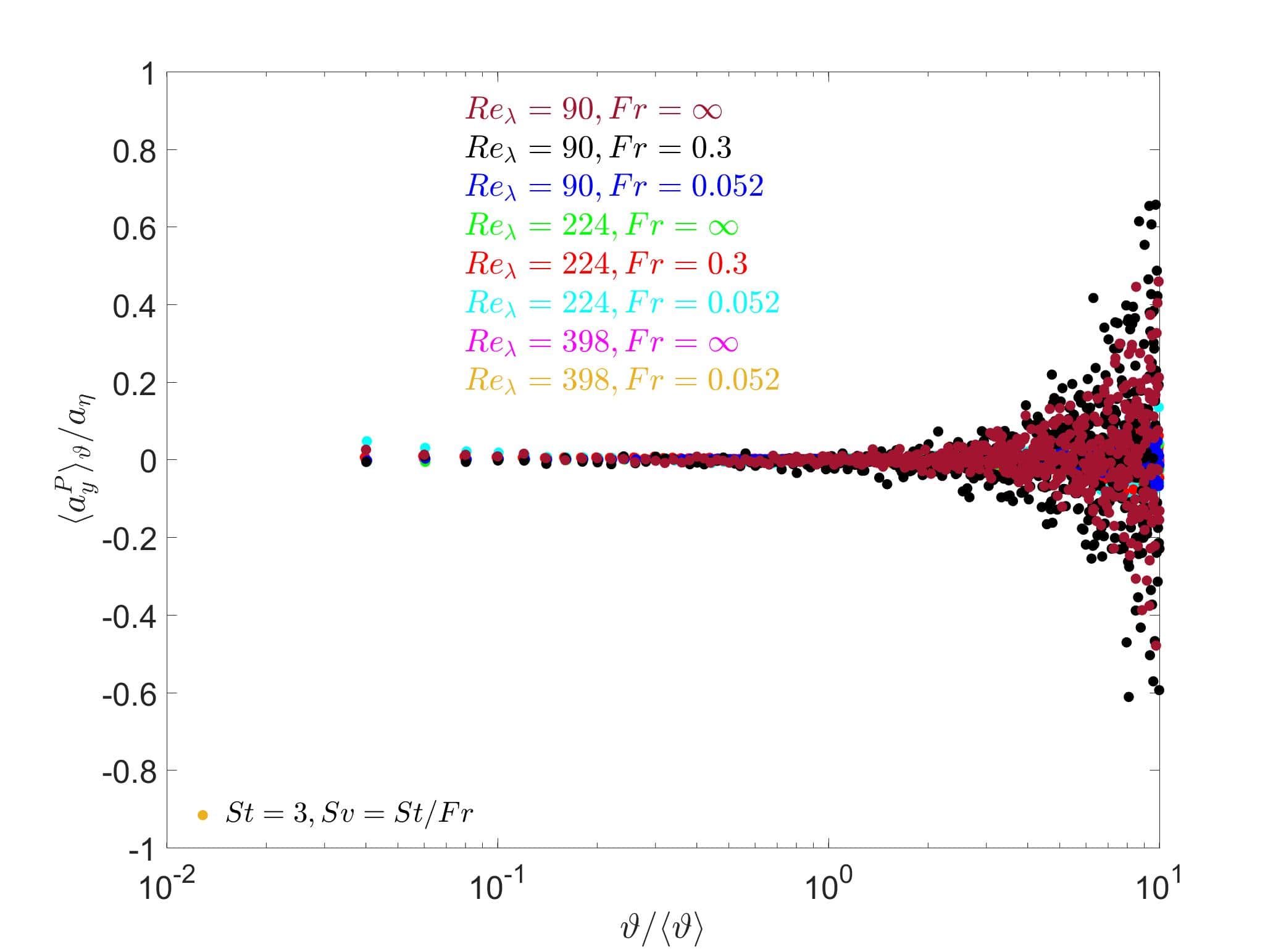}
		\caption{$St=3$ }
	\end{subfigure}
	\caption{Average acceleration of particles, in the plane normal to the gravity direction, conditioned on the PDF of the Vorono\text{\"i} volumes at different cases of $Fr$ and $R_\lambda$ combinations for  (a) $St=0$, (b) $St=0.2$, (c) $St=0.5$, (d) $St=0.7$,(e) $St=1$, and (f) $St=3$. Different colors represents different cases.}\label{fig:VT_PDF_PAccel_y_Cnd_St}
\end{figure}

Figures \ref{fig:VT_PDF_PAccel_x_Cnd_St} and \ref{fig:VT_PDF_PAccel_y_Cnd_St} show results for the particle acceleration ($\mathbf{a}^p(t)\equiv\dot{\mathbf{v}}^p(t)$) conditioned on the local Vorono\text{\"i} volume, in the direction parallel and perpendicular to gravity, respectively. 
According to these results, in the absence of gravity ($Fr=\infty$), the particle accelerations are zero (to within statistical noise) in both directions, independent of $St$ and $R_\lambda$. In the presence of gravity, this same behavior is observed for the accelerations in the direction normal to gravity, however, in the direction of gravity the quantity becomes finite for finite $St$. In particular, $\langle a^p_x\rangle_\vartheta$ becomes negative at small volumes and positive for large volumes. Moreover, $\langle a^p_x\rangle_\vartheta$ becomes increasingly negative as $R_\lambda$ is increased. Some of this behavior may be understood by considering the following identity
\begin{align}
\langle a^p_x\rangle\equiv\int_0^\infty \langle a^p_x\rangle_\vartheta P(\vartheta)\,d\vartheta.\label{Accd}
\end{align}
Since $P(\vartheta)\geq 0$, and $\langle a^p_x\rangle=0$ for statistically stationary, homogeneous turbulence, then the only way for \eqref{Accd} to be satisfied is if either $\langle a^p_x\rangle_\vartheta=0\,\forall \vartheta$, or else its sign varies with $\vartheta$. The latter situation is what we observe in the presence of gravity, with particles in high concentration regions (small $\vartheta$) exhibiting a net acceleration in the direction of gravity even though the net acceleration of all the particles is zero, i.e. $\langle a^p_x\rangle=0$. However, we are not able to explain why $\langle a^p_x\rangle_\vartheta$ is negative, rather than positive at small $\vartheta$.

\FloatBarrier
\FloatBarrier

\begin{figure}
	\vspace{-0.8in}
	\centering
	\begin{subfigure}[b]{0.5\linewidth}
		\includegraphics[width=\linewidth]{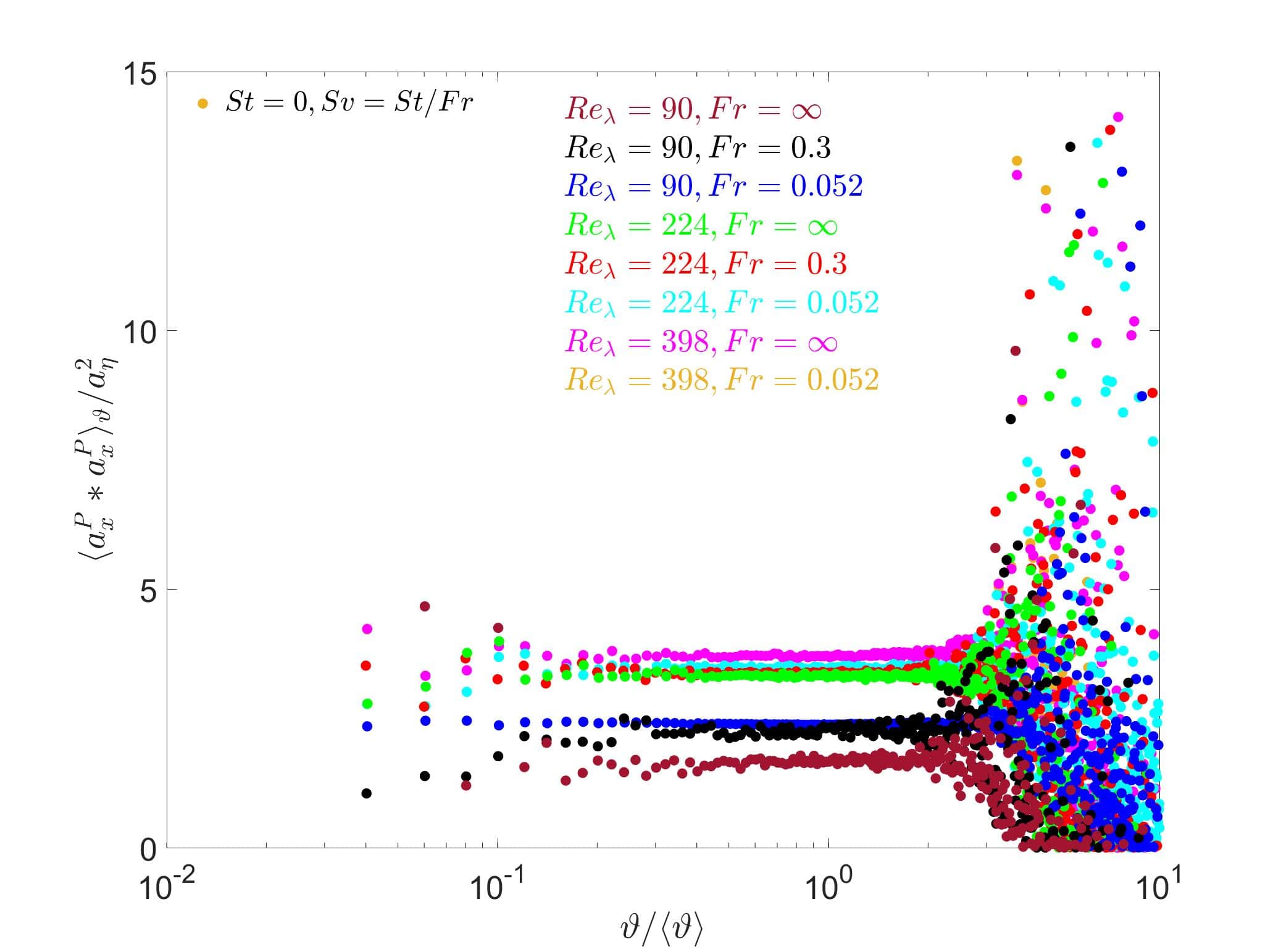}
		\caption{$St=0$}
	\end{subfigure}%
	\begin{subfigure}[b]{0.5\linewidth}
		\includegraphics[width=\linewidth]{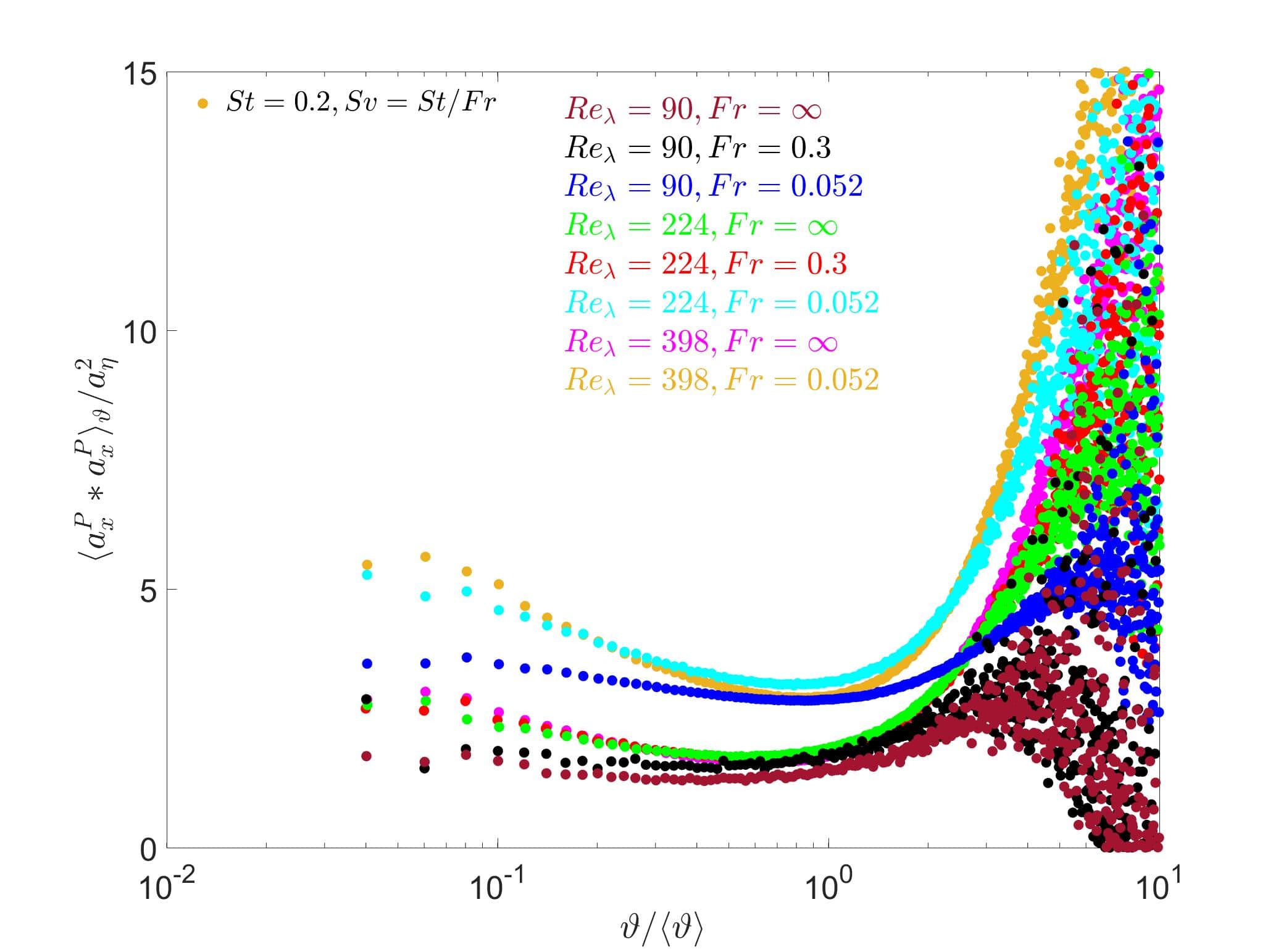}
		\caption{$St=0.2$ }
	\end{subfigure}	
	\begin{subfigure}[b]{0.5\linewidth}
		\includegraphics[width=\linewidth]{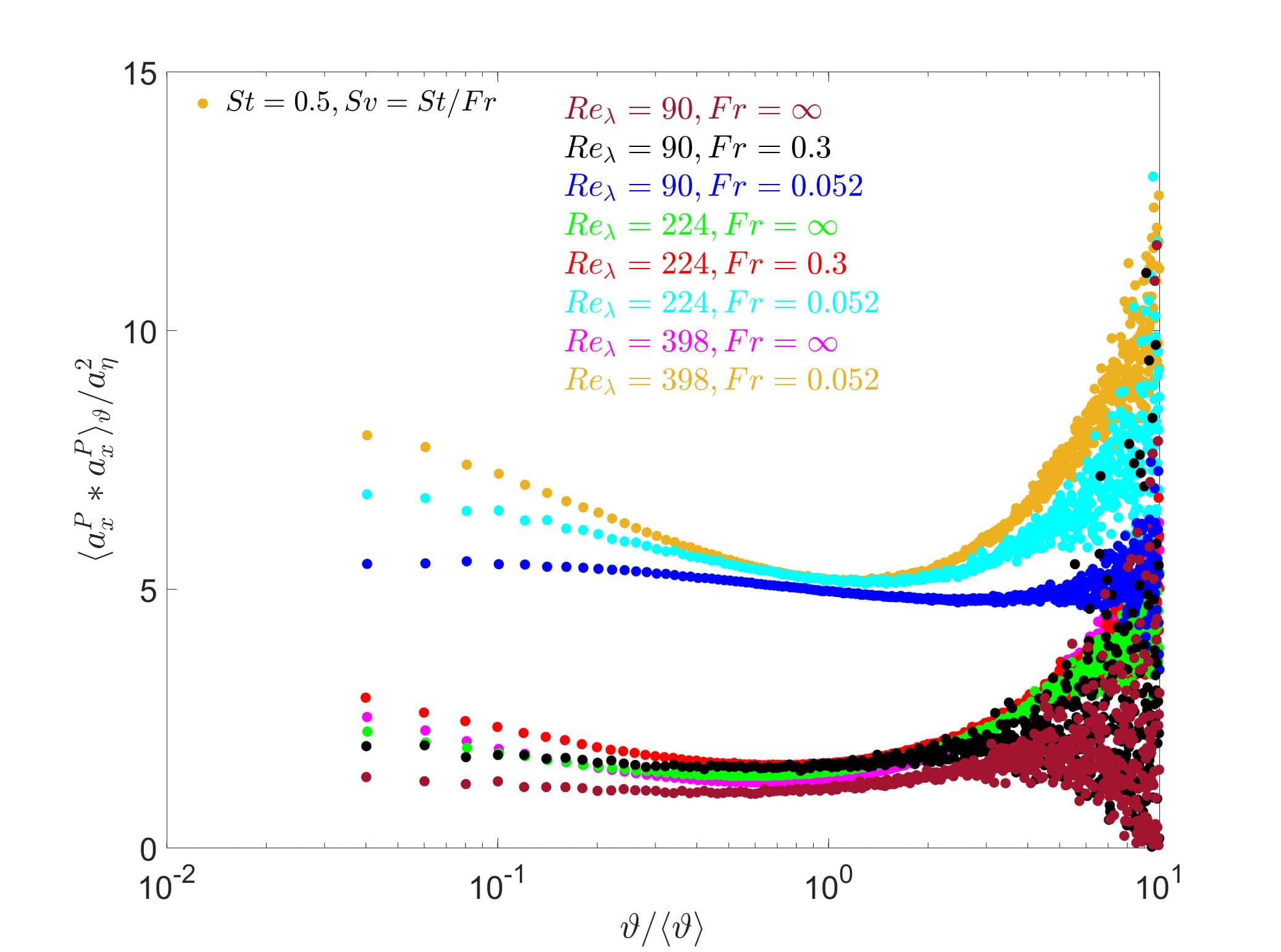}
		\caption{$St=0.5$}
	\end{subfigure}%
	\begin{subfigure}[b]{0.5\linewidth}
		\includegraphics[width=\linewidth]{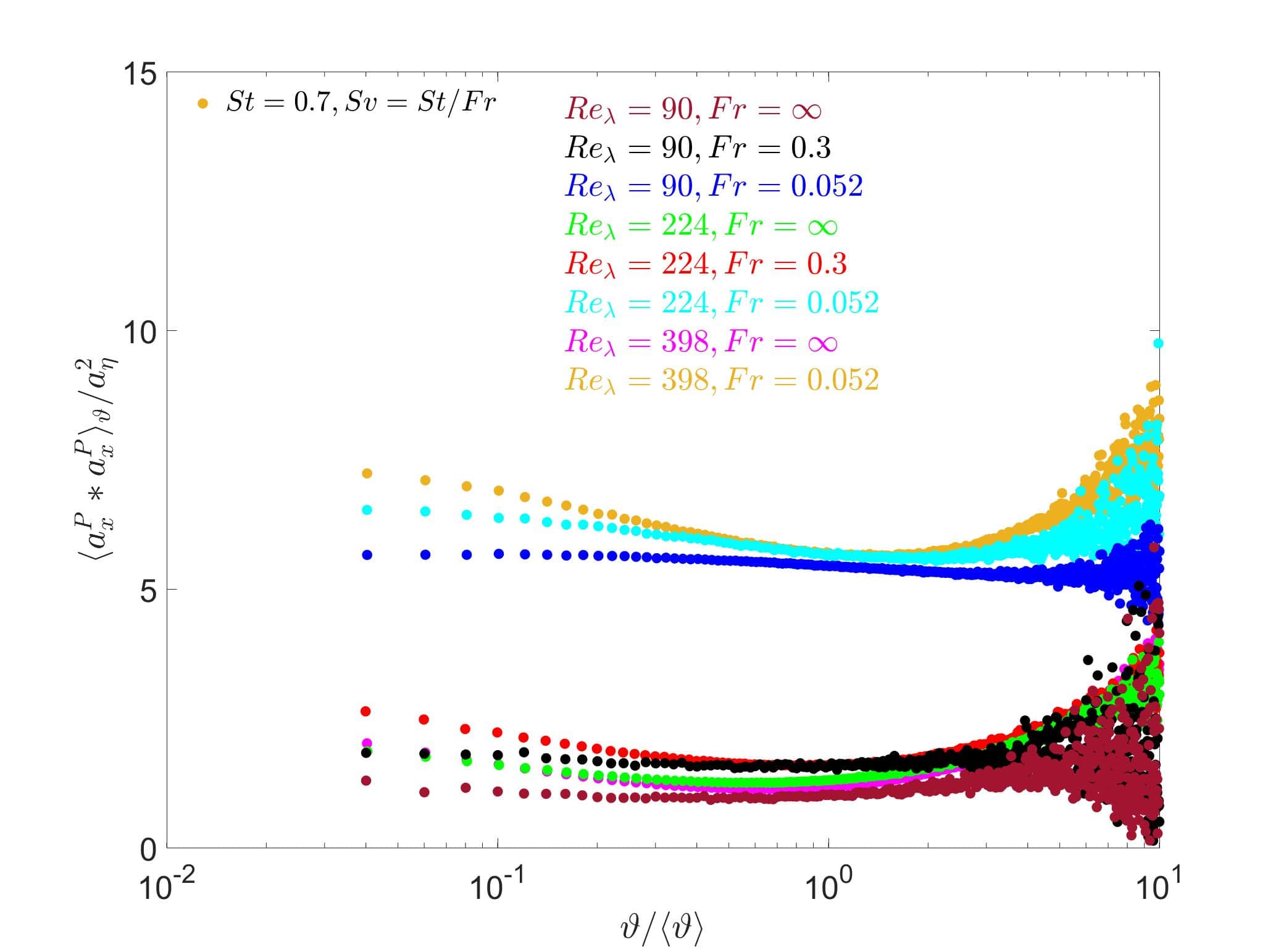}
		\caption{$St=0.7$ }
	\end{subfigure}
	\begin{subfigure}[b]{0.5\linewidth}
		\includegraphics[width=\linewidth]{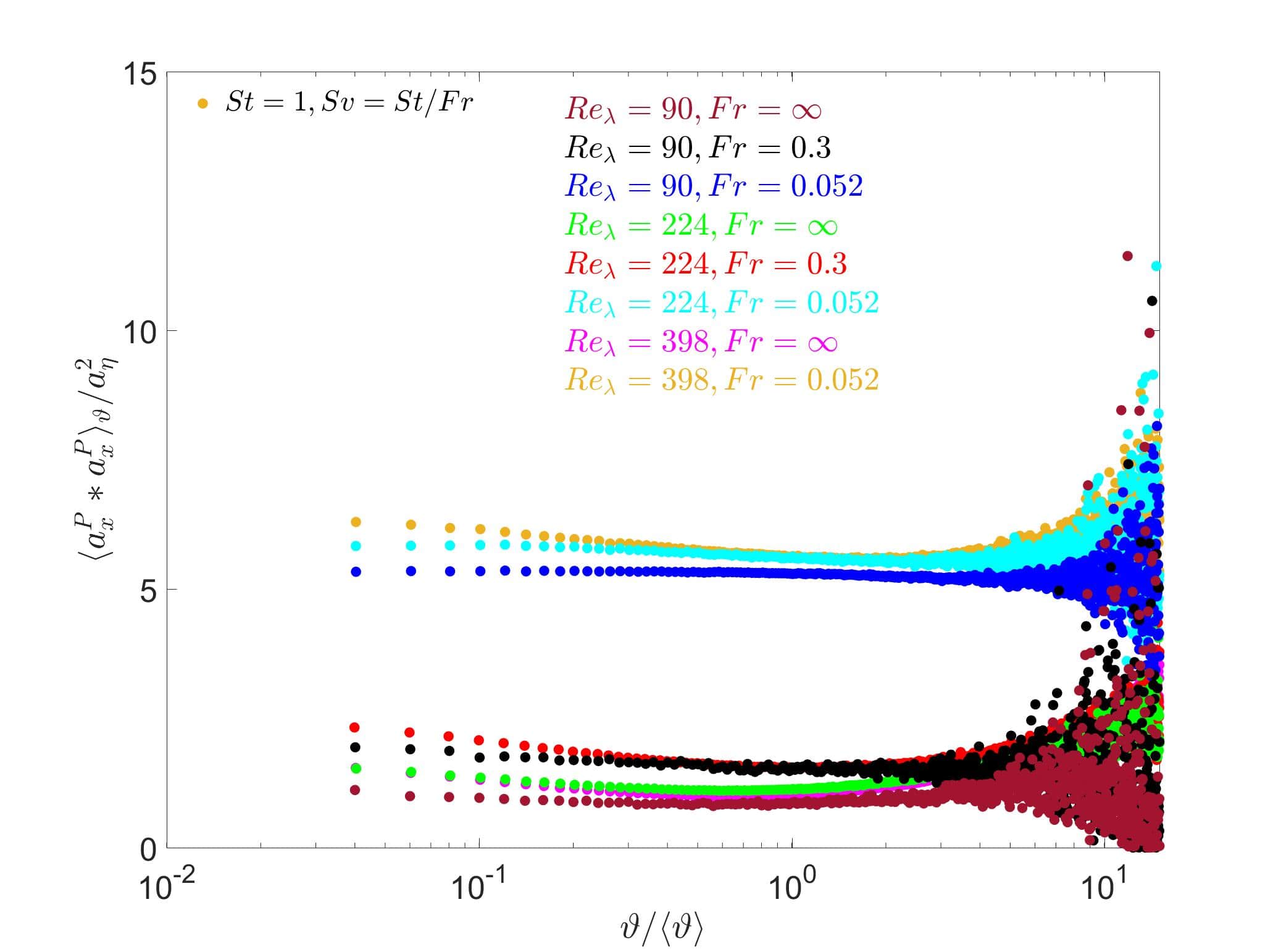}
		\caption{$St=1$ }
	\end{subfigure}%
	\begin{subfigure}[b]{0.5\linewidth}
		\includegraphics[width=\linewidth]{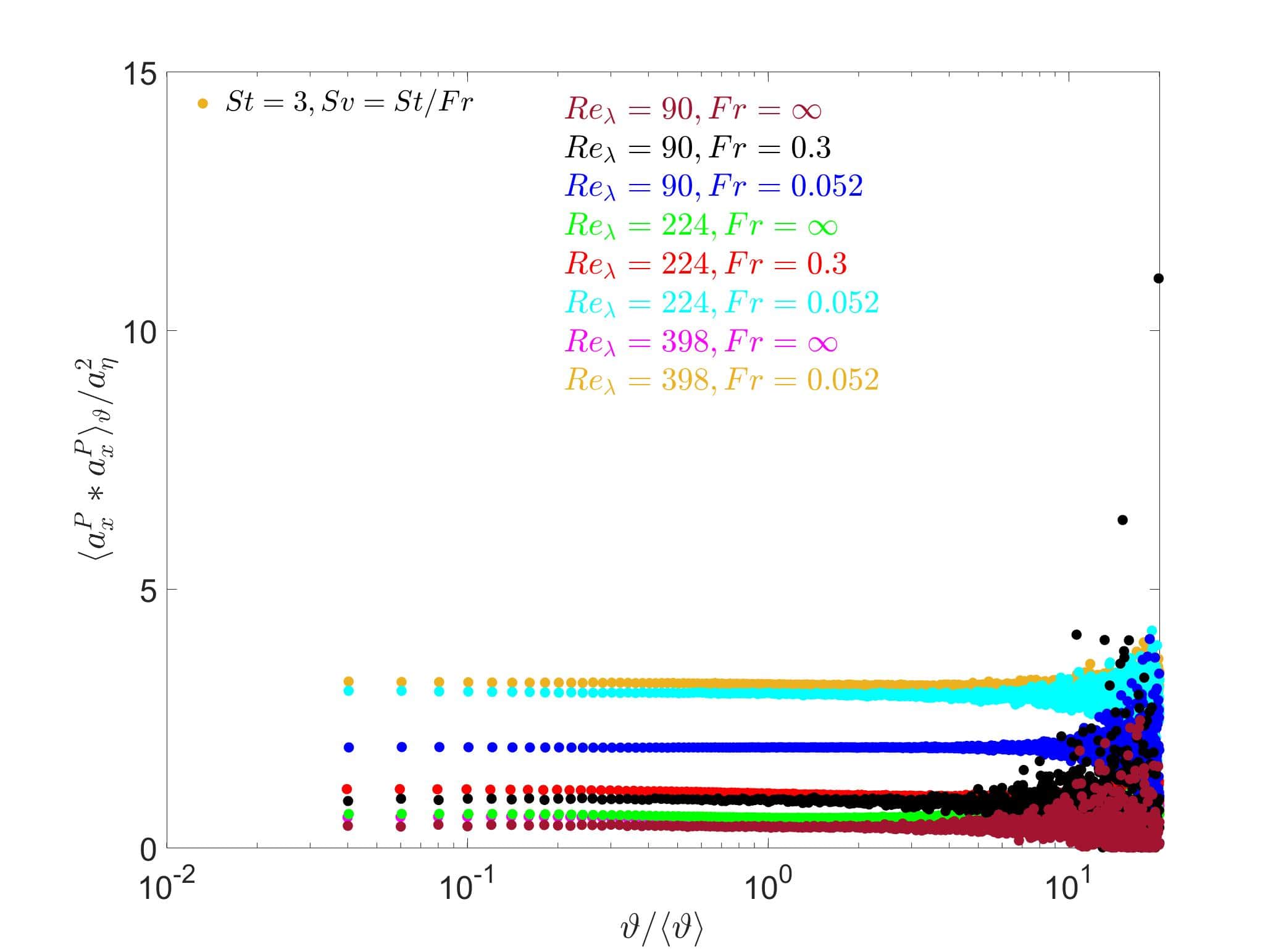}
		\caption{$St=3$ }
	\end{subfigure}
	\caption{Second moment of particle accelerations, in the gravity direction, conditioned on the PDF of the Vorono\text{\"i} volumes at different cases of $Fr$ and $R_\lambda$ combinations for  (a) $St=0$, (b) $St=0.2$, (c) $St=0.5$, (d) $St=0.7$,(e) $St=1$, and (f) $St=3$. Different colors represents different cases.}\label{fig:VT_Avg_PAccel_x_squred_Cnd_Vol__St}
\end{figure}

\begin{figure}
	\vspace{-0.8in}
	\centering
	\begin{subfigure}[b]{0.5\linewidth}
		\includegraphics[width=\linewidth]{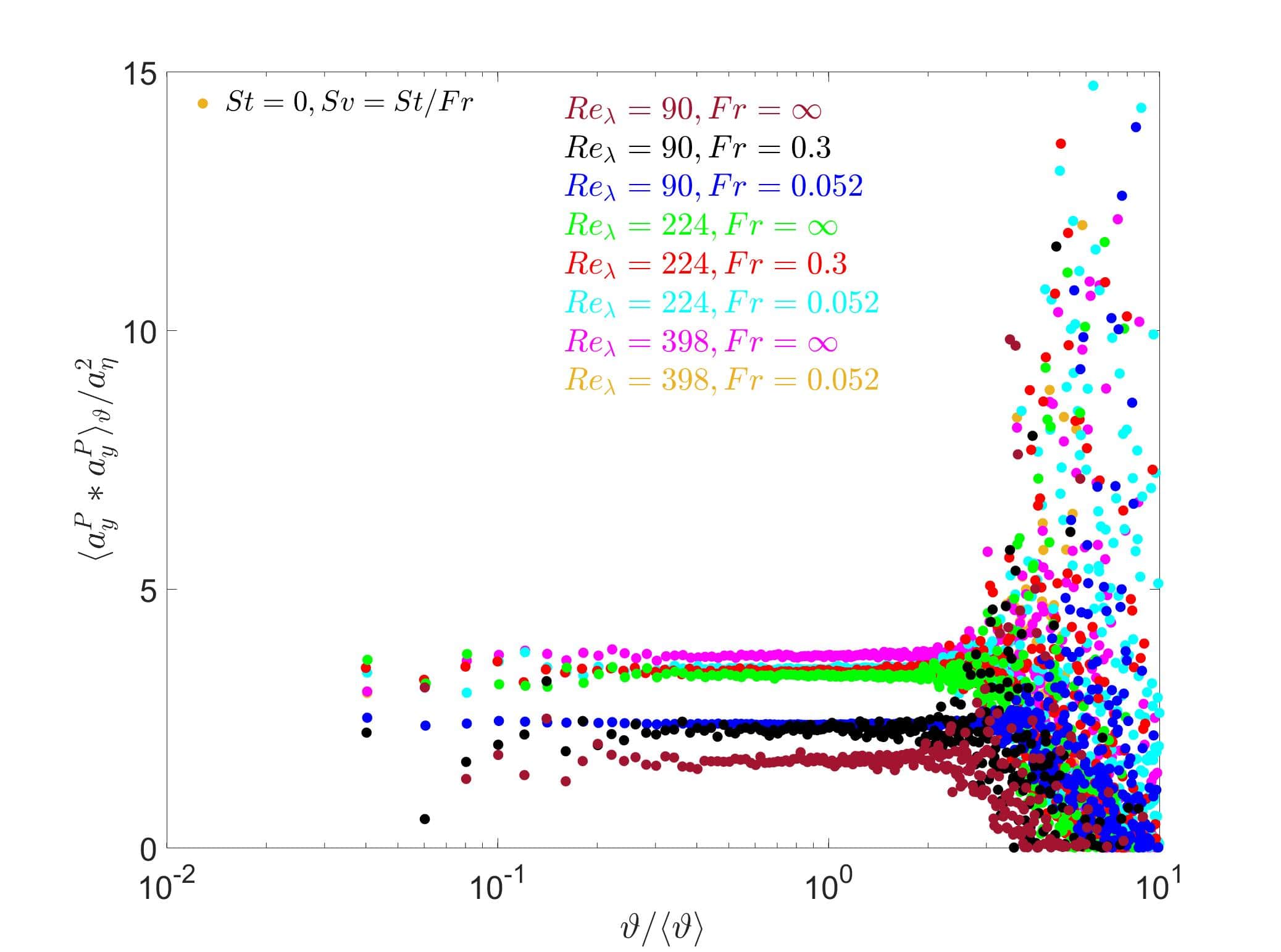}
		\caption{$St=0$}
	\end{subfigure}%
	\begin{subfigure}[b]{0.5\linewidth}
		\includegraphics[width=\linewidth]{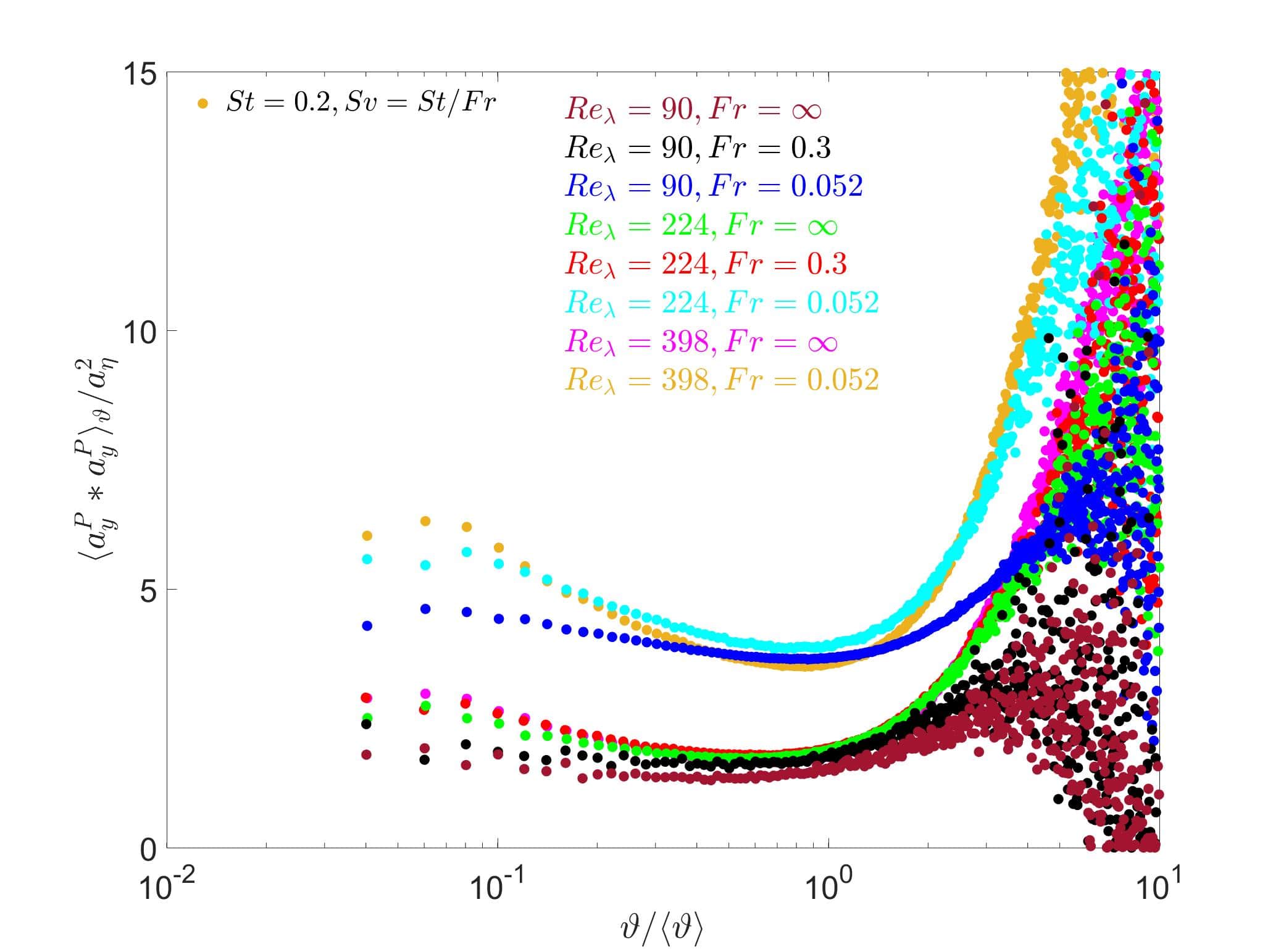}
		\caption{$St=0.2$ }
	\end{subfigure}	
	\begin{subfigure}[b]{0.5\linewidth}
		\includegraphics[width=\linewidth]{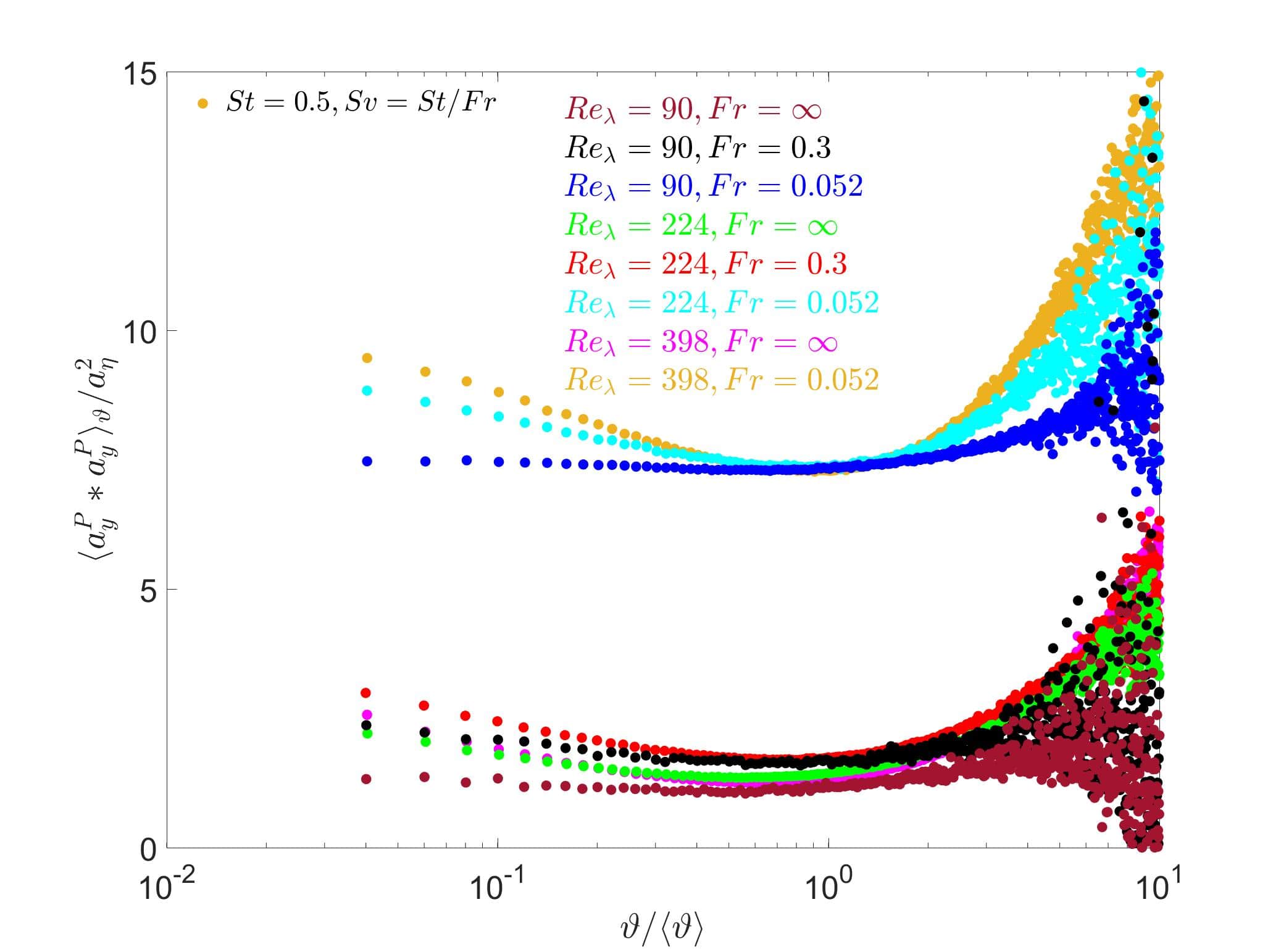}
		\caption{$St=0.5$}
	\end{subfigure}%
	\begin{subfigure}[b]{0.5\linewidth}
		\includegraphics[width=\linewidth]{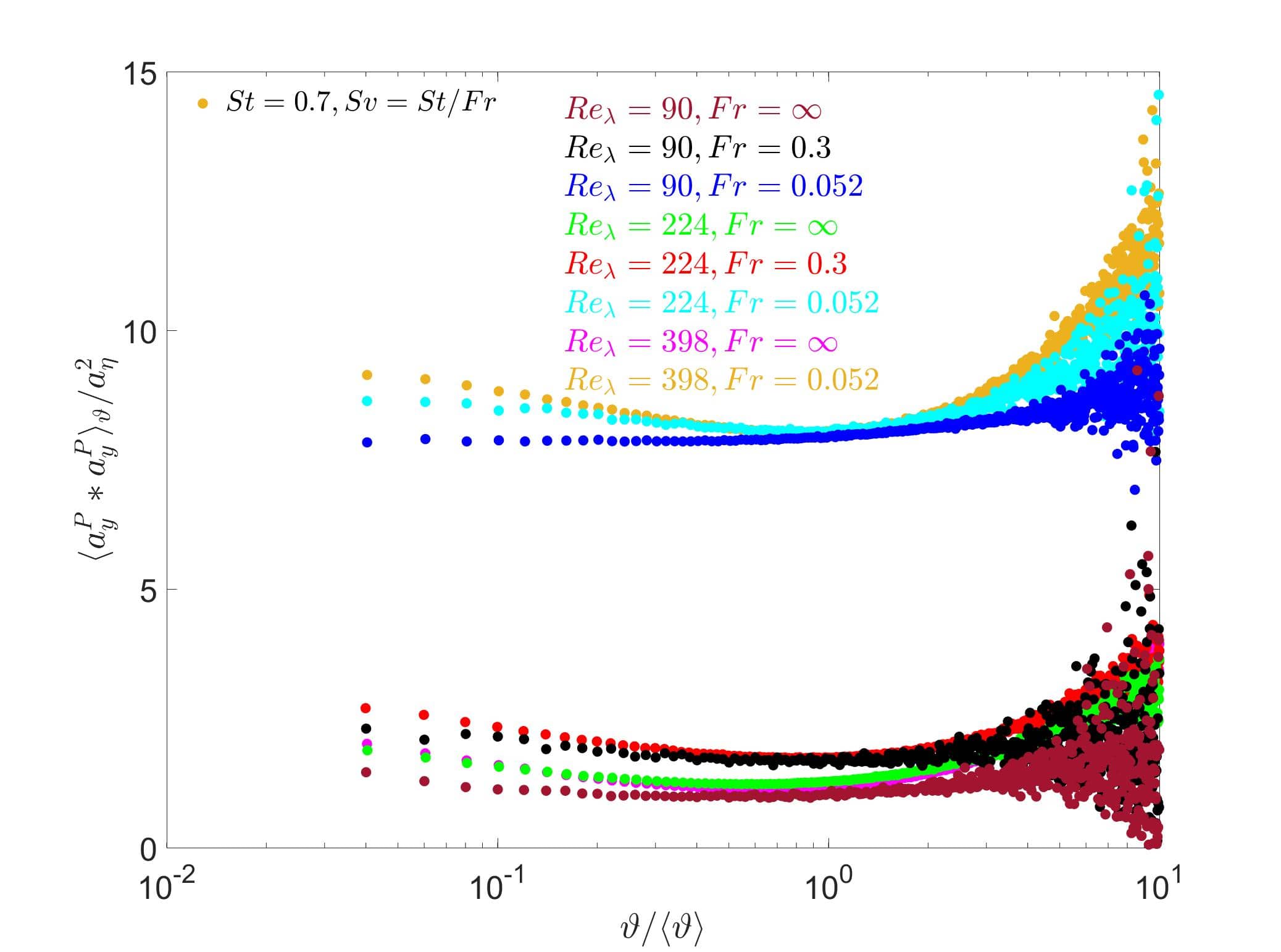}
		\caption{$St=0.7$ }
	\end{subfigure}
	\begin{subfigure}[b]{0.5\linewidth}
		\includegraphics[width=\linewidth]{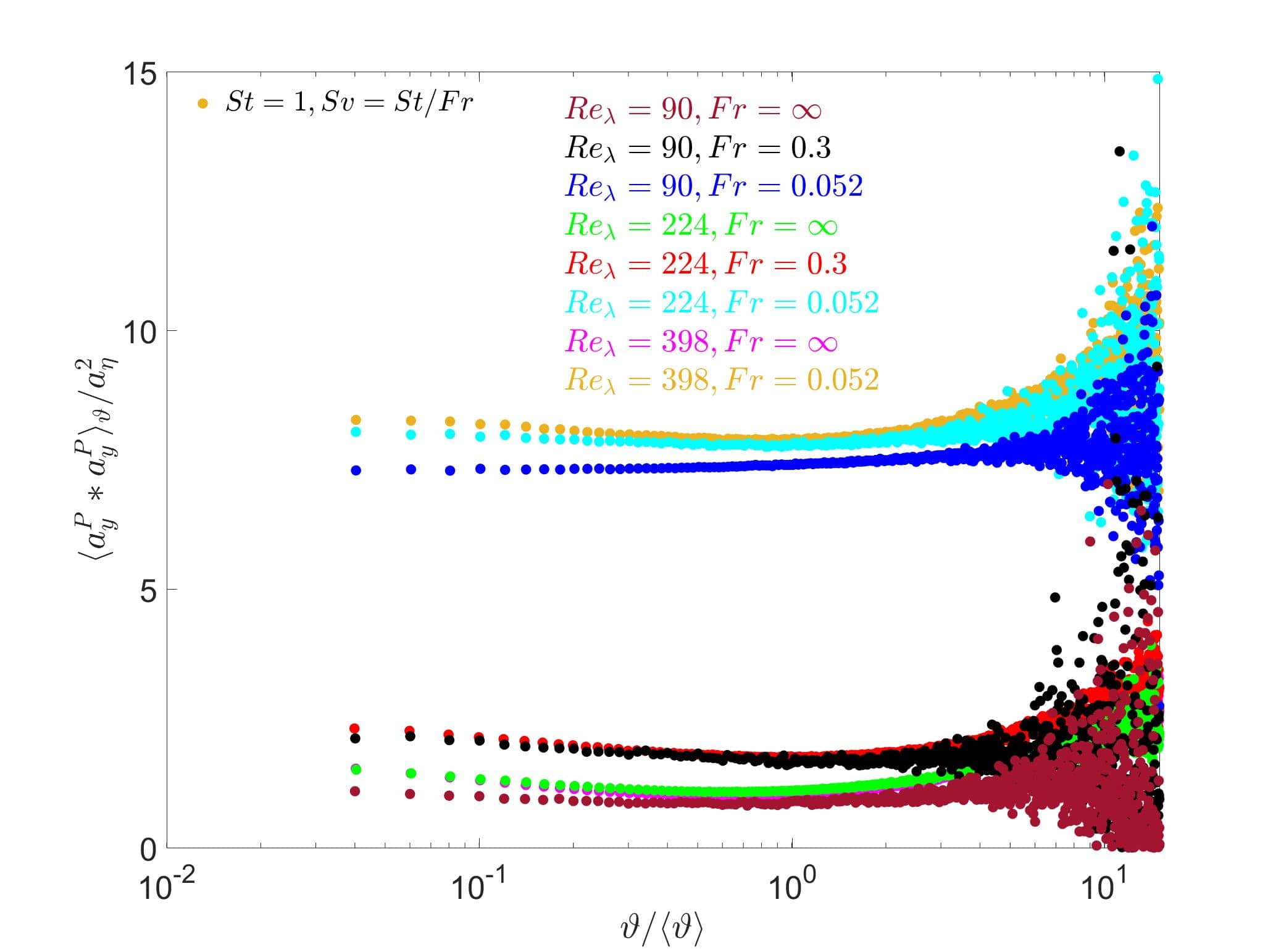}
		\caption{$St=1$ }
	\end{subfigure}%
	\begin{subfigure}[b]{0.5\linewidth}
		\includegraphics[width=\linewidth]{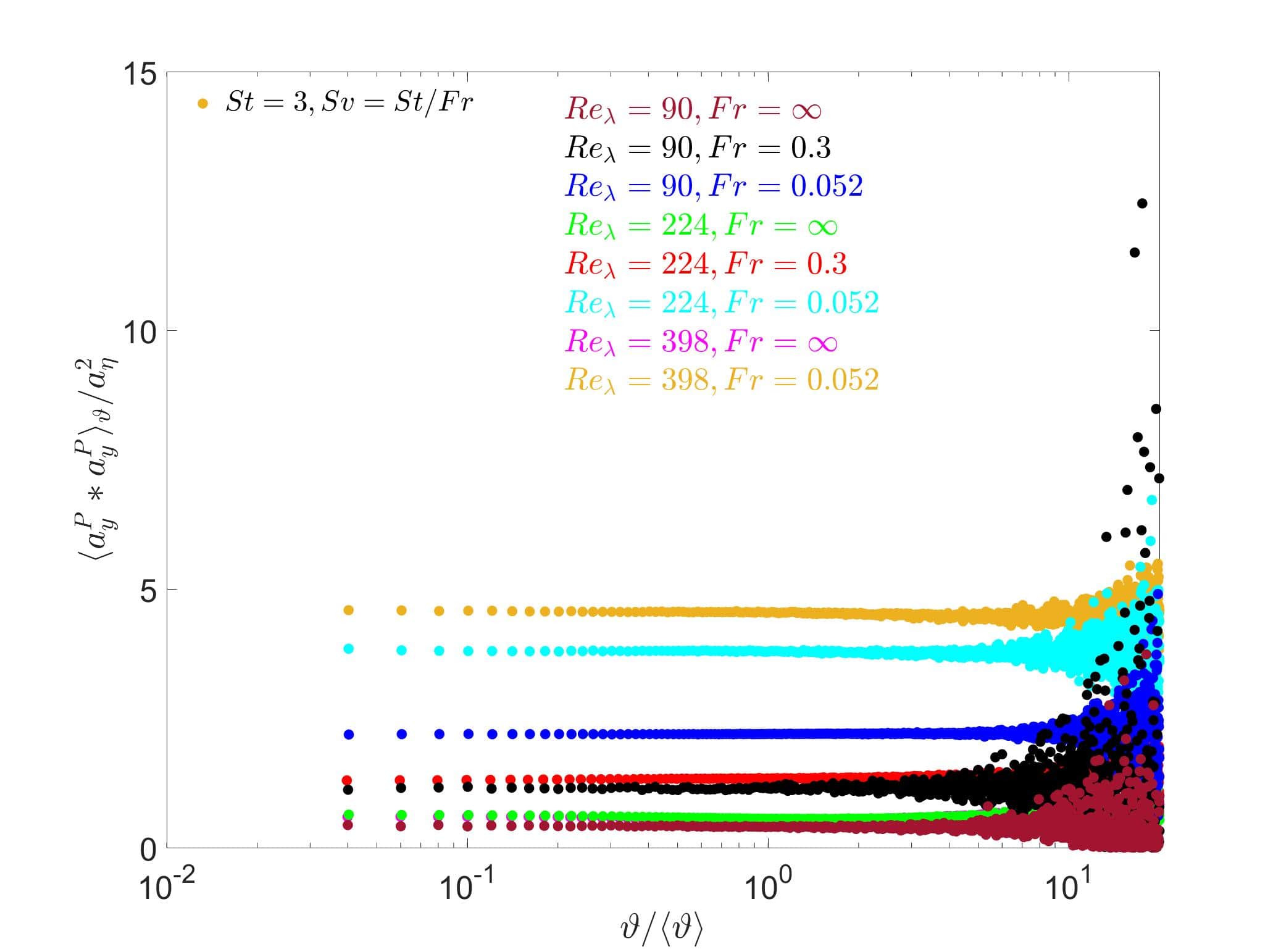}
		\caption{$St=3$ }
	\end{subfigure}
	\caption{Second moment of particle accelerations, in the plane normal to the gravity direction, conditioned on the PDF of the Vorono\text{\"i} volumes at different cases of $Fr$ and $R_\lambda$ combinations for  (a) $St=0$, (b) $St=0.2$, (c) $St=0.5$, (d) $St=0.7$,(e) $St=1$, and (f) $St=3$. Different colors represents different cases.}\label{fig:VT_Avg_PAccel_y_squred_Cnd_Vol__St}
\end{figure}

Figures \ref{fig:VT_Avg_PAccel_x_squred_Cnd_Vol__St} and \ref{fig:VT_Avg_PAccel_y_squred_Cnd_Vol__St} show results for the second moment of the particle accelerations conditioned on the Vorono\text{\"i} volumes, in the direction parallel and perpendicular to gravity, respectively. The results reveal a surprising nonlinear dependence on $\vartheta$, that is present irrespective of $Fr, R_\lambda$, that disappears for sufficiently small or large $St$. If one adopts a standard view that the Vorono\text{\"i} volumes are connected to the different range of scales in a turbulent flow, with small volumes associated with small turbulent flow scales etc, then this would explain why the acceleration variances increase with decreasing $\vartheta$, since accelerations are largest at the smallest scales of a turbulent flow. However, this explanation cannot account for why at sufficiently large $\vartheta$, the acceleration variances increase with increasing $\vartheta$. The results in figures \ref{fig:VT_Avg_PAccel_x_squred_Cnd_Vol__St} and \ref{fig:VT_Avg_PAccel_y_squred_Cnd_Vol__St} therefore perhaps indicate that there is not a simple relationship between the size of the Vorono\text{\"i} volumes and the size of the turbulent flow scales. Furthermore, the strong dependence of this behavior on $St$ indicates that the nonlinear dependence of the acceleration variances on $\vartheta$ is related to the spatial clustering of the particles. 

This nonlinear dependence of the acceleration variances on $\vartheta$ is not qualitatively affected by $Fr, R_\lambda$, however, the results show that decreasing $Fr$ and/or increasing $R_\lambda$ lead to an enhancement of the particle accelerations at all $\vartheta$ (to within statistical noise).  This would seem to imply, in agreement with the physical arguments in \cite{ireland2016effectb}, that the enhancement of the inertial particle accelerations due to gravitational settling are not fundamentally associated with a change in how gravity causes the particles to cluster. Rather, it is caused by the fact that gravity causes the particles to experience rapid changes in the fluid velocity along their trajectory, which in turn leads to large particle accelerations. This effect, described more fully in \cite{dhariwal2018small}, would operate even if the particles were uniformly distributed throughout the flow. Indeed, our results in figures \ref{fig:VT_Avg_PAccel_x_squred_Cnd_Vol__St} and \ref{fig:VT_Avg_PAccel_y_squred_Cnd_Vol__St} for $St=3$ reveal strong enhancements of the particle accelerations due to gravity, even though the acceleration variances are almost independent of $\vartheta$.

\begin{figure}
	\vspace{-0.8in}
	\centering
	\begin{subfigure}[b]{0.5\linewidth}
		\includegraphics[width=\linewidth]{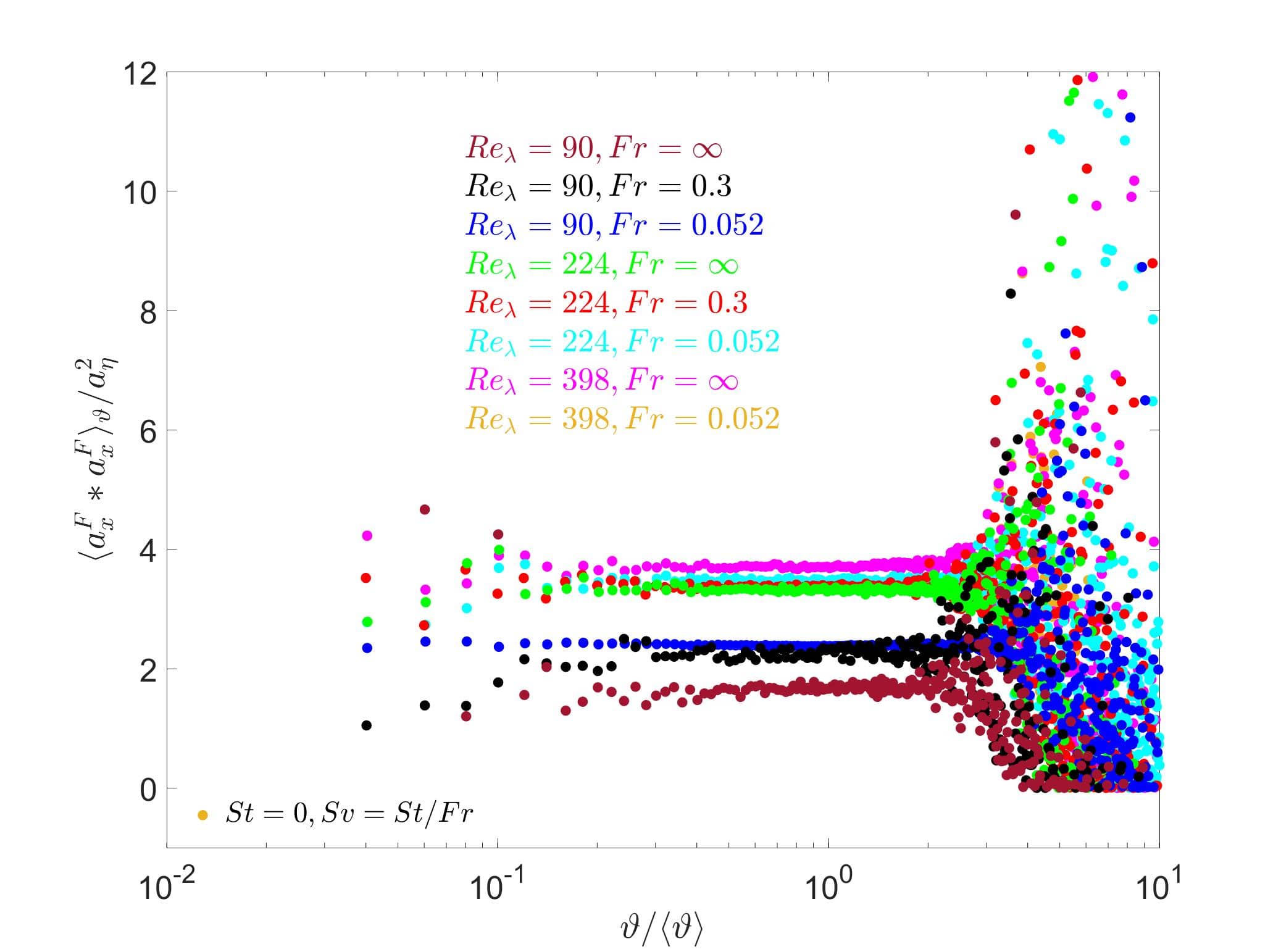}
		\caption{$St=0$}
	\end{subfigure}%
	\begin{subfigure}[b]{0.5\linewidth}
		\includegraphics[width=\linewidth]{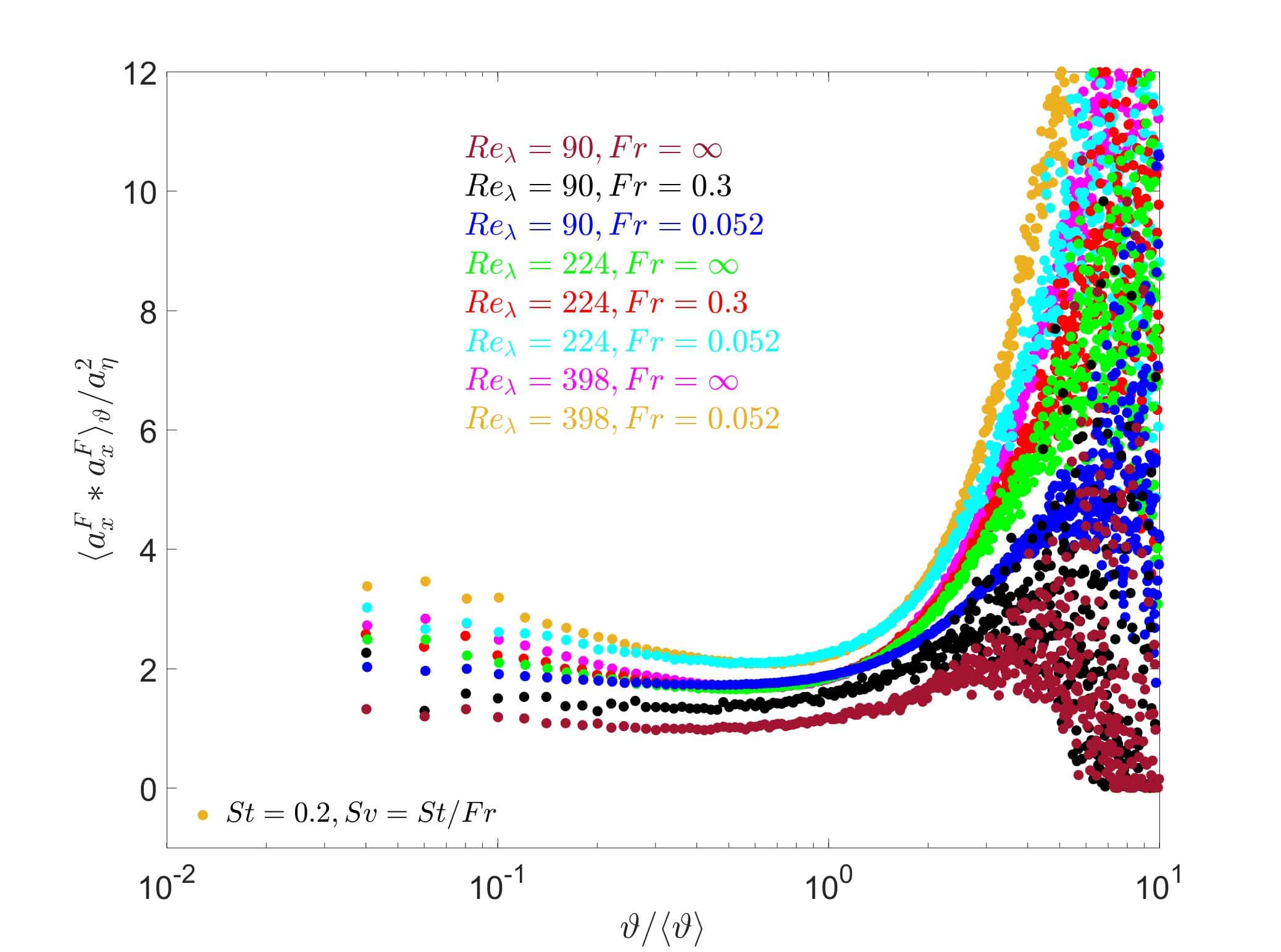}
		\caption{$St=0.2$ }
	\end{subfigure}	
	\begin{subfigure}[b]{0.5\linewidth}
		\includegraphics[width=\linewidth]{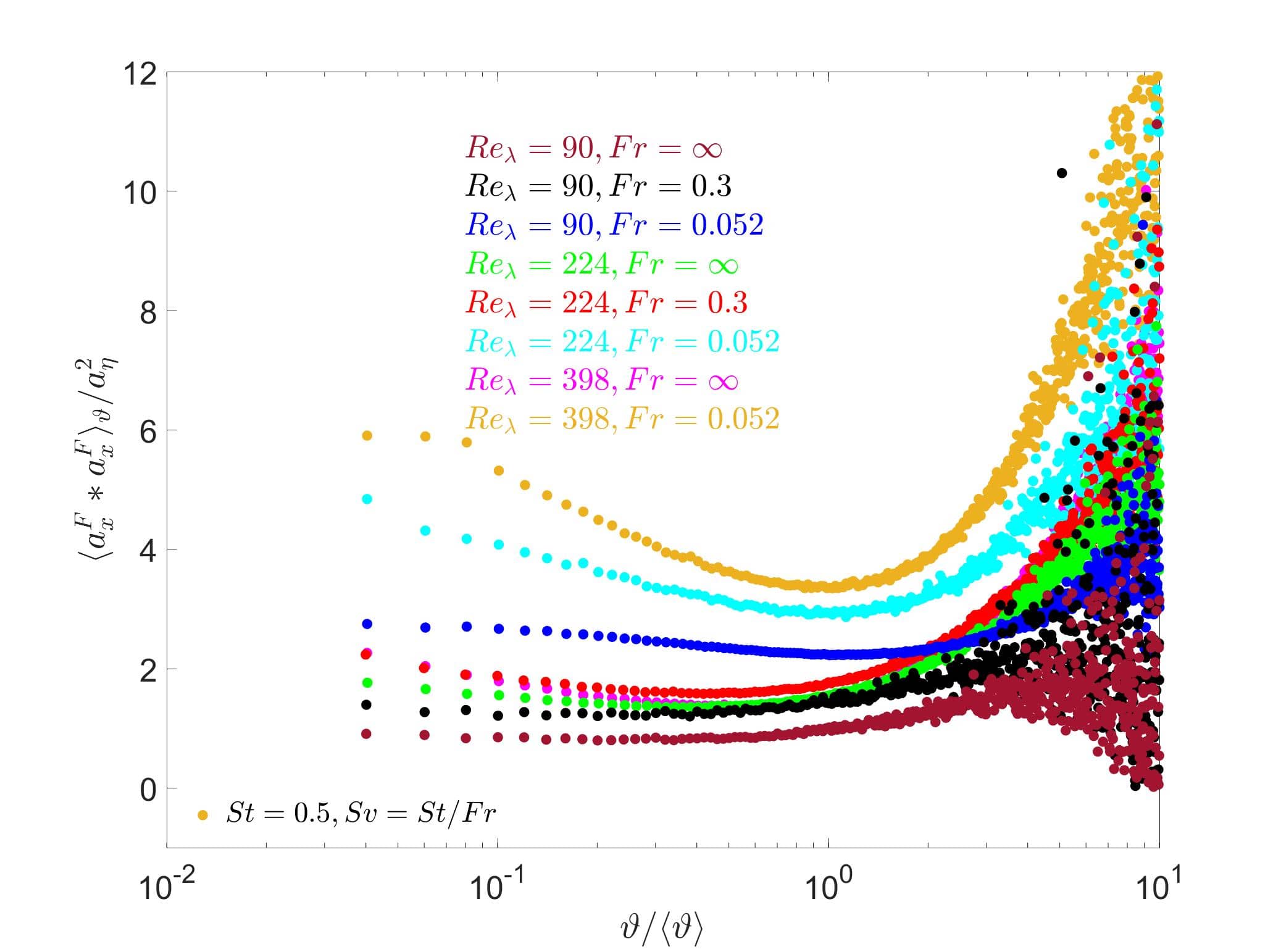}
		\caption{$St=0.5$}
	\end{subfigure}%
	\begin{subfigure}[b]{0.5\linewidth}
		\includegraphics[width=\linewidth]{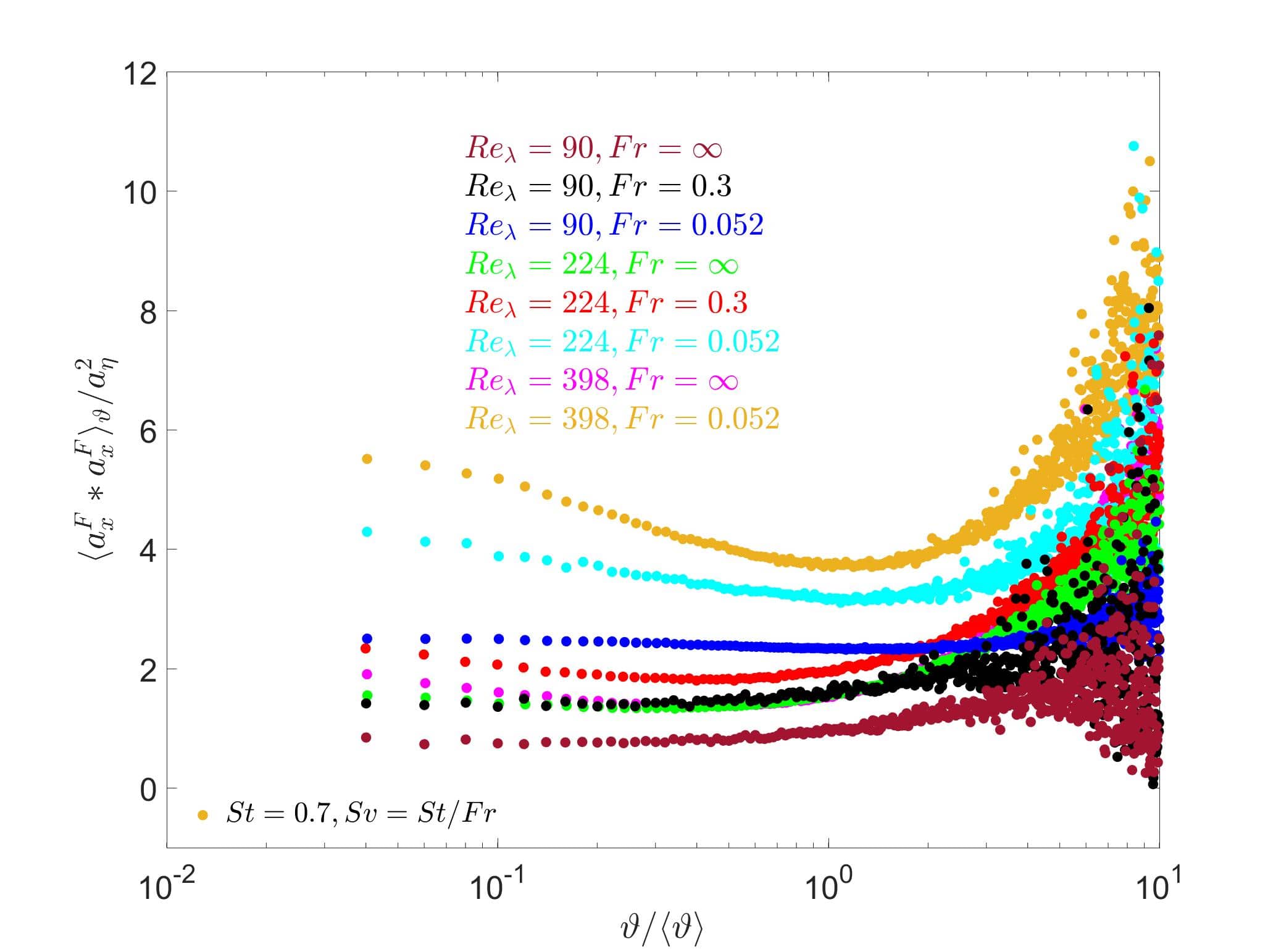}
		\caption{$St=0.7$ }
	\end{subfigure}
	\begin{subfigure}[b]{0.5\linewidth}
		\includegraphics[width=\linewidth]{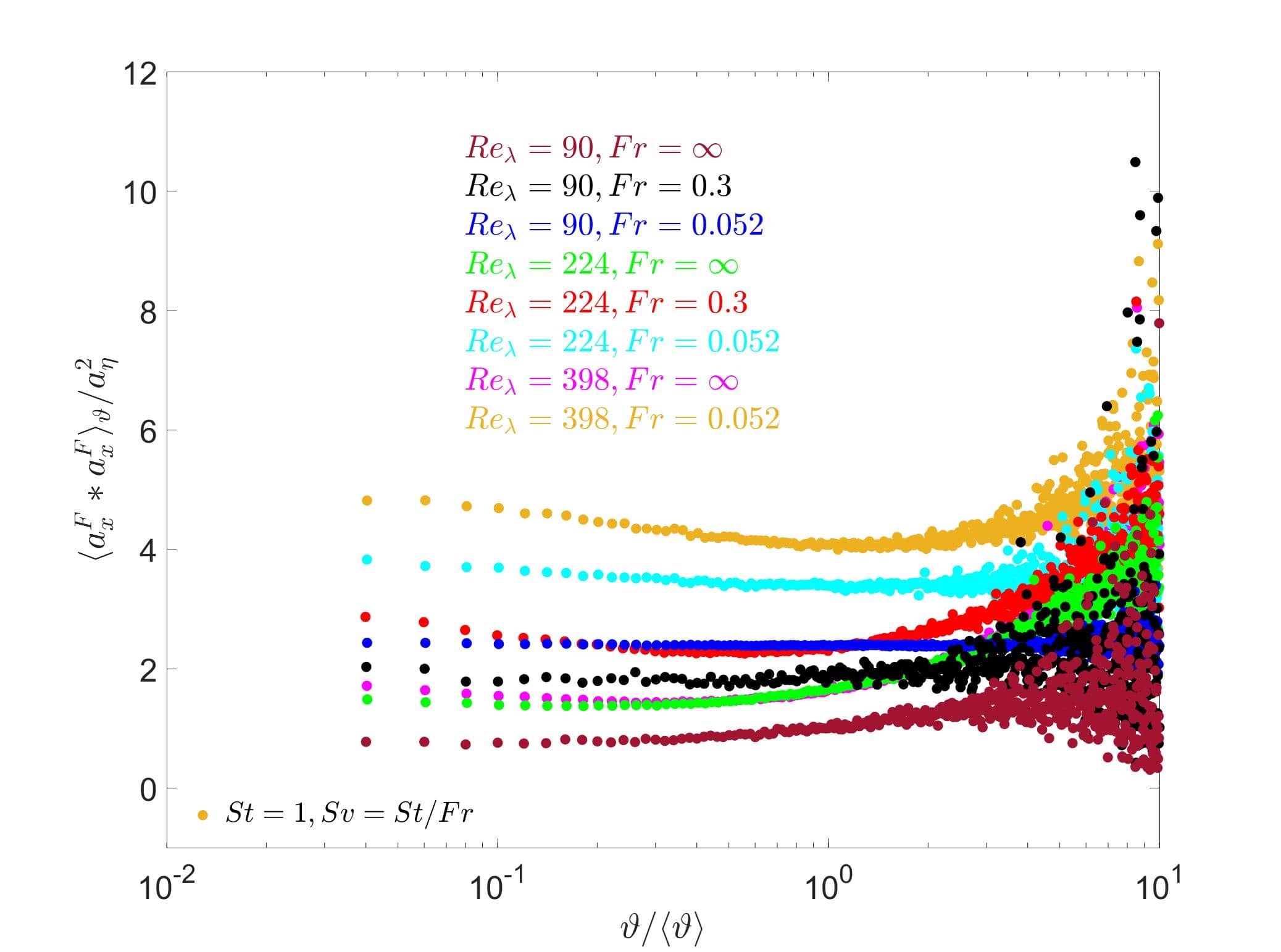}
		\caption{$St=1$ }
	\end{subfigure}%
	\begin{subfigure}[b]{0.5\linewidth}
		\includegraphics[width=\linewidth]{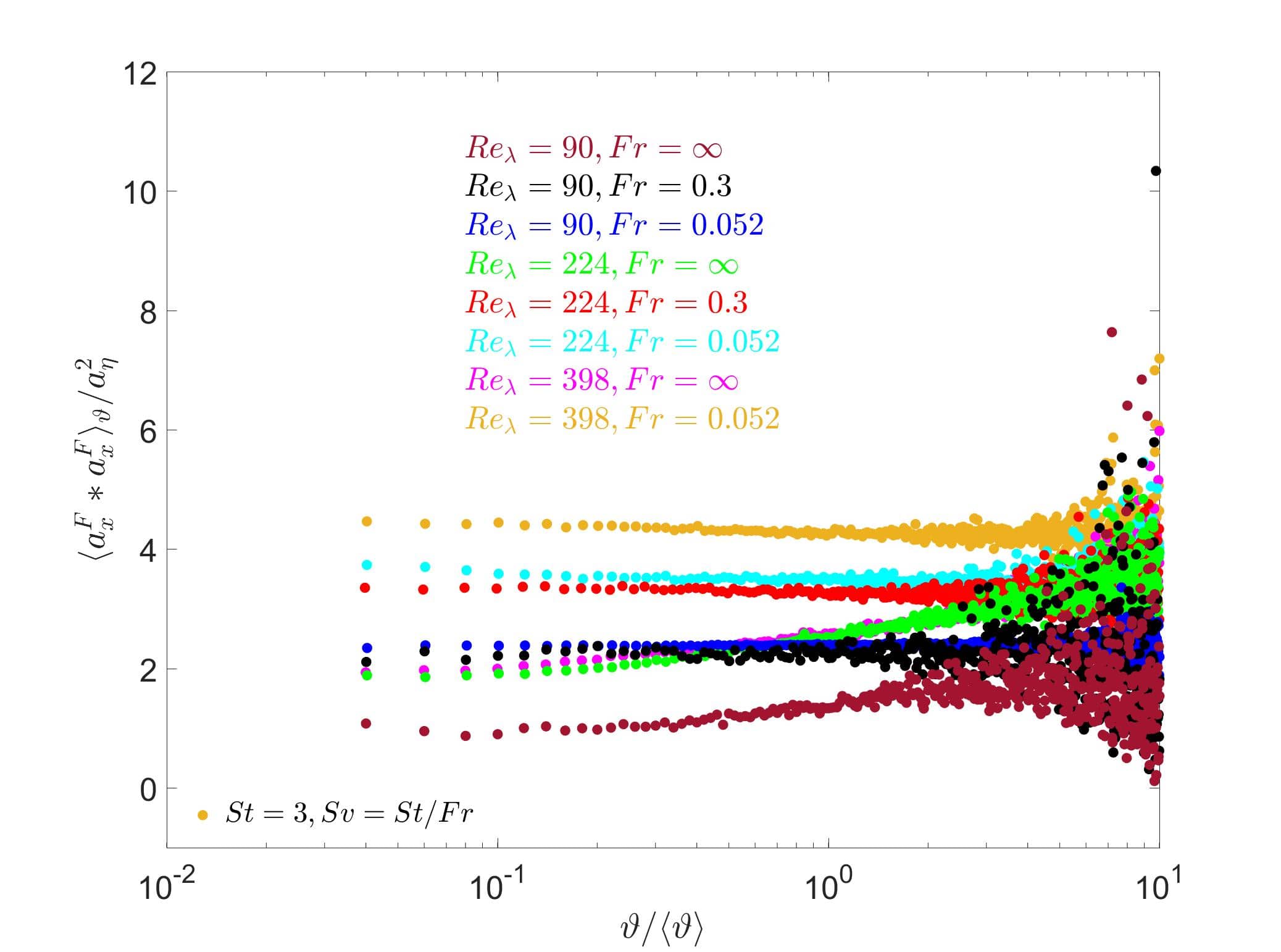}
		\caption{$St=3$ }
	\end{subfigure}
	\caption{Second moment of fluid accelerations at the particle position, in the gravity direction, conditioned on the PDF of the Vorono\text{\"i} volumes at different cases of $Fr$ and $R_\lambda$ combinations for  (a) $St=0$, (b) $St=0.2$, (c) $St=0.5$, (d) $St=0.7$,(e) $St=1$, and (f) $St=3$. Different colors represents different cases.}\label{fig:VT_Avg_FAccel_x_squred_Cnd_Vol__St}
\end{figure}

In figure \ref{fig:VT_Avg_FAccel_x_squred_Cnd_Vol__St} we show results for the variances of the fluid acceleration ($\bm{a}^F$) at the inertial particle positions, conditioned on $\vartheta$, and for the direction of gravity. These results also exhibit, in general, a convex functional shape with respect to $\vartheta$. They also show that as $Fr$ is decreased, the accelerations become stronger, even for cases where the $St=0$ are independent of $Fr$ (in general the fluid particle accelerations should be independent of $Fr$, however since the different $Fr$ cases correspond to different simulations, the results can depend on $Fr$, for example due to different domain sizes in the simulations). Since the fluid acceleration measured at the inertial particle positions can only differ from the fluid acceleration at fluid particle positions due to preferential sampling of the flow by the particles, this indicates that the settling causes the inertial particles to sample regions of the flow exhibiting strong accelerations.

The results in figure \ref{fig:VT_Avg_FAccel_x_squred_Cnd_Vol__St} also present a challenge to the ``sweep-stick'' mechanism that has been proposed to explain particle clustering in turbulent flows \cite{coleman09}. The sweep-stick mechanism states that inertial particles stick to points in the flow where the fluid acceleration is zero, and since these acceleration stagnation points are themselves clustered, this therefore explains the clustering of the inertial particles. In \cite{bragg15} a number of theoretical arguments were given that call into question the validity of this proposed mechanism for particle clustering, supported by results from DNS. If the sweep-stick mechanism were correct, we should expect to find that in regions where the particles are clustered (i.e. where $\vartheta$ is small), the fluid acceleration at the particle position should be zero, or at least very small compared with $a_\eta$. The results in figure \ref{fig:VT_Avg_FAccel_x_squred_Cnd_Vol__St} show, however, that in the regions where the particles are clustered, the fluid accelerations at the inertial particle positions are large, i.e. their magnitudes are $\geq\mathcal{O}(a_\eta)$. Moreover, over a significant range of $\vartheta$, the accelerations are almost independent of $\vartheta$. These results therefore strongly call into question the validity of the sweep-stick mechanism. We refer the reader to \cite{bragg15} for a detailed discussion concerning theoretical issues with the sweep-stick mechanism.
\FloatBarrier
\FloatBarrier
\FloatBarrier

\begin{figure}
	\vspace{-0.7in}
	\centering
	\begin{subfigure}[b]{0.5\linewidth}
		\includegraphics[width=\linewidth]{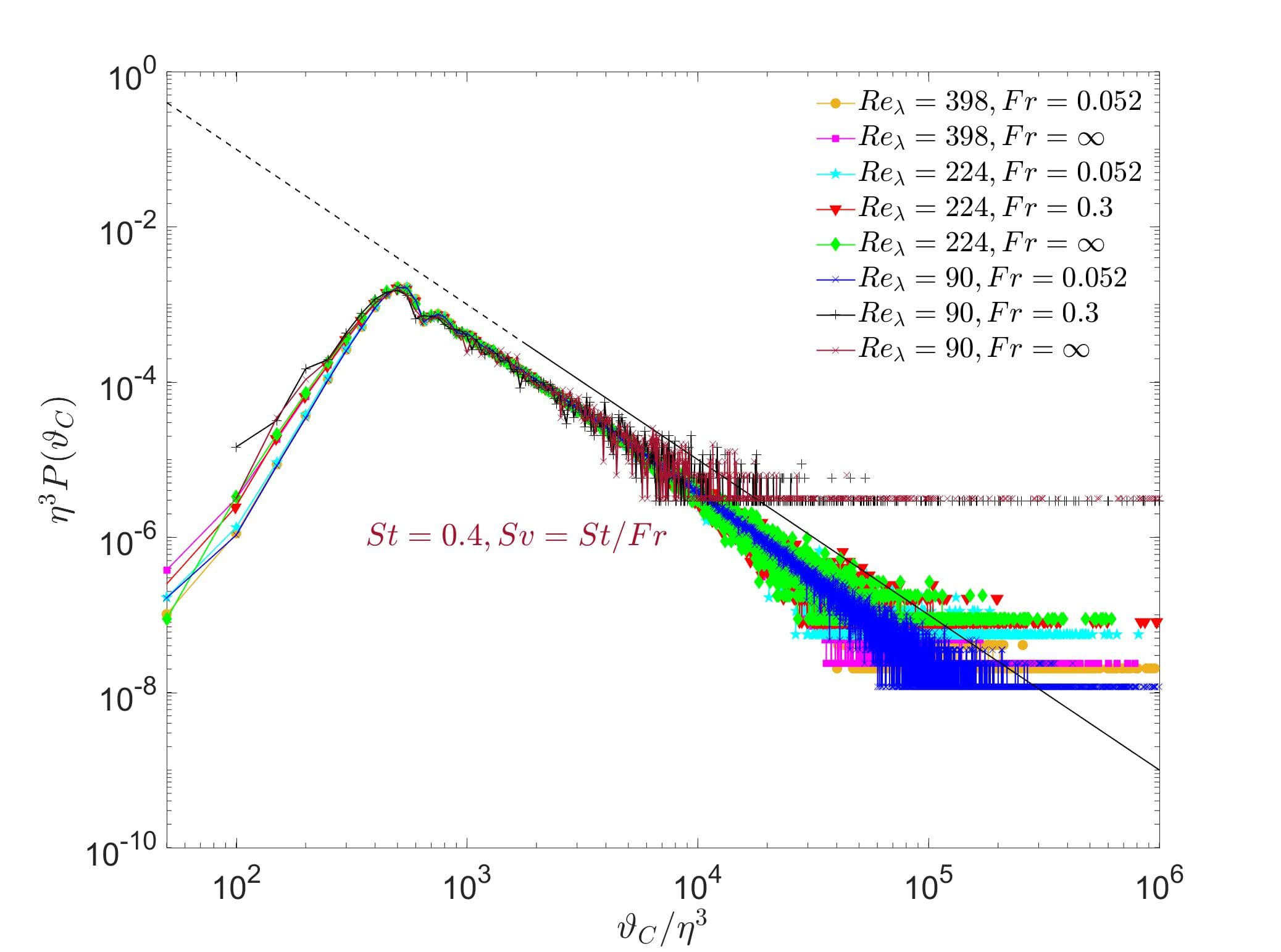}
		\caption{$St=0.4$}
	\end{subfigure}%
	\begin{subfigure}[b]{0.5\linewidth}
		\includegraphics[width=\linewidth]{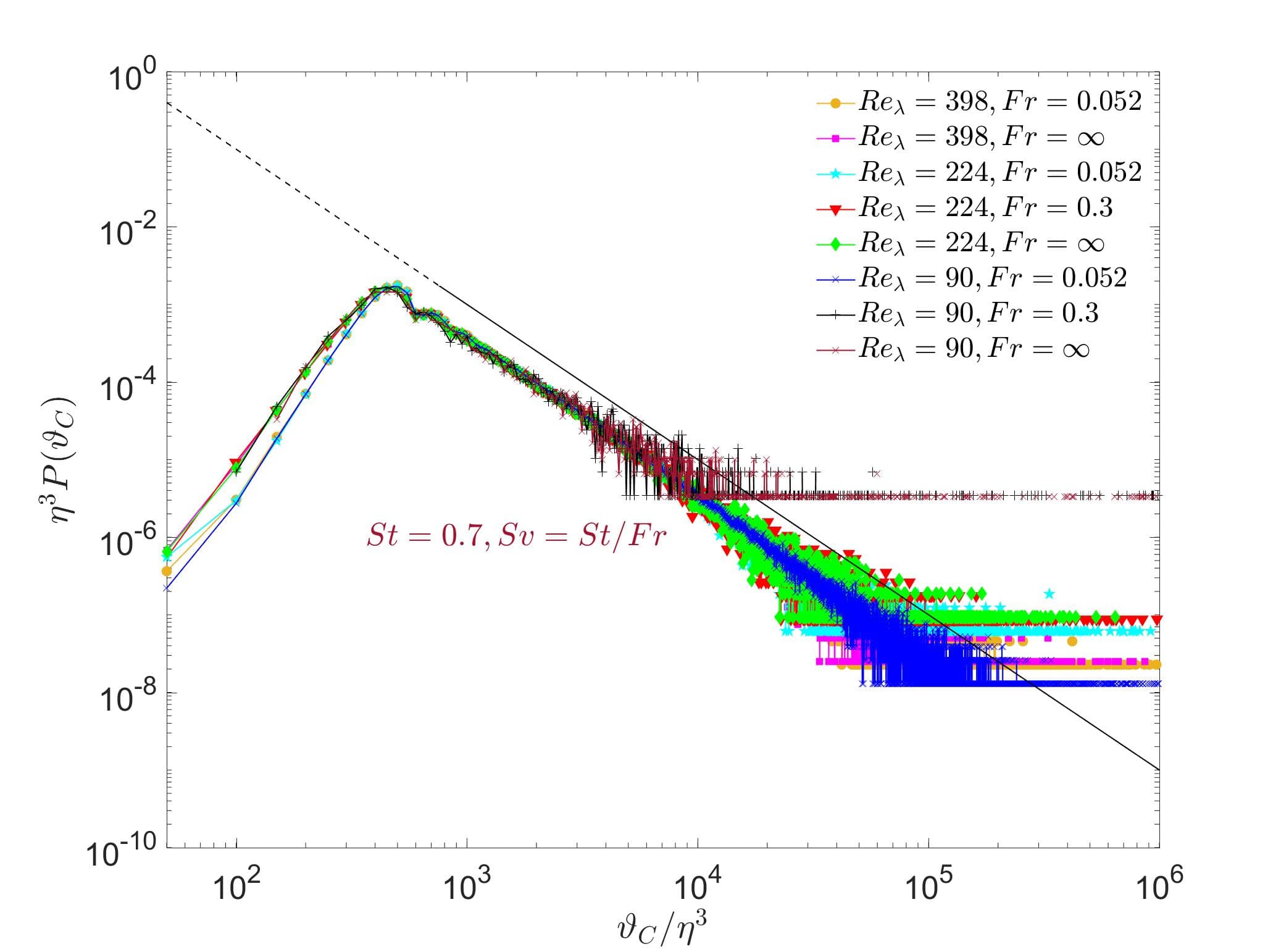}
		\caption{$St=0.7$ }
	\end{subfigure}
	
	\begin{subfigure}[b]{0.5\linewidth}
		\includegraphics[width=\linewidth]{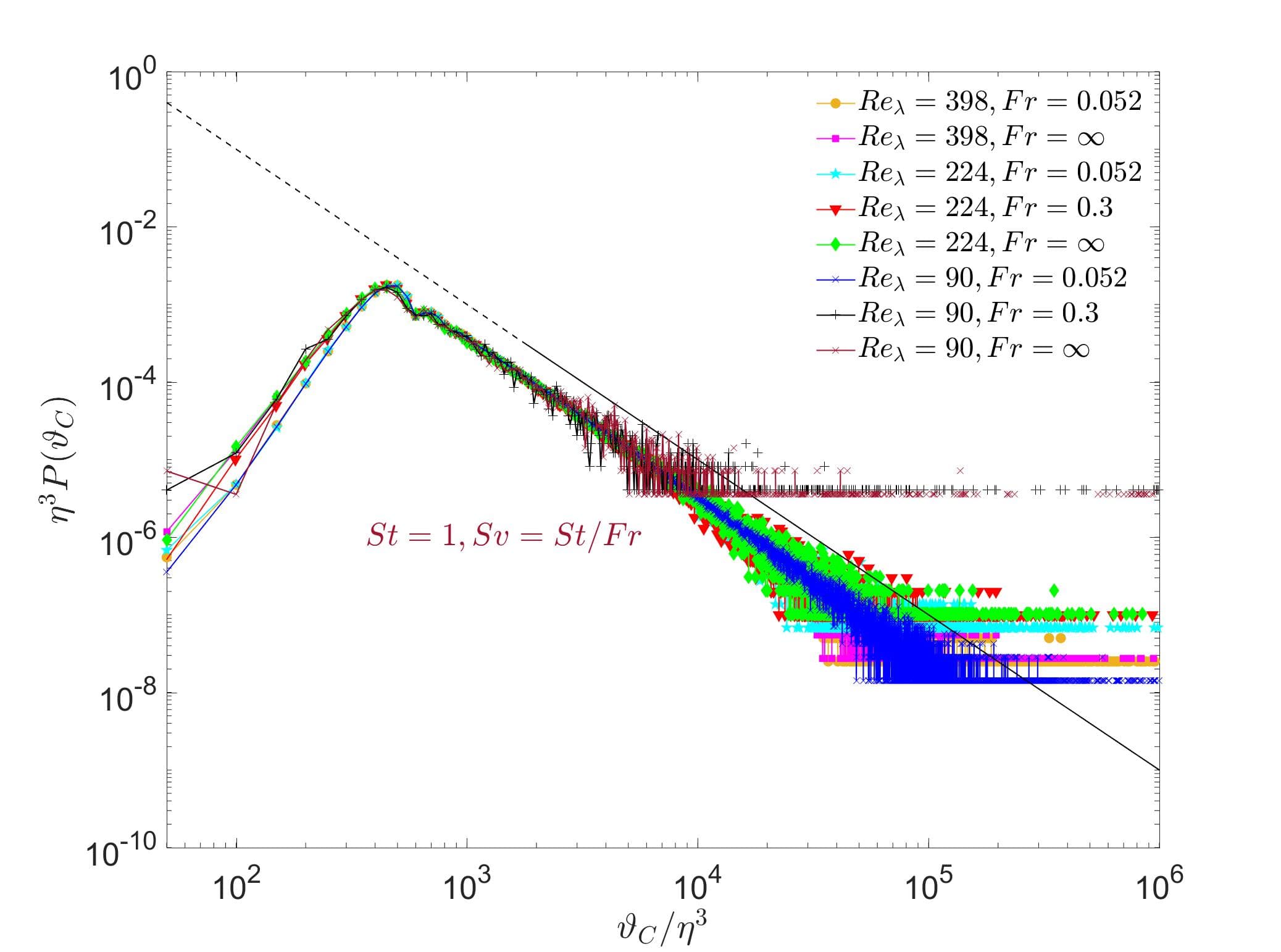}
		\caption{$St=1$}
	\end{subfigure}%
	\begin{subfigure}[b]{0.5\linewidth}
		\includegraphics[width=\linewidth]{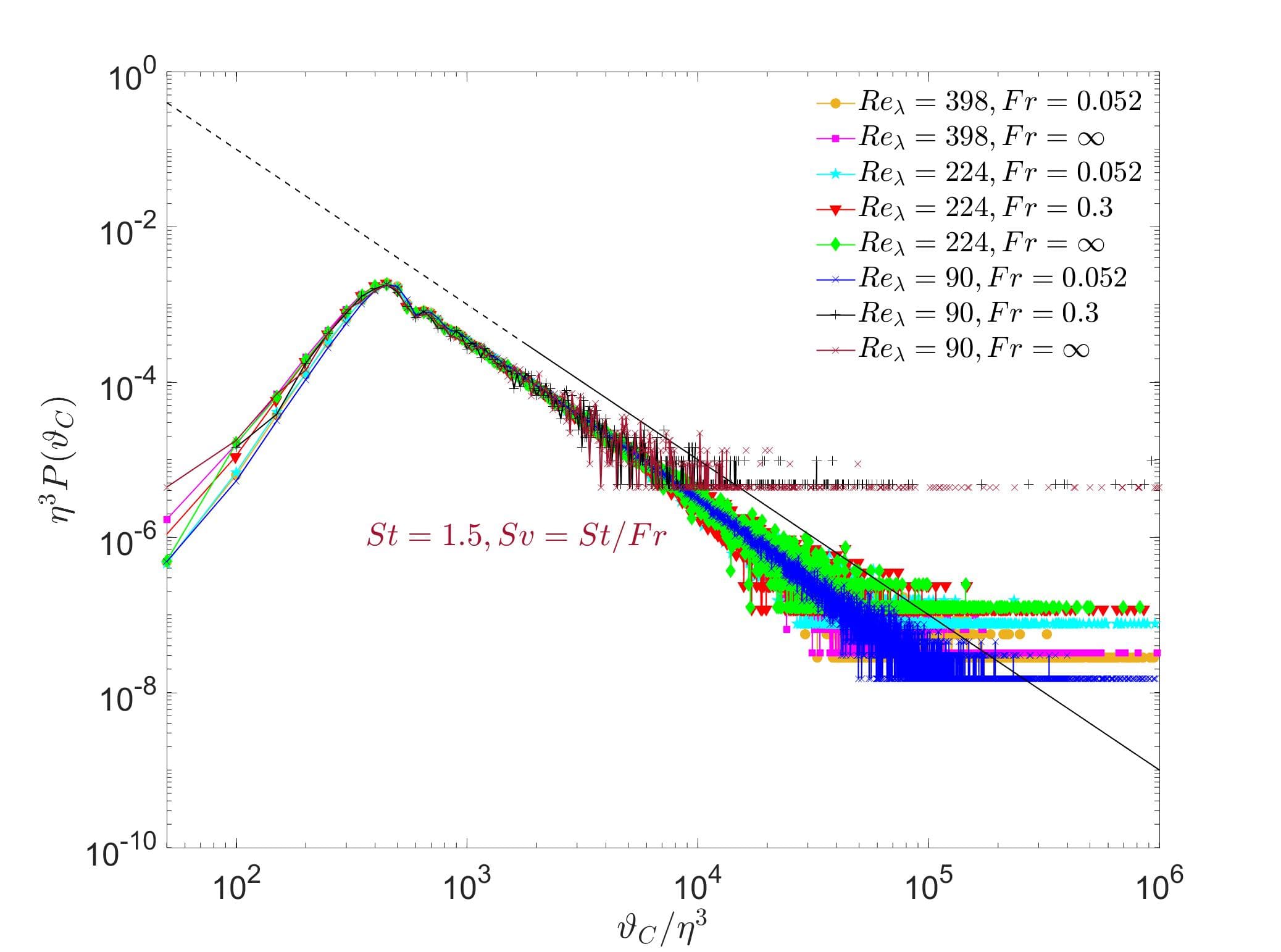}
		\caption{$St=1.5$ }
	\end{subfigure}
	\begin{subfigure}[b]{0.5\linewidth}
		\includegraphics[width=\linewidth]{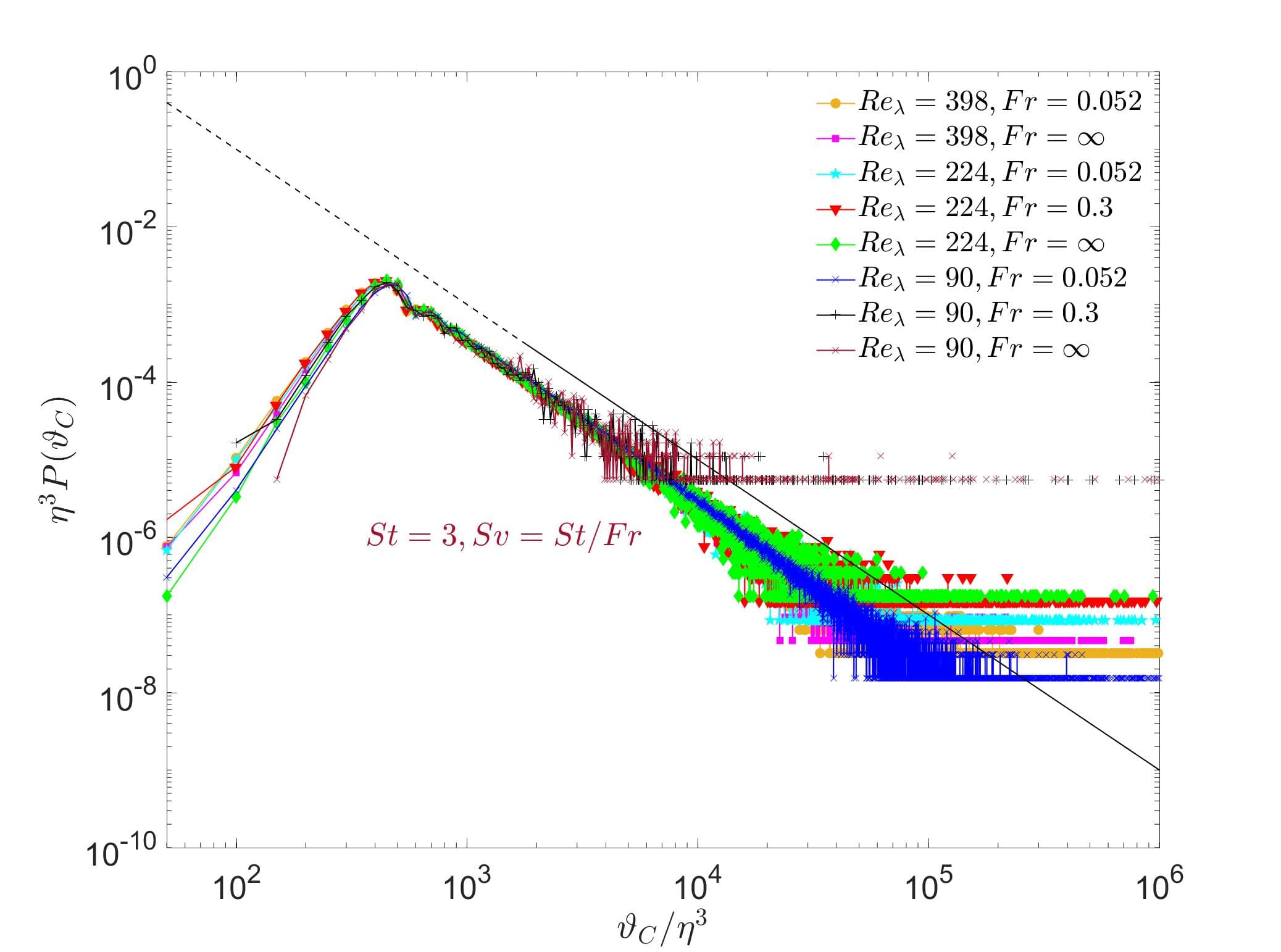}
		\caption{$St=3$ }
	\end{subfigure}%
	\caption{PDF of the coherent cluster volumes (normalized by the Kolmogorov scale) at different cases of $Fr$ and $R_\lambda$ combinations for (a) $St=0.4$, (b) $St=0.7$, (c) $St=1$, (d) $St=1.5$, and (e) $St=3$. Different colors represents different cases. The dashed line represents a power law with exponent $-2$.}\label{fig:VT_CC_PDF_Vol_Normalized_eta3_AllCaseFigure_St}
\end{figure}

We now turn our attention to the results for clusters of particles, which are defined as connected Vorono\text{\"i} cells that have volumes $\leq \vartheta_C$ \cite{baker2017coherent}. Figure \ref{fig:VT_CC_PDF_Vol_Normalized_eta3_AllCaseFigure_St} shows the PDF of cluster volumes normalized by the Kolmogorov scale. The results show that cluster sizes are distributed over a wide range of scales, from the dissipation up to integral range scales. In agreement with previous studies (\cite{frankel2016settling,baker2017coherent,monchaux2010preferential, obligado2014preferential, monchaux2017settling}), we find that the right hand side of the PDF is well described as a power law with exponent $-2$, which implies self-similarity of the clusters, in line with results in \cite{yoshimoto2007self,monchaux2010preferential}. This slope was found to be $\sim -5/3$ in the experimental study of \cite{sumbekova2017preferential}.   
\FloatBarrier

\begin{figure}
	\vspace{-0.7in}
	\centering
	\begin{subfigure}[b]{0.5\linewidth}
		\includegraphics[width=\linewidth]{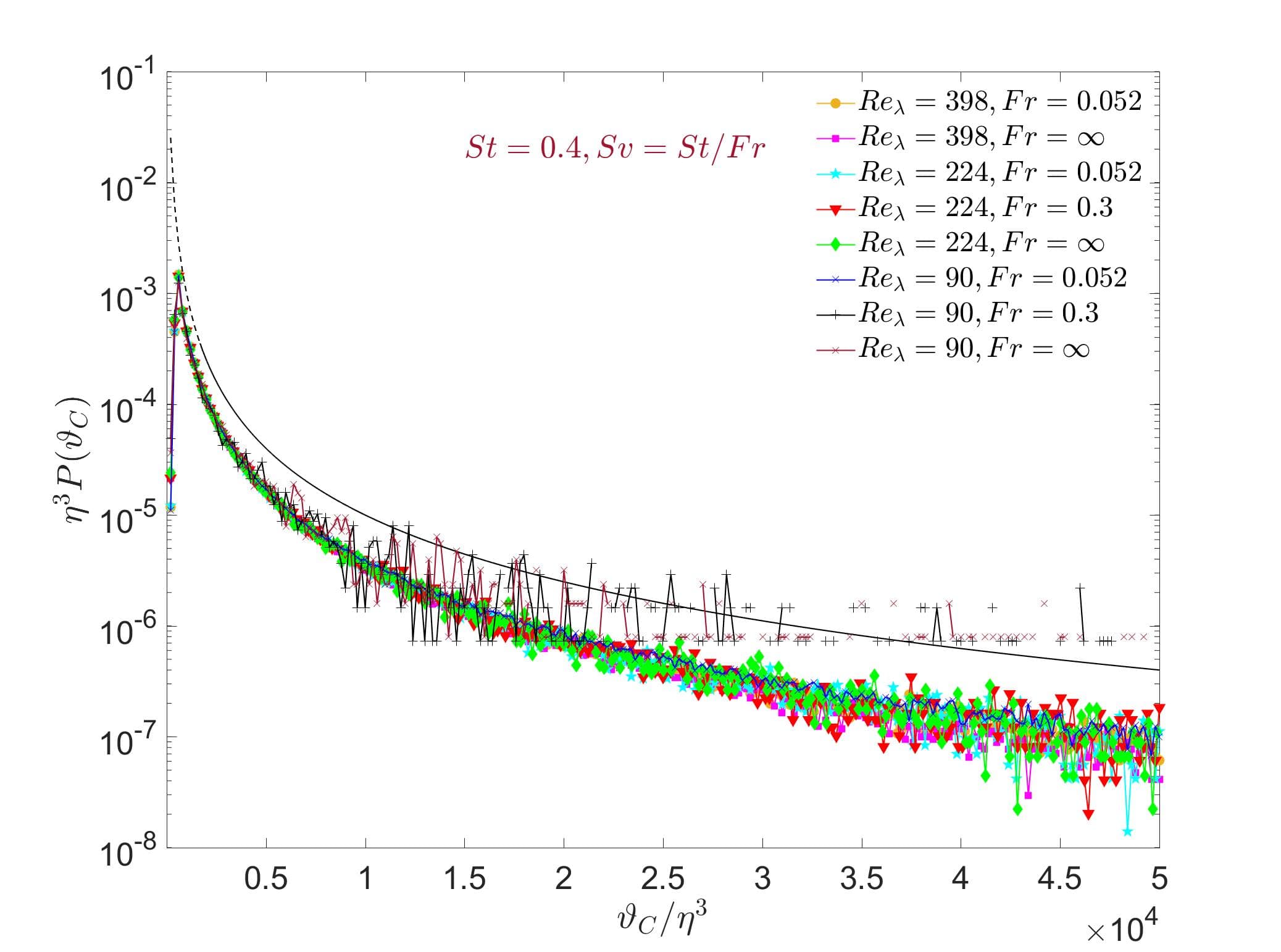}
		\caption{$St=0.4$}
	\end{subfigure}%
	\begin{subfigure}[b]{0.5\linewidth}
		\includegraphics[width=\linewidth]{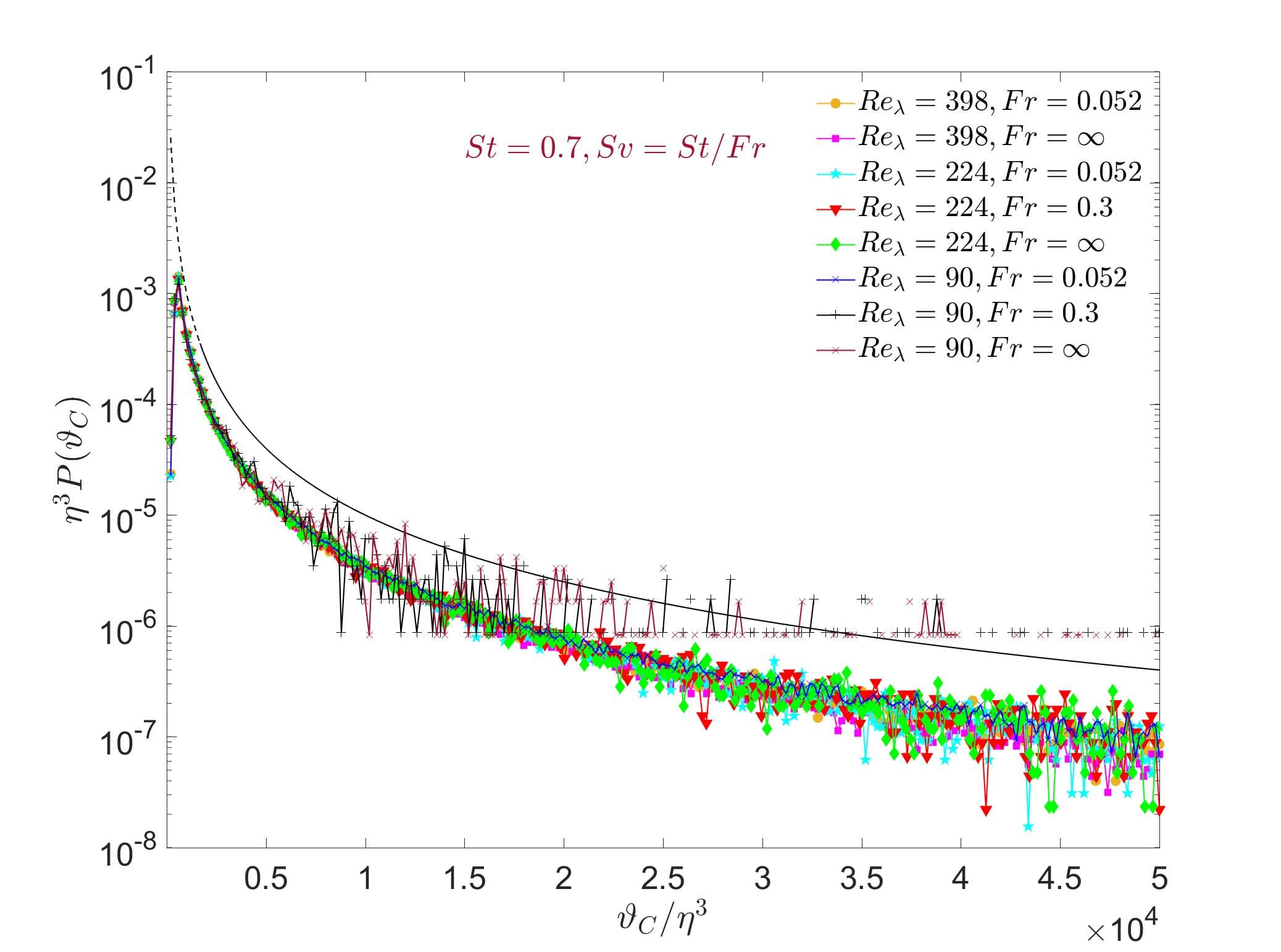}
		\caption{$St=0.7$ }
	\end{subfigure}
	
	\begin{subfigure}[b]{0.5\linewidth}
		\includegraphics[width=\linewidth]{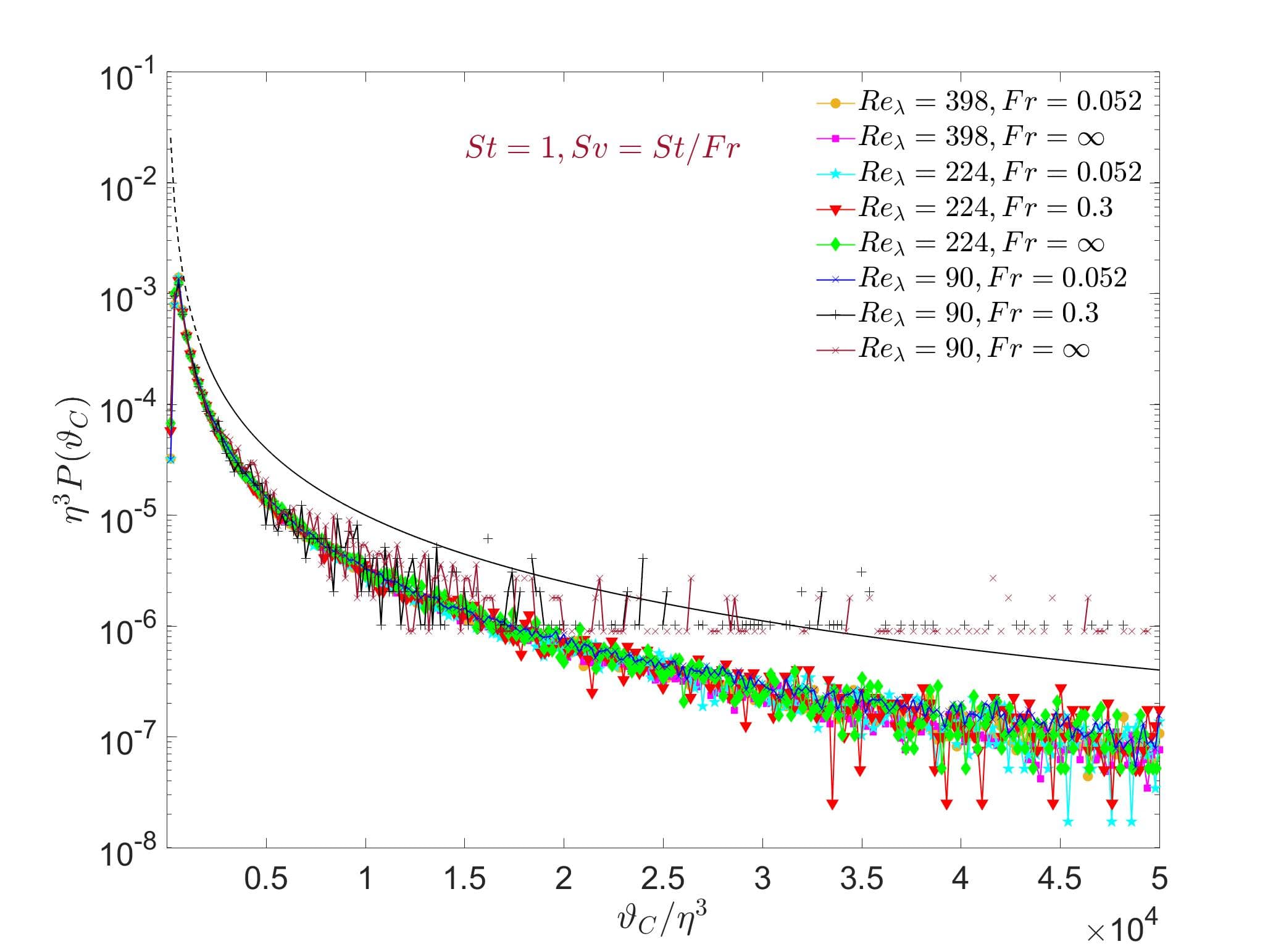}
		\caption{$St=1$}
	\end{subfigure}%
	\begin{subfigure}[b]{0.5\linewidth}
		\includegraphics[width=\linewidth]{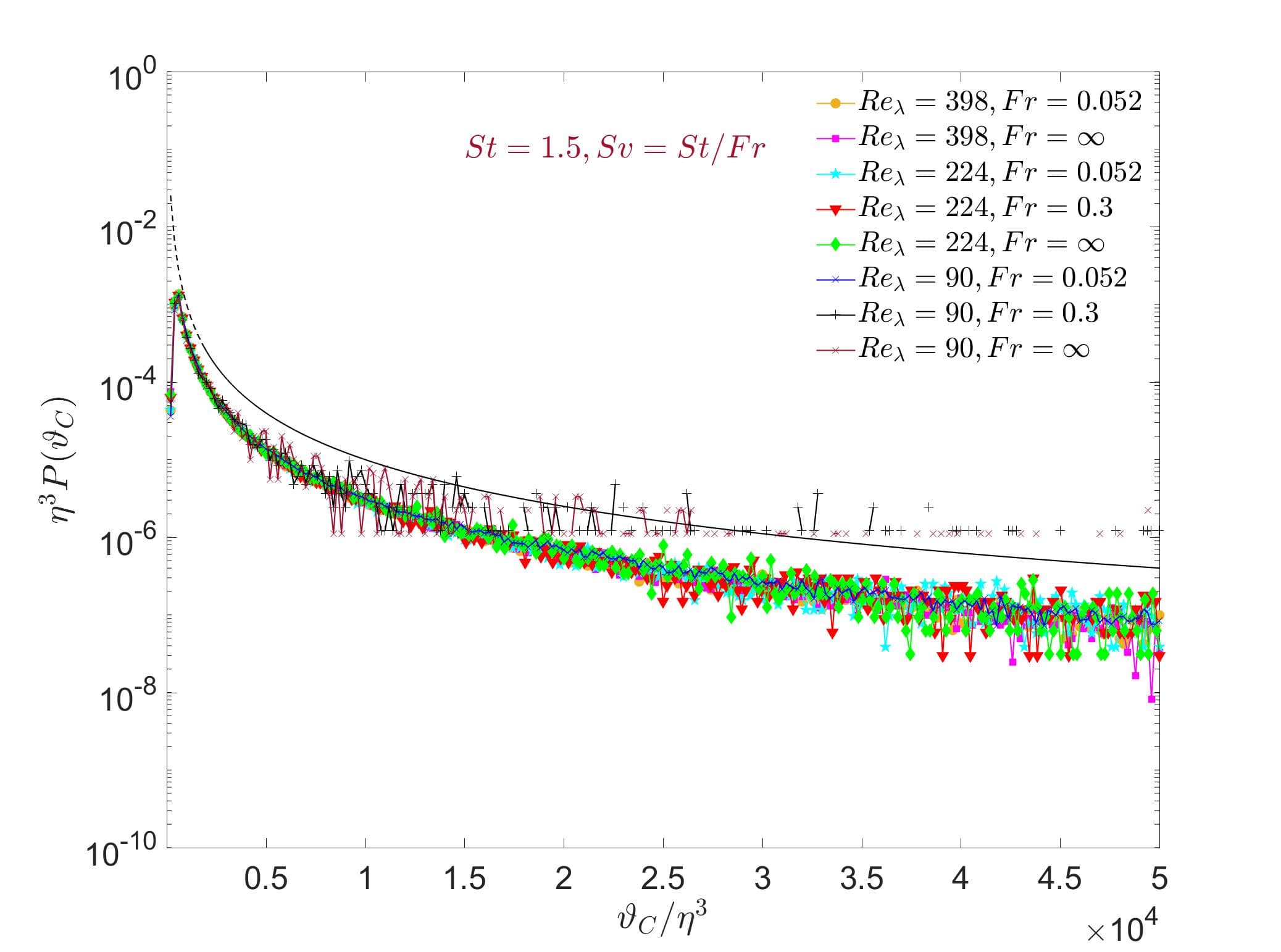}
		\caption{$St=1.5$ }
	\end{subfigure}
	\begin{subfigure}[b]{0.5\linewidth}
		\includegraphics[width=\linewidth]{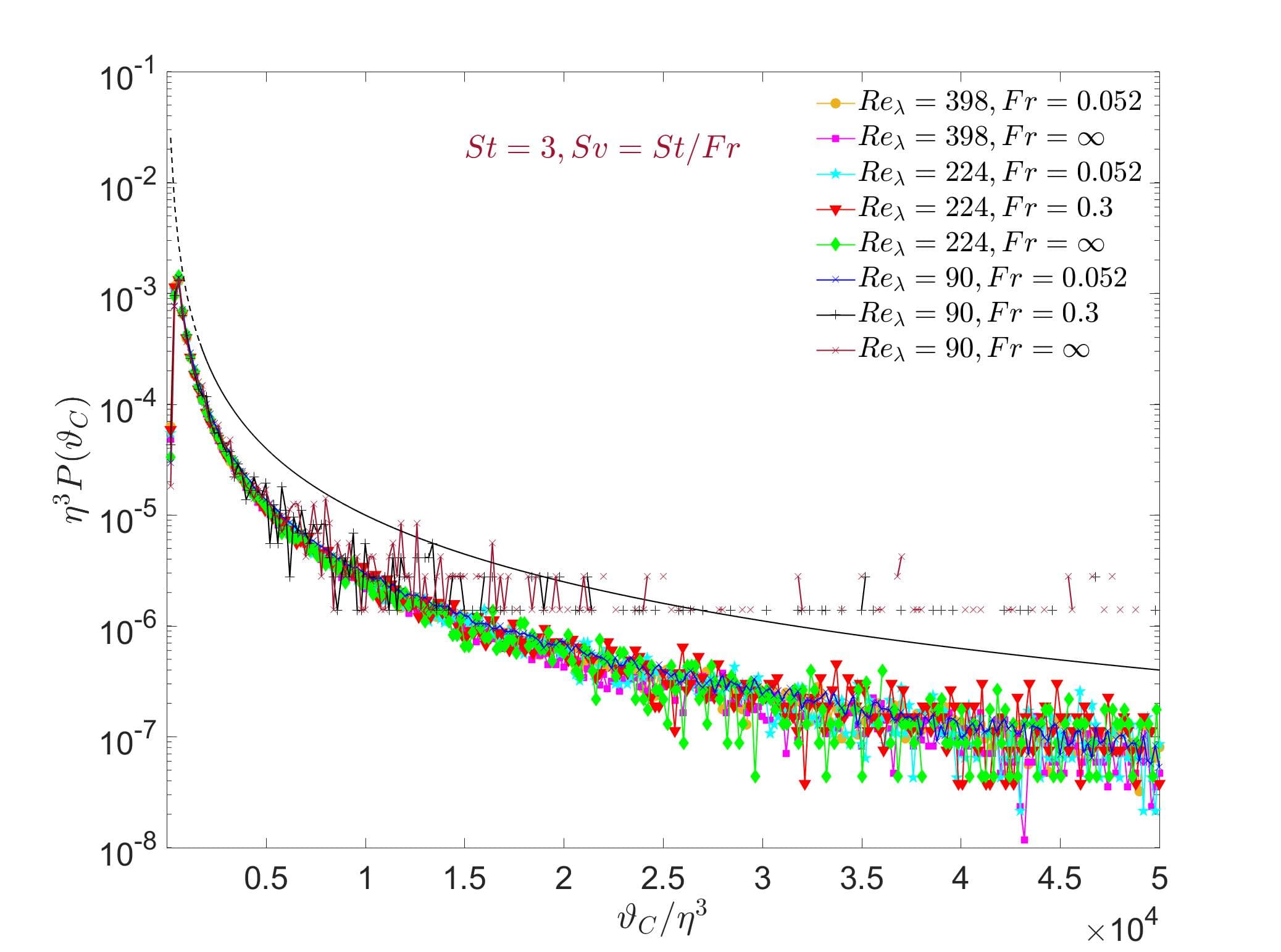}
		\caption{$St=3$ }
	\end{subfigure}%
	\caption{PDF of the coherent cluster volumes (normalized by the Kolmogorov scale) at different cases of $Fr$ and $R_\lambda$ combinations for (a) $St=0.4$, (b) $St=0.7$, (c) $St=1$, (d) $St=1.5$, and (e) $St=3$. Different colors represents different cases. The dashed line represents a power law with exponent $-2$.}\label{fig:VT_CC_PDF_Vol_Normalized_eta3_semilogy_AllCaseFigure_St}
\end{figure}

The results in Figure \ref{fig:VT_CC_PDF_Vol_Normalized_eta3_AllCaseFigure_St} reveal a much weaker dependence on $St, R_\lambda$ and $Fr$ than the PDF of the Vorono\text{\"i} volumes considered earlier. Figure \ref{fig:VT_CC_PDF_Vol_Normalized_eta3_semilogy_AllCaseFigure_St} shows the same results in a log-lin plot to emphasize the behavior for larger clusters, and again we find a weak dependence on $St, R_\lambda$ and $Fr$. One possible reason for the weak dependence is related to the fact that given that the average inter-particle distance is $7.9\eta$, the clusters are quite large and so the effects of inertia are weaker than they would be for smaller clusters. Indeed, the study of \cite{baker2017coherent} had a much smaller average inter-particle distance of $\approx 2.2\eta$, and in their results the strongest effects of the particle inertia were found for the smallest clusters, as expected.

\FloatBarrier

\begin{figure}
\hspace{15mm}
	\includegraphics[width=0.7\linewidth]{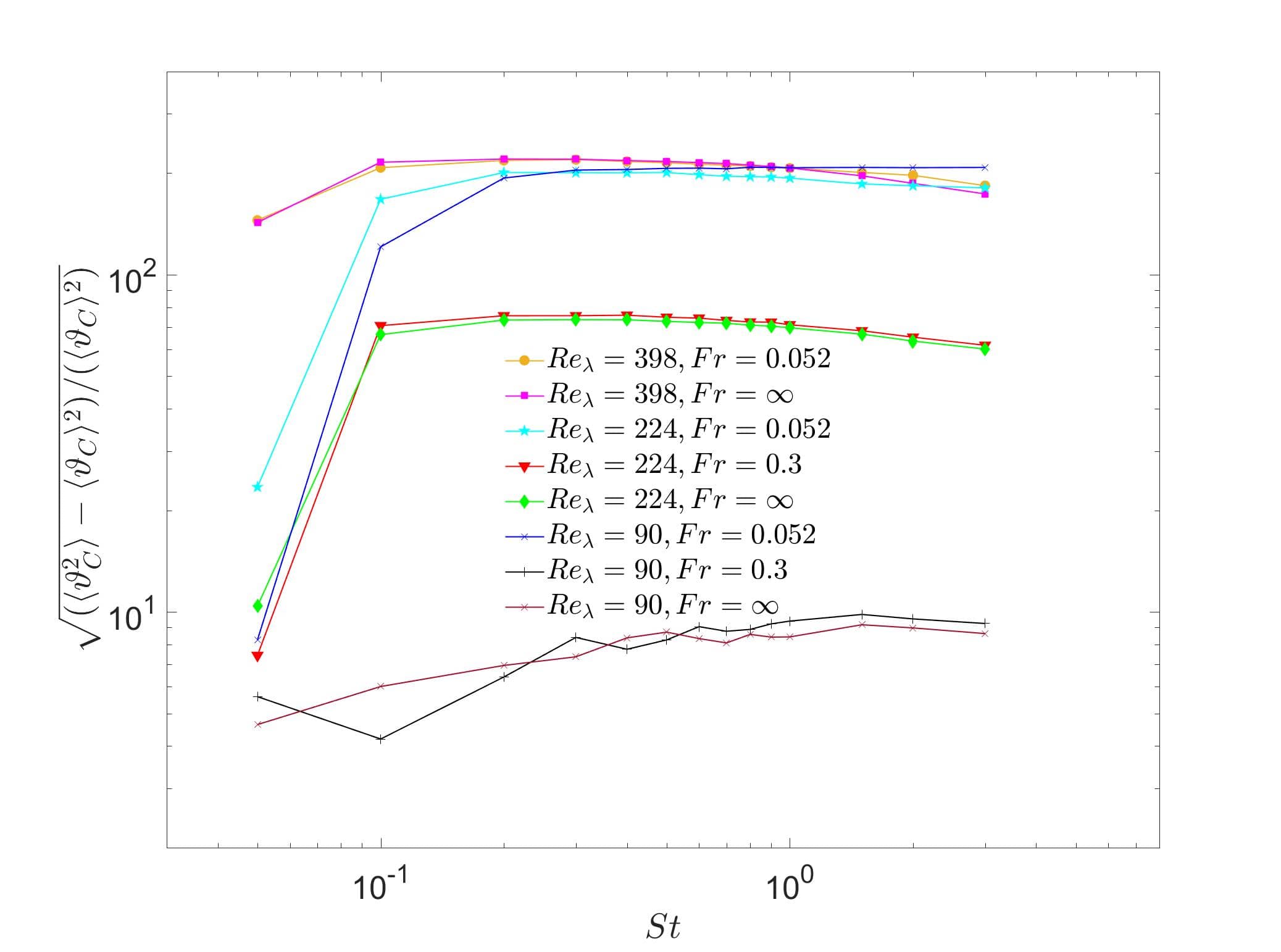}
	\caption{Standard deviation of Vorono\text{\"i} cells of clusters as a function of $St$.}
	\label{fig:CF_50p00_VT_CC_SD_Vol_Normalized_mean2}
\end{figure}

To obtain clearer quantitative insight, the standard deviation (s.d.) of the coherent cluster volumes is shown in the figure \ref{fig:CF_50p00_VT_CC_SD_Vol_Normalized_mean2}. The results reveal a quite weak dependence on $St$, which was also reported in the experimental study of \cite{sumbekova2017preferential}. This is in striking contrast to the results for the s.d. of the Vorono\text{\"i} volumes shown in Figure~\ref{fig:VT_Var_Vol_norm_mean}, for which a very strong dependence on $St$ was observed. However, the effect of $Fr$ and $R_\lambda$ on the s.d. of the coherent cluster volumes is much stronger. In particular, the s.d. increases strongly with increasing $R_\lambda$, while it increases only slightly in going from $Fr=\infty$ to 0.3, and then increases substantially in going from $Fr=0.3$ to 0.052.


%
\FloatBarrier

\begin{figure}
\hspace{15mm}
	\includegraphics[width=0.7\linewidth]{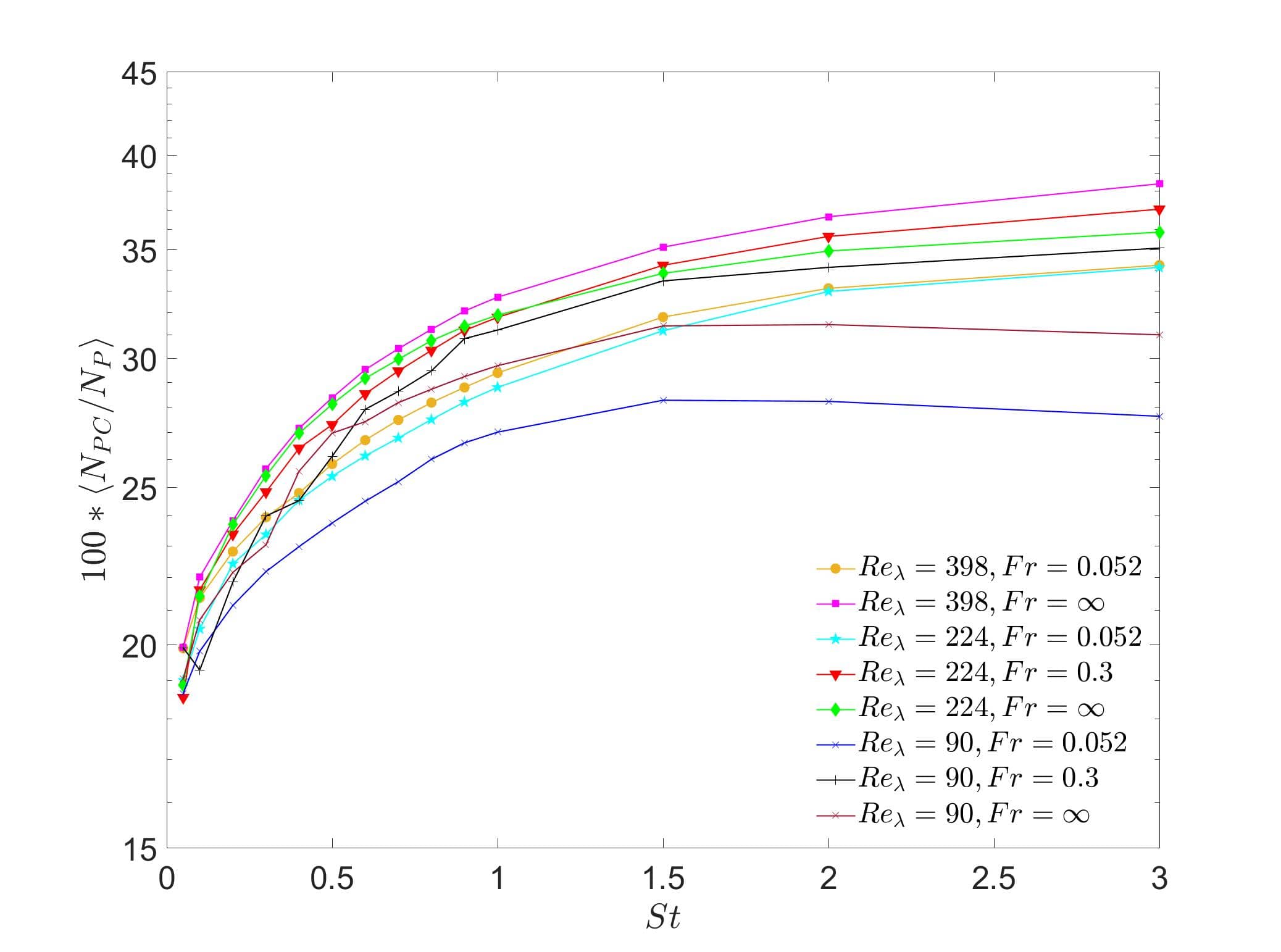}
	\caption{Average percentage of total particles in clusters to the total number of particles in the whole domain, as a function of $St$ and for different cases}
	\label{fig:Fraction_of_particles_belong_to_CC}
\end{figure}

In figure \ref{fig:Fraction_of_particles_belong_to_CC}, the average percentage of particles within clusters compared with the number in the whole domain is shown. In agreement with \cite{baker2017coherent}, we observe that this percentage increases with increasing $St$, though for $R_\lambda$ this reaches a maximum at $St\approx 2$ and then reduces. We expect that a maximum would also be observed for the higher $R_\lambda$, but at larger $St$ (a maximum must occur since in the limits $St\to 0$ and $St\to\infty$ the percentage must go to zero). The results show that generally, increasing $R_\lambda$ leads to an increase of the percentage, while the dependence on $Fr$ is non-monotonic, with the percentage generally being largest for the intermediate value $Fr=0.3$. It is important to note, however, that the percentages we observe are much larger than those observed by \cite{baker2017coherent}. This difference is likely due to the fact that in their simulations the average inter-particle distance was $2.2\eta$, while in ours it has the significantly larger value $7.9\eta$. If this indeed the explanation, then it shows that the significance of the particle clusters, in terms of the number of particles contained within them as a fraction of the total number in the flow, depends essentially on the particle loading in the flow.

\FloatBarrier

\begin{figure}
	\vspace{-0.7in}
	\centering
	\begin{subfigure}[b]{0.5\linewidth}
		\includegraphics[width=\linewidth]{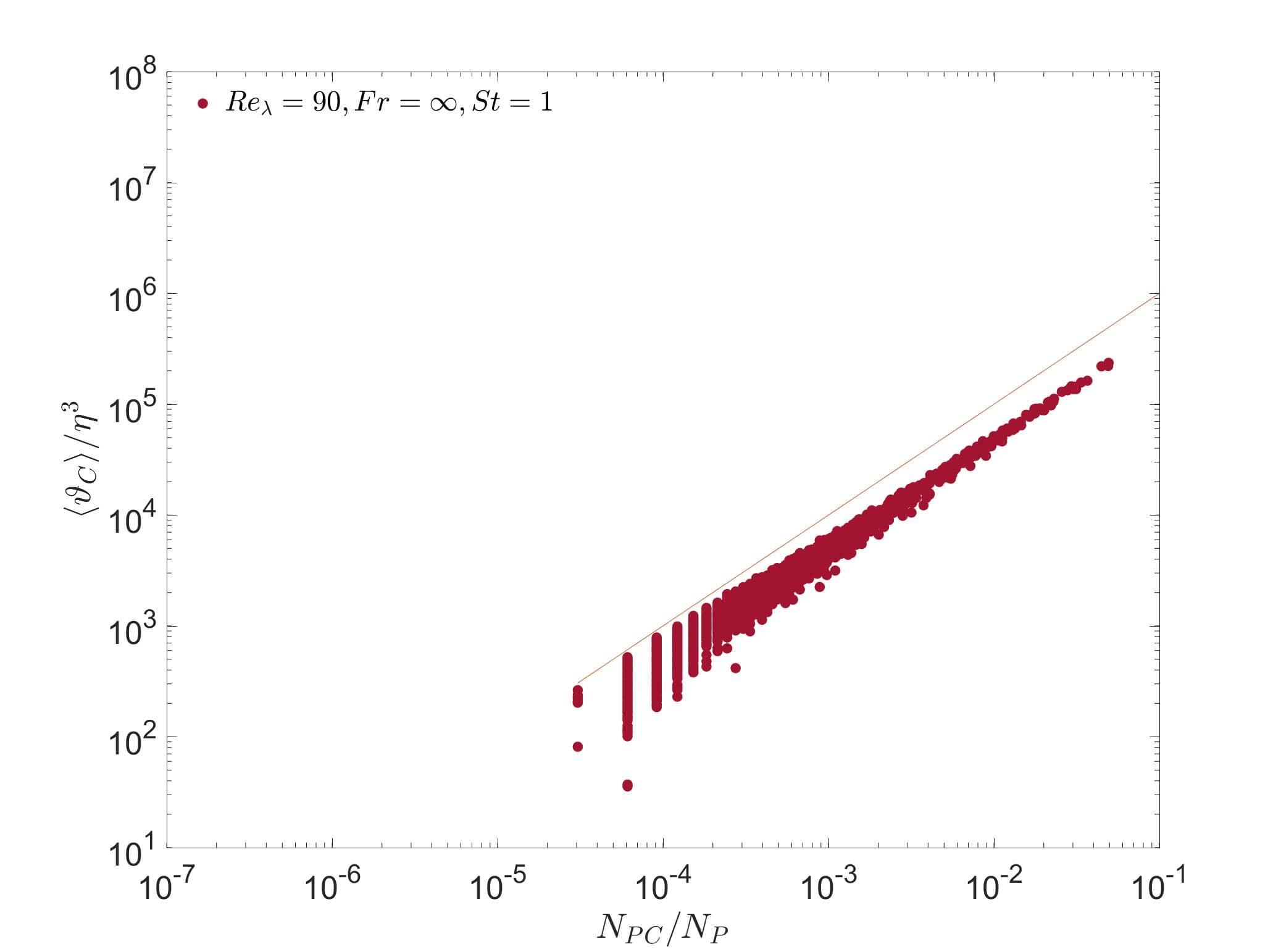}
		\caption{ $Re=90 \: \& \: Fr=\infty$}
	\end{subfigure}%
	\begin{subfigure}[b]{0.5\linewidth}
		\includegraphics[width=\linewidth]{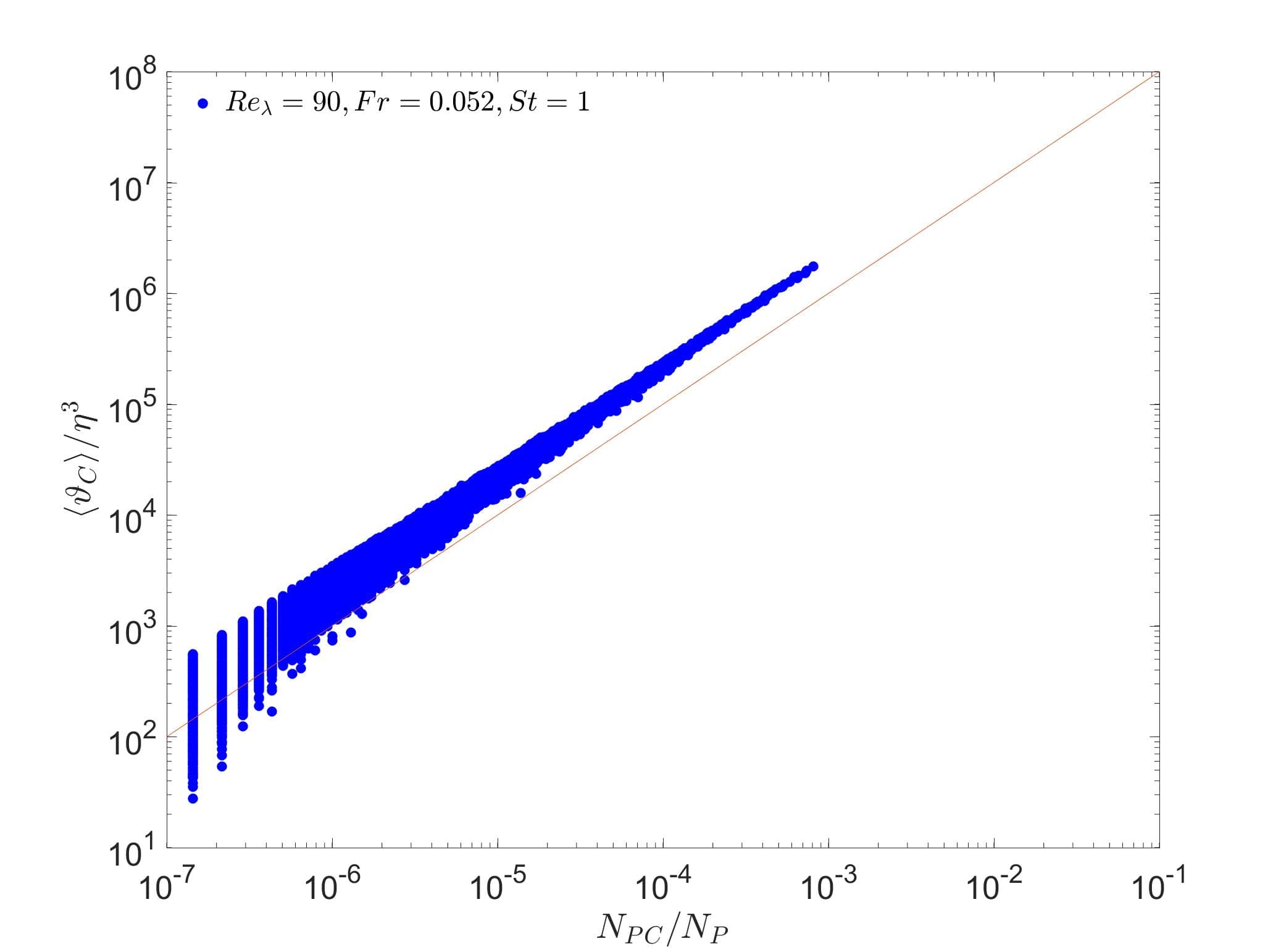}
		\caption{ $Re=90 \: \& \: Fr=0.052$ }
	\end{subfigure}
	
	\begin{subfigure}[b]{0.5\linewidth}
		\includegraphics[width=\linewidth]{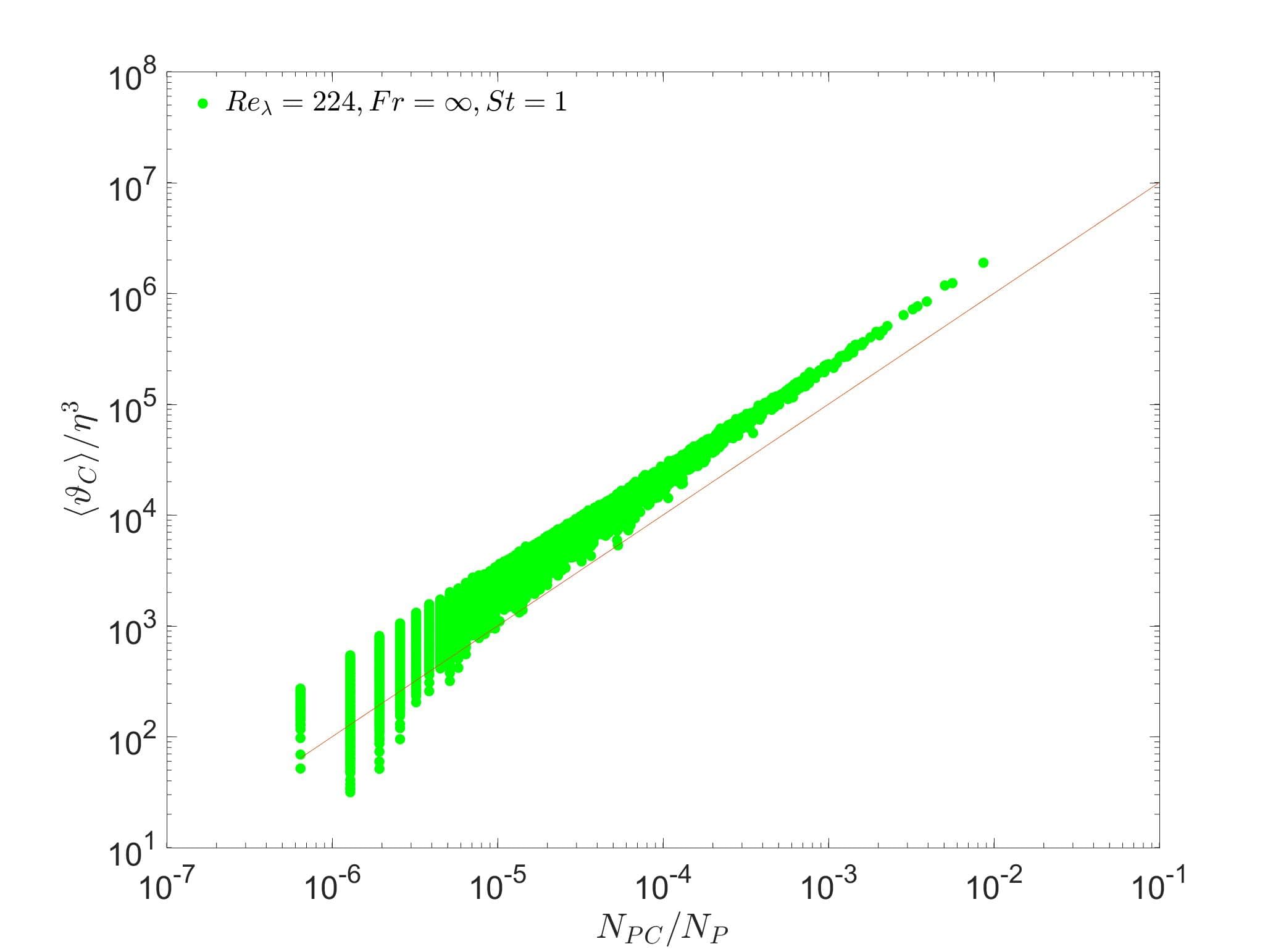}
		\caption{$Re=224 \: \& \: Fr=\infty$}
	\end{subfigure}%
	\begin{subfigure}[b]{0.5\linewidth}
		\includegraphics[width=\linewidth]{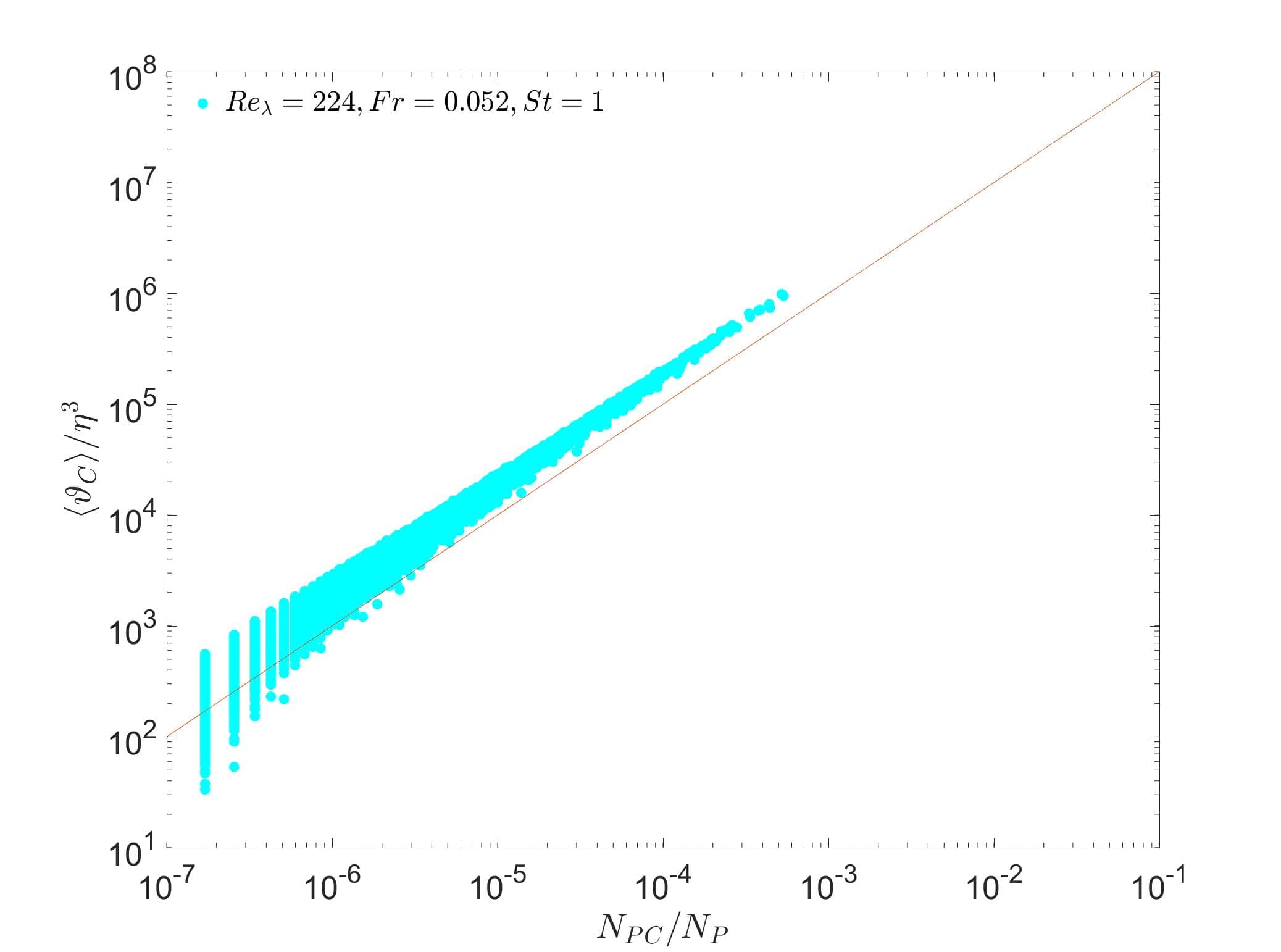}
		\caption{ $Re=224 \: \& \: Fr=0.052$ }
	\end{subfigure}
	\begin{subfigure}[b]{0.5\linewidth}
		\includegraphics[width=\linewidth]{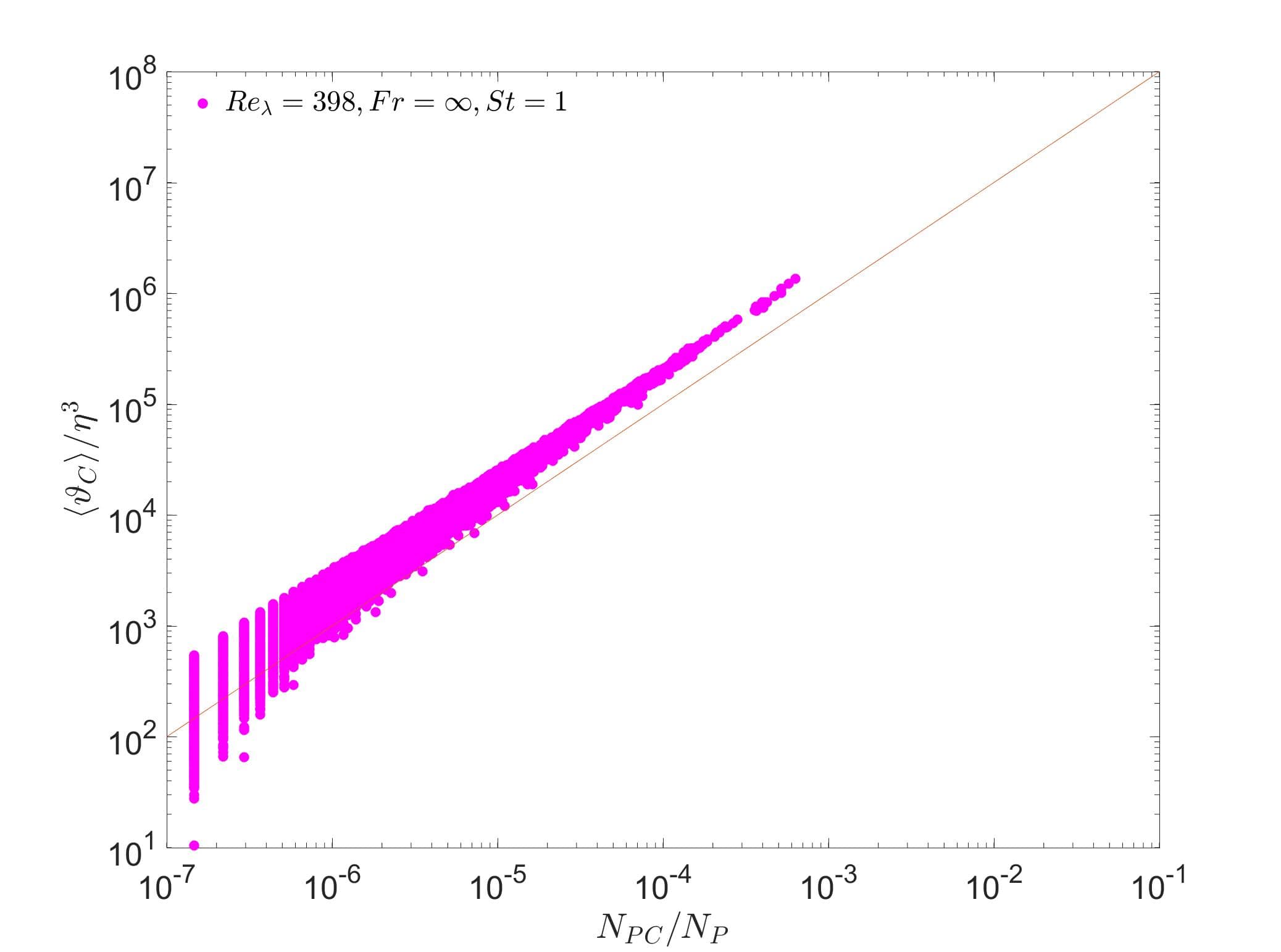}
		\caption{$Re=398 \: \& \: Fr=\infty$ }
	\end{subfigure}%
	\begin{subfigure}[b]{0.5\linewidth}
		\includegraphics[width=\linewidth]{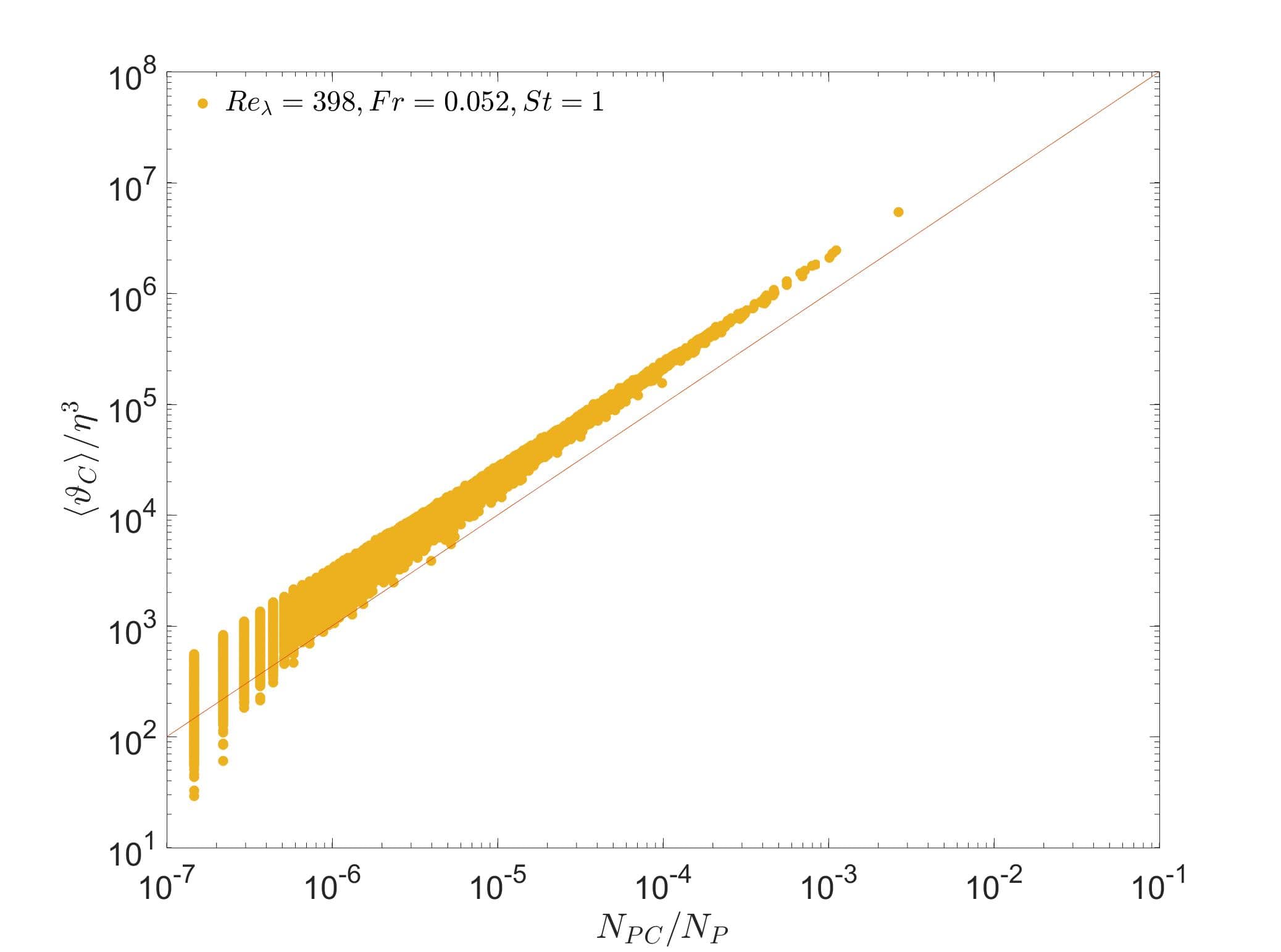}
		\caption{ $Re=398 \: \& \: Fr=0.052$ }
	\end{subfigure}
	\caption{The cluster volume normalized by the Kolmogorov scale versus particle counts in clusters normalized by the whole particles in the domain, for $St=1$.}\label{fig:St_1_ClusterVol_Norm_etacubed_ParticlePerCluster_Norm_TPD_OneCaseAllSt_loglog}
\end{figure}

Figure \ref{fig:St_1_ClusterVol_Norm_etacubed_ParticlePerCluster_Norm_TPD_OneCaseAllSt_loglog} shows results for the average cluster size as a function of the number of particles in the clusters, which provides information on the relationship between the coherent cluster sizes and the particle concentrations within them. (For brevity we only show the results for $St=1$, but the behavior is the same for the other $St$ cases). This relationship is found to be approximately linear, and is independent of all the control parameters over the range we have explored. Such a linear relationship, and its lack of sensitivity to $St$ and $Fr$, were already reported in \cite{baker2017coherent} for $R_\lambda=65$, however, our results confirm that this quantity is also independent of $R_\lambda$. 
\FloatBarrier
\begin{figure}
	\includegraphics[width=0.7\linewidth]{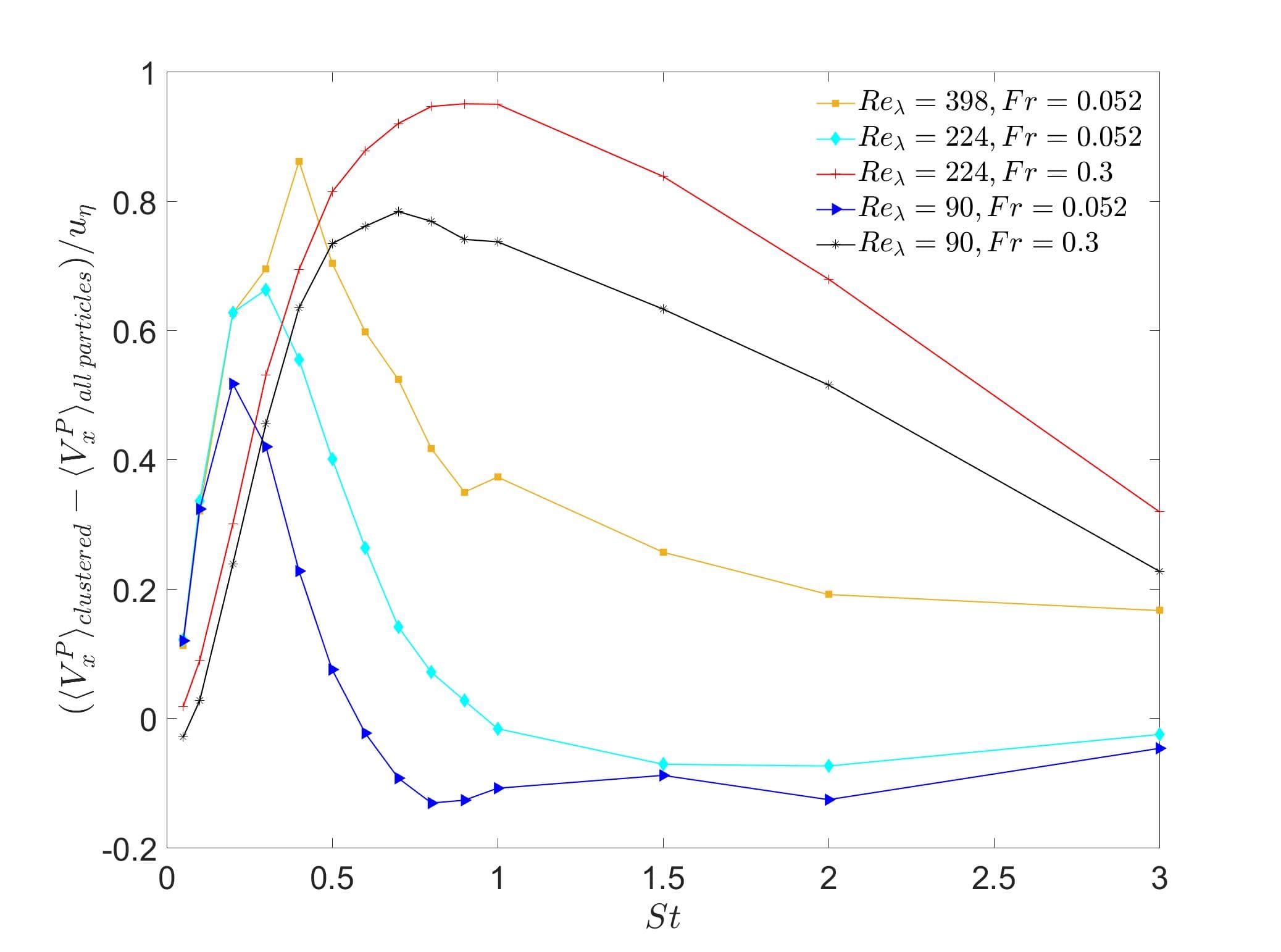}
	\caption{Difference between the average velocity of particles within the clusters and the whole particles in the domain.}
	\label{fig:VT_CC_PVel_x_Normalized_NotAll_Difference}
\end{figure}
\FloatBarrier
Finally, we turn to consider the velocity statistics of the particles in coherent clusters as compared with those in the entire domain. In figure \ref{fig:VT_CC_PVel_x_Normalized_NotAll_Difference} we show results for the difference in the average velocity of particles in clusters to that based on all particles in the flow.  In agreement with those of \cite{baker2017coherent}, our results show that the mean settling velocity of particles in clusters can be significantly larger than than based upon averaging over all particles in the flow. Our results also show that this difference increases with increasing $R_\lambda$, with a non-monotonic dependence on $Fr$ (since the results are zero for $Fr=\infty$). The average settling velocity of all particles is known to increase with increasing $R_\lambda$, which is principally due to the enhanced range of scales for the particles to interact with as $R_\lambda$ grows \cite{tom2019}. However, the $R_\lambda$ dependence of the results in figure \ref{fig:VT_CC_PVel_x_Normalized_NotAll_Difference} would seem to be due to something more subtle to this, since all particles in the flow, whether clustered or not, would be subject to the enhanced range of scales available for the particles to interact with as $R_\lambda$ grows. 

\begin{figure}
\hspace{15mm}
	\includegraphics[width=0.7\linewidth]{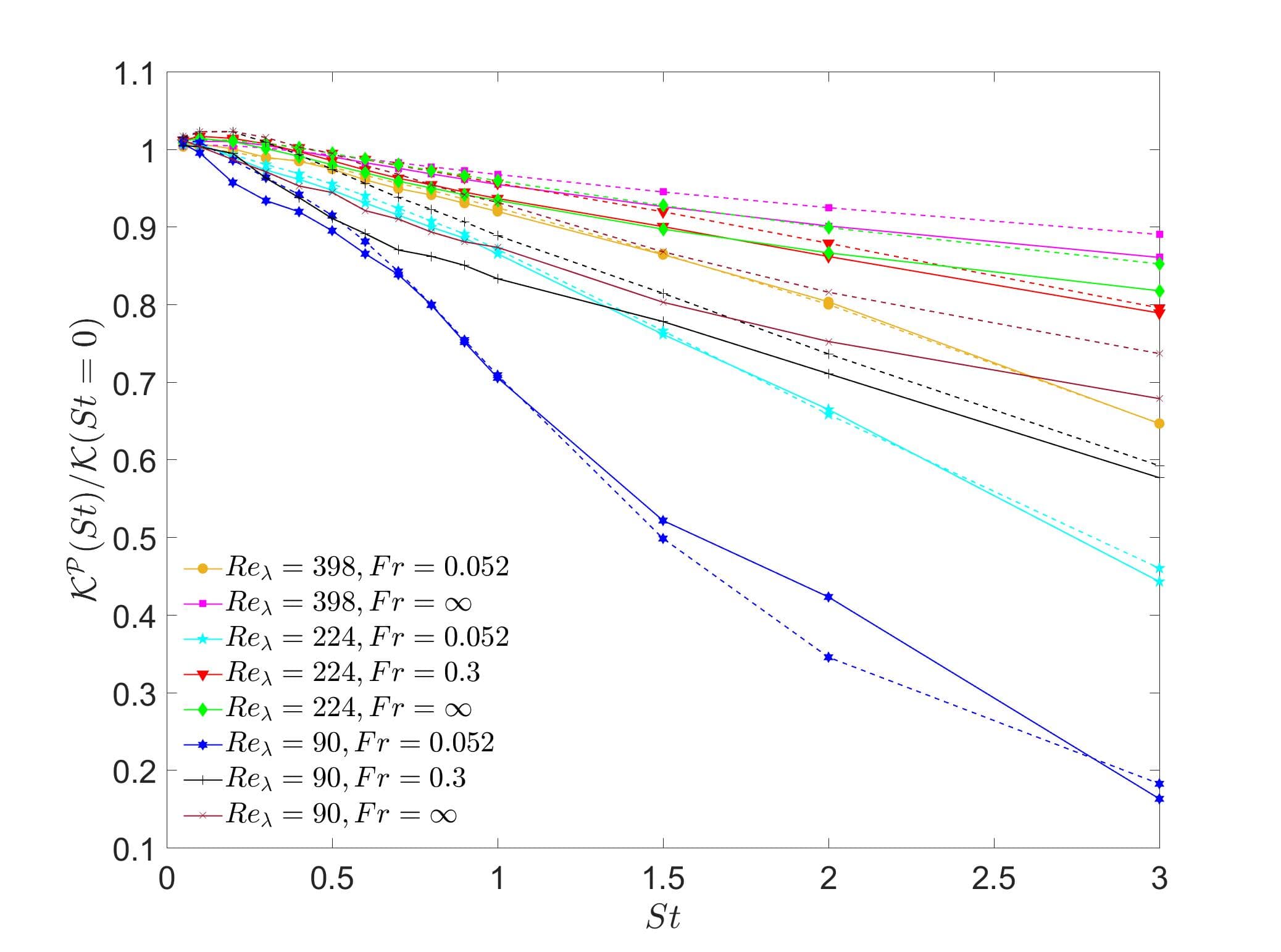}
	\caption{Average kinetic energy of particles within the clusters (solid lines) and the whole particles in the domain (dashed lines).}
	\label{fig:VT_CC_PKE_Correct_Normalized_PKE_St0}
\end{figure}

The results for the particle fluctuating kinetic energy $\mathcal{K}(St)\equiv (1/2)\langle \|\mathbf{v}^p(t)-\langle\mathbf{v}^p(t)\rangle\|^2\rangle$ are shown in figure \ref{fig:VT_CC_PKE_Correct_Normalized_PKE_St0}. The results show that while the settling velocity of particles in coherent clusters is significantly different from that of all the particles in the flow, there is only a small difference for their kinetic energy. In the absence of gravity, the particles in coherent clusters have slightly less kinetic energy compared with the average based on all particles in the flow. The most likely explanation for this is that the difference is caused by the preferential sampling of the turbulent flow, which is known to be suppressed in the presence of gravity \cite{ireland2016effectb}. As explained in \cite{ireland2016effecta}, the decrease of the kinetic energy with increasing $St$ is due to the inertial particles filtering out fluctuations in the turbulent flow, while the increase of the kinetic energy with increasing $R_\lambda$ (for a fixed $St$ and $Fr$) is due to a reduction of the filtering effect owing to the increased timescale separation in the turbulent flow as $R_\lambda$ is increased. Finally, the decrease of the kinetic energy as $Fr$ is decreased (for a fixed $St$ and $R_\lambda$) is because the filtering effect is enhanced due to the timescale of the fluid velocity seen by the particle becoming shorter as the particles settle faster. 

\FloatBarrier
%

%
%
%
%
%
%
%
%
%
%

%
\section{Conclusions}

In this work, we have used DNS and three-dimensional Vorono\text{\"i} analysis to explore the local distribution, settling velocity and acceleration of inertial particles (in the point particle limit) in statistically stationary, isotropic turbulence. In these simulations, we independently varied the Taylor Reynolds number $R_\lambda \in[90,398]$, Froude number $Fr\equiv a_\eta/g\in[0.052,\infty]$ (where $a_\eta$ is the Kolmogorov acceleration, and $g$ is the acceleration due to gravity), and Kolmogorov scale Stokes number $St\equiv\tau_p/\tau_\eta\in[0,3]$. Independently varying these parameters was not possible in previous experimental investigations on the problem. The average inter-particle distance was held fixed in all the DNS for consistency when comparing the Vorono\text{\"i} analysis results for different $R_\lambda$.

In agreement with previous results using the Radial Distribution Function (RDF) to quantify particle clustering \cite{ireland2016effecta,ireland2016effectb}, which is a global measure, we find that for small Vorono\text{\"i} volumes (corresponding to the most clustered particles), the behavior is strongly dependent upon $St$ and $Fr$, but only weakly dependent upon $R_\lambda$, unless $St>1$. However, larger Vorono\text{\"i} volumes (void regions) exhibit a much stronger dependence on $R_\lambda$, even when $St\leq 1$, and we show that this, rather than the behavior at small volumes, is the cause of the sensitivity of the standard deviation of the Vorono\text{\"i} volumes that has been previously reported. Our results also show that the standard deviation of the Vorono\text{\"i} volumes is dominated by the behavior of the void regions, rather than the clustered regions. As a result, the results show that the standard deviation of the Vorono\text{\"i} volumes is reduced by gravity at all $St$, even though the Probability Density Function (PDF) results show that gravity can enhance the small-scale clustering, as was also observed using the RDF analysis in \cite{ireland2016effectb}. This is because gravity seems to always suppress the void regions. This highlights that using the standard deviation of Vorono\text{\"i} volumes as a measure of particle clustering (which has been done in many previous studies) can be somewhat misleading, since its properties are dominated by void, rather than small-scale clustered regions of the flow. We also show that the PDF of Vorono\text{\"i} volumes exhibits a quasi-lognormal behavior over a certain range of volumes, as previously reported. However, we show that this is only approximate, and does not describe well Vorono\text{\"i} volumes that are sufficiently large or small. The validity of the log-normal assumption also depends in a rather complex way on $R_\lambda$ and $Fr$.

We find that the average settling velocities conditioned on the Vorono\text{\"i} volumes exhibits a non-monotonic dependence on the Vorono\text{\"i} volumes, with the largest contribution to the particle settling velocities being associated with increasingly larger Vorono\text{\"i} volumes as the settling parameter $Sv\equiv St/Fr$ is increased. This non-monotonic behavior does not seem to have been previously reported, likely because the $Sv$ considered were not sufficiently large. It can be explained by the recent work of \cite{tom2019} who show theoretically and numerically that as $Sv$ is increased, the scales responsible for the particle settling velocities shift to larger scales.

Even though the globally averaged particle acceleration is zero in the system we are considering, our local analysis of the acceleration statistics of settling inertial particles shows that clustered particles experience a net acceleration in the direction of gravity, while particles in void regions experience the opposite. In the direction normal to gravity, and in the absence of gravity, the average particle accelerations are independent of the Vorono\text{\"i} volumes. The particle acceleration variance, however, is a convex function of the Vorono\text{\"i} volumes, with or without gravity, which seems to indicate a non-trivial relationship between the Vorono\text{\"i} volumes and the sizes of the turbulent flow scales. Results for the variance of the fluid acceleration at the inertial particle positions are of the order of the square of the Kolmogorov acceleration and depend only weakly on Vorono\text{\"i} volumes. These results call into question the ``sweep-stick'' mechanism for particle clustering in turbulence which would lead one to expect that clustered particles reside in regions where the fluid acceleration is zero \cite{coleman09}. 

We then consider the properties of particles in clusters, which are regions of connected Vorono\text{\"i} cells whose volume is less than a certain threshold. The results for the PDFs of the cluster volumes reveal significant self-similarity of the clusters. The standard deviation of the cluster volumes provides further insight and shows that the statistics of the cluster volumes depends only weakly on $St$, with a stronger dependance on $Fr$ and $R_\lambda$. The weak dependence upon $St$ may however be due to the fact that given the average inter-particle distance in our simulations, the cluster sizes are comparable to scales of the flow where one might expect the effects of particle inertia to be weak anyway. Finally, we compared the average settling velocities of all particles in the flow with those in clusters, and showed that those in the clusters settle much faster, in agreement with previous work. However, we also find that this difference grows significantly with increasing $R_\lambda$ and exhibits a non-monotonic dependence on $Fr$. The kinetic energy of the particles, on the other hand, is almost the same for particles in and not in the clusters, especially in the presence of gravity.

\section{Acknowledgments}
This work used the Extreme Science and Engineering Discovery Environment (XSEDE), which is supported by National Science Foundation grant number ACI-1548562 \cite{xsede}. Specifically, the Comet cluster was used under allocation CTS170009.
The Comet cluster and the Duke Computing Cluster (DCC) were used extensively for the computations in this study. The authors would like to express their sincere thanks and gratitude to Tom Milledge and Roberto E. Gonzalez for their generous helps with computational challenges of this work. We also would like to thank Yuanqing Liu for all his input during the discussion on the coding issues.

\bibliographystyle{unsrt}
\bibliography{jfm-instructions}

\end{document}